\documentclass{article}
\usepackage{amsmath}
\usepackage{amsfonts}
\usepackage{graphicx}
\setkeys{Gin}{width=\linewidth,totalheight=\textheight,keepaspectratio}
\graphicspath{{figures/}}
\usepackage{bm}
\usepackage{booktabs}
\usepackage{units}
\usepackage{tabularx}
\usepackage{fancyvrb}
\fvset{fontsize=\normalsize}
\newcommand\blfootnote[1]{%
  \begingroup
  \renewcommand\thefootnote{}\footnote{#1}%
  \addtocounter{footnote}{-1}%
  \endgroup
}

\usepackage{algorithm}
\usepackage{algpseudocode}

\algblock{Input}{EndInput}
\algnotext{EndInput}
\algblock{Output}{EndOutput}
\algnotext{EndOutput}
\newcommand{\Desc}[2]{\State \makebox[10em][l]{#1}#2}

\usepackage{multicol}
\usepackage{xcolor,soul}
\usepackage{caption,subcaption,authblk}
\usepackage[
backend=biber,
style=numeric-comp,
sorting=ynt
]{biblatex}
\AtEveryBibitem{
    \clearfield{urlyear}
    \clearfield{urlmonth}
    \clearfield{urldate}
}
\usepackage{filecontents}

\usepackage{subfiles}

\newcommand{\norm}[1]{\left|\left|#1\right|\right|}
\newcommand{\abs}[1]{\left|#1\right|}
\newcommand{\prt}[1]{\left(#1\right)}

\newcommand{\pd}[2]{\dfrac{\partial #1}{\partial #2}}

\newcommand{\dg}{$^\circ$}
\newcommand{\eifs}{\left(\Phi,\Psi,U\right)}

\newcommand{\sample}[1]{\{#1\}_{i=1}^{N_{p}}}
\newcommand{\eval}[1]{\{#1\}_{i=1}^{N_{e}}}
\newcommand{\Kcases}[1]{\{#1\}_{k=1}^{K}}

\newcommand{\plf}{p_{LF}}
\newcommand{\ulf}{u_{LF}}
\newcommand{\vlf}{v_{LF}}

\newcommand{\vecqhf}{\bm{q}}

\newcommand{\vecqlf}{\bm{q}_{LF}}

\newcommand{\vecqml}{\tilde{\bm{q}}}
\newcommand{\deltavecqml}{\delta{\tilde{\bm{q}}}}
\newcommand{\vecqfs}{\bm{q}_{FS}}
\newcommand{\vecqpofu}{\bm{q}_{POFU}}
\newcommand{\uml}{$\tilde{u}$}
\newcommand{\vml}{$\tilde{v}$}

\newcommand{\pml}{$\tilde{p}$}

\newcommand{\Rey}{$Re$}
\newcommand{\lossl}{\mathcal{L}}
\newcommand{\lossldata}{\mathcal{L_{\rm{data}}}}
\newcommand{\losslgrad}{\mathcal{L_{\rm{grad}}}}

\newcommand{\x}{\mathbf{x}}

\newcommand{\atxi}{\left(\x_{i}\right)}

\DeclareCaptionFormat{algor}{%
  \hrulefill\par\offinterlineskip\vskip1pt%
  \textbf{#1#2}#3\offinterlineskip\hrulefill}
\DeclareCaptionStyle{algori}{singlelinecheck=off,format=algor,labelsep=space}
\newcounter{parentalgorithm}

\makeatletter
\newenvironment{subalgorithms}{%
  \refstepcounter{algorithm}%
  \protected@edef\theparentalgorithm{\thealgorithm}%
  \setcounter{parentalgorithm}{\value{algorithm}}%
  \setcounter{algorithm}{0}%
  \def\thealgorithm{\theparentalgorithm\alph{algorithm}}%
  \ignorespaces
}{%
  \setcounter{algorithm}{\value{parentalgorithm}}%
  \ignorespacesafterend
}
\makeatother

\begin{document}
\title{Multi-Fidelity Machine Learning Applied to Steady Fluid Flows}

\author{Kazuko W. Fuchi$^1$, Eric M. Wolf$^2$, David S. Makhija$^3$, Christopher R. Schrock$^4$, and Philip S. Beran$^3$ }
\date{%
    $^1$University of Dayton Research Institute\\%
    $^2$Ohio Aerospace Institute\\%
    $^3$Lateral Unbounded\\%
    $^4$Air Force Research Laboratory\\[2ex]%
}

\maketitle

\abstract{A machine learning method to predict steady external fluid flows using elliptic input features is introduced. Using data from as few as one high-fidelity simulation, the proposed method produces models generalizable under changes to boundary geometry by using solutions to elliptic boundary value problems over the flow domain as the model input, instead of Cartesian coordinates of the domain. Training data is generated through pointwise evaluation of flow features at points selected through a quad-tree adaptive sampling method to concentrate training points in areas with large field gradients. Models are trained within a training window around the body, while predictions are smoothly extended to freestream conditions using a Partition-of-Unity extension. Predictive capabilities of the machine learning model are demonstrated in steady-state flow of incompressible fluid around a cylinder and a Joukowski airfoil. The predicted flow field is used to warm-start CFD simulations to achieve acceleration in solver convergence.
}


\blfootnote{\noindent~Distribution Statement A: Approved for Public Release; Distribution is Unlimited. PA\# AFRL-2022-2754.}
\blfootnote{\noindent~This is an Accepted Manuscript of an article published by Taylor \& Francis in International Journal of Computational Fluid Dynamics on 10 February 2023, available at: https://doi.org/10.1080/10618562.2022.2154758.}
\blfootnote{\noindent~Please cite as: Fuchi, K. W., Wolf, E. M., Makhija, D. S., Schrock, C. R., \& Beran, P. S. (2022). Multi-Fidelity Machine Learning Applied to Steady Fluid Flows. \textit{International Journal of Computational Fluid Dynamics}, 36(7), 618–640. https://doi.org/10.1080/10618562.2022.2154758} 


\section{Introduction}\label{sec:intro}

It is of great interest to develop machine learning (ML) methods to aid the use of computational fluid dynamics (CFD) in application to aerospace design optimization. In this context, a high-fidelity CFD solution must be obtained at potentially many points across a large design space. The overall computational cost of the optimization may be lowered by finding a low-dimension design space representation, by reducing the number of high-fidelity solutions required by the optimizer, or by reducing the cost of individual solutions, with various ML approaches in each category discussed in the literature~\cite{Li2022}. In this work, we propose an approach falling in the third category, where an ML model reduces the cost of a high-fidelity solution by making flow field predictions used for CFD initialization. Initial conditions are crucial in non-linear and time dependent problems. Naive selection of initial conditions through freestream idealizations or previous design solution fields often lead to slow non-linear solver convergence, non-linear solver divergence, or unnecessarily long time integration to steady conditions. This research provides a mathematical paradigm to efficiently predict initial condition fields before the solution is computed, thus mitigating non-linear solver and time-integration difficulties. 

Machine learning has been applied to the representation of fluid flow fields using data-driven and physics-informed approaches~\cite{Brunton2020,Karniadakis2021}. In data-driven approaches, supervised learning is applied using data from CFD simulations or experiments to train ML models. In physics-informed approaches, un- or semi-supervised learning is applied using residuals for model partial differential equations (PDEs) along with boundary and initial conditions. So-called physics-informed neural networks (PINNs)~\cite{Raissi2019,Cai2021,Cuomo2022} have been applied to PDE models of fluid flows including the incompressible Navier-Stokes~\cite{Jin2021,Rao2020a,Sun2020}, compressible Euler~\cite{Mao2020a}, and RANS equations~\cite{Eivazi2022a}. In a purely physics-informed approach, PINNs act as a replacement for conventional numerical solver, but PINNs have not yet demonstrated superiority over conventional CFD solvers in accuracy, efficiency or robustness, with reports of difficulties encountered in applying PINNs to fluid flow problems~\cite{Chuang2022}. While a great deal of work continues to improve the performance of PINNs as PDE solvers using a variety of means~\cite{Jagtap2020,Krishnapriyan2021,Yu2022,Basir2022a,Wu2022b,Chiu2022}, other works have developed hybrid approaches that combine data-driven and physics-informed methods, assimilating data and physical laws in training the model to solve inverse problems~\cite{Raissi2020,Jagtap2022,Molnar2022} or provide super-resolution of fluid flow fields~\cite{Eivazi2022}. In the present work, we choose to pursue a data-driven approach using solutions from a trusted CFD solver to avoid the difficulties of a purely physics-informed approach, while allowing for the possibility of future extension to a hybrid approach. Considering the intended use in the design context, our goal is to construct an ML model that can be trained with data from a very small number of CFD simulations and evaluated at nearby points in design space. 

One of the fundamental questions in the construction of models of physical systems using ML is that of the proper form and structure of the inputs and outputs of the model. Among the approaches presented in the literature for the prediction of physical fields are function-type and operator-type models. Function-type models map pointwise coordinate-like inputs to pointwise scalar or vector outputs; operator-type models map inputs in one function space onto outputs in another function space. The literature presents many function-type ML models with pointwise inputs in the form of Cartesian coordinates, which is typical of PINNs. Generalizability of these models - their applicability to cases not seen in the training process - can be achieved to a degree by augmenting the inputs with additional scalar parameters~\cite{White2020}, such as Mach number, Reynolds number, or angle of attack (AoA). However, use of design variables as explicit model inputs is undesirable when operating in a large design space, as it may require training data at many points in design space. Recently, transfer learning methods have also been applied to aid the generalizability of PINNs~\cite{Penwarden2021a,Desai2021,Chen2021j,Wang2022d}. Other research has emulated image processing applications of deep learning and constructed operator-type ML models with convolutional neural networks (CNNs) using pixelated or voxelated inputs and outputs~\cite{Guo2016,Thuerey2020}. Further operator-type models have been constructed with unstructured representations of inputs and outputs using graph neural networks (GNNs)~\cite{Alet2019,Li2020b,Wu2021a}, DeepONets~\cite{Lu2021b}, PointNets~\cite{Kashefi2021}, and Fourier Neural Operators~\cite{Li2020a}. While operator-type approaches have been shown to be powerful in representing a variety of systems, they could require a large number of simulations to provide sufficient training data. It remains a challenge to construct a generalizable ML model applicable to CFD-related needs with high accuracy and low data requirements. In order to avoid the large data requirements of operator-type models, we will construct function-type models in the present work that will map suitable input features to outputs in a pointwise manner.

Generalizability of ML models for external flow problems may be achieved by using input features that are mapped appropriately under changes to body geometry. The solutions of elliptic boundary value problems (BVPs) enjoy favorable properties including strong regularity properties and maximum/minimum principles and may be easily obtained by standard numerical methods at an acceptable computational cost, which makes them attractive as means of constructing smooth mappings to augment ML models. An example of the use of elliptic BVPs in conjunction with ML is the PhyGeoNet method~\cite{Gao2021} that uses elliptic coordinate mappings from curved problem domains to a square reference domain where CNNs may be naturally applied. Potential flow quantities are derived from solutions to particular Laplace BVPs, which are a subset of general elliptic BVPs, and previous works have used ML models trained on potential flow features to perform flow field regression~\cite{fuchi_enhancement_2020,Maulik2020b}. In the present work, we construct elliptic input features from potential flow solution quantities, leading to ML models that are generalizable under changes to body geometry.

Several works have considered the acceleration of CFD solver convergence to a steady state through initialization with an ML model prediction. Operator-type models for this purpose were constructed through GNN-based upscaling of coarse-mesh CFD results~\cite{AvilaBelbutePeres2020}, through the application of CNNs on a logically Cartesian mesh to infer converged states from the results of a few ``warm-up" CFD iterations~\cite{ObiolsSales2020,ObiolsSales2021} or from mesh and airfoil geometry directly~\cite{Tsunoda2022}. As previously noted, these operator-type methods require many (potentially hundreds or thousands of) CFD simulations worth of training data. Another approach aimed to provide a surrogate model for eddy viscosity in a RANS equation solver based on potential flow features, and discovered that a substantial CFD solver speed-up could be achieved due to accelerated solver convergence tolerating more aggressive solver parameters~\cite{Maulik2020b}. That work adopted a function-type pointwise mapping, which reduced data requirements relative to operator-type approaches, but included small number of design variables as explicit inputs to their models, which may limit applicability to a large design space. 

The present work demonstrates an ML model construction that preserves local features expressed in the input-to-output map across different problems by using features derived from solutions to elliptic BVPs solved in the fluid domain as model inputs. The ML model output is taken to be the additive discrepancy between the high-fidelity flow solutions and a low-fidelity solution that may be easily and inexpensively evaluated, leading to a multi-fidelity approach. Pointwise evaluations of the input features and output are used to prepare the training data, allowing for an arbitrary data size to be generated using a small number of high-fidelity simulations. The predicted flow field is used to initialize CFD simulations at nearby design points, leading to accelerated solver convergence. Details of the ML model construction, data preparation and the model training are described in the next section, followed by case study results demonstrating the proposed method and its effectiveness in CFD acceleration.

\section{Methodology}\label{sec:method}
\begin{figure}[h!]
  \centering
    \begin{subfigure}{0.8\textwidth}
        \centering
        \includegraphics[width=4 in]{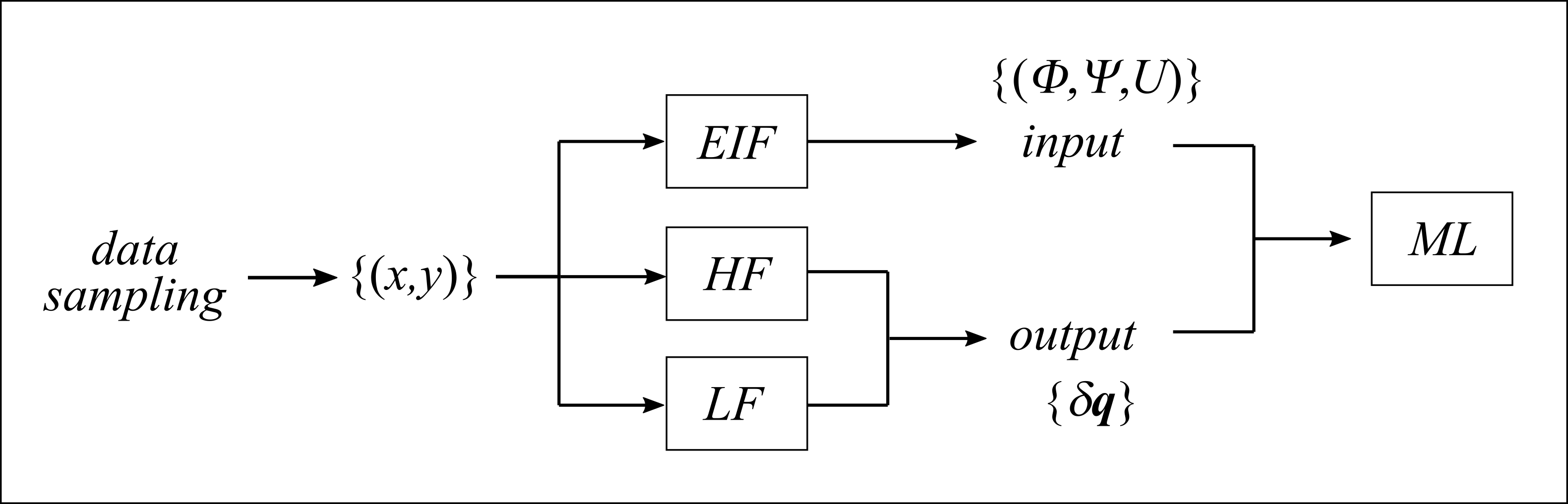}
        \caption{MF-LEIF model training process (Algorithm \ref{alg:ml-training}, Appendix \ref{sec:appendix-alg})}
    \end{subfigure}
    \\
   \begin{subfigure}{0.8\textwidth}
        \centering
        \includegraphics[width=4 in]{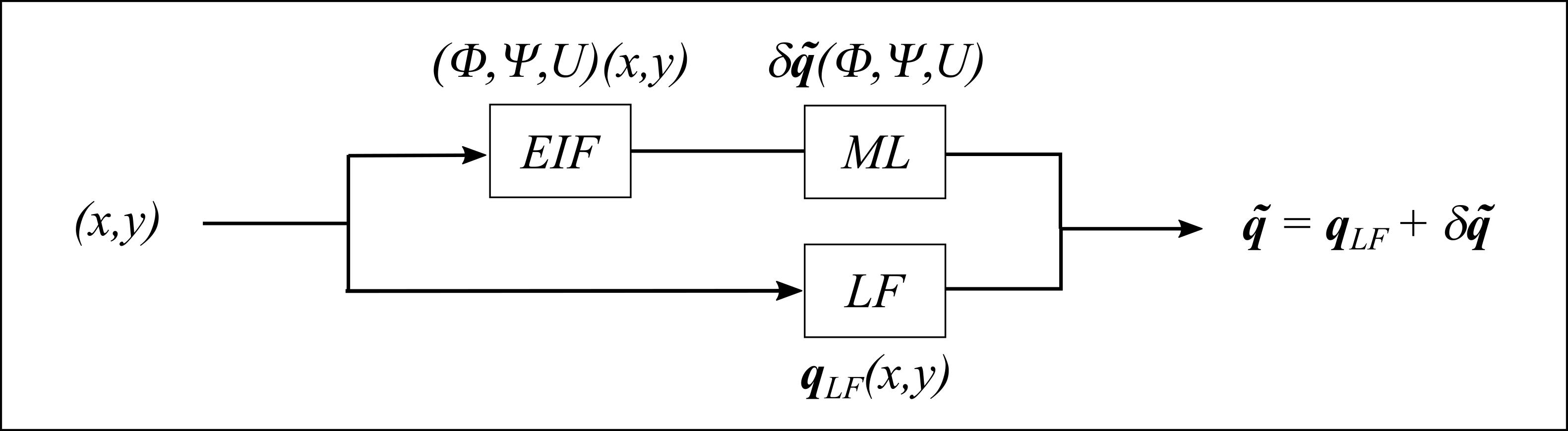}
        \caption{MF-LEIF model prediction process (Algorithm \ref{alg:prediction}, Appendix \ref{sec:appendix-alg})}
   \end{subfigure}
 \caption{Flow charts of the MF-LEIF training and prediction processes.}
  \label{fig:flowchart}
\end{figure}

The proposed ML method for flow prediction constructs a pointwise mapping of elliptic input features (EIFs) to predicted fluid flow field vectors. We call our method Learning with Elliptic Input Features (LEIF). The overall LEIF method is summarized in the flow chart in Fig. \ref{fig:flowchart}, which illustrates the training and prediction phases of operation. First, high-fidelity solutions to one or more reference flow problems under specified conditions are found, and the associated data are stored. Solutions to elliptic BVPs that correspond to the reference problems are then found and stored (see Algorithm \ref{alg:eifs} in Appendix \ref{sec:appendix-alg}); combinations of the solution fields and secondary quantities, referred to as EIFs, will be used as inputs to an ML model. Datasets for ML regression are generated from the evaluation of EIFs and the high-fidelity solution fields inside of a user-defined training window at locations determined by a selected data sampling procedure. By using the EIFs, the proposed formulation eliminates the use of Cartesian coordinates as explicit inputs to the ML model. 

The ML model output is used to predict a flow field vector, $\vecqml$, approximating the true flow field vector, $\vecqhf$. To improve training efficiency and prediction accuracy, a multi-fidelity model is constructed. A low-fidelity fluid model (e.g., potential flow), which may be evaluated at a given design point with a much lower cost than a high-fidelity solution, is used as an additive baseline model for the ML prediction of the flow field, so that the output of the ML model, $\deltavecqml$, approximates the discrepancy between the low and high-fidelity solutions: 
\begin{equation}
    \vecqml = \vecqlf + \deltavecqml \label{eqn:mf_model_def}.
\end{equation}
We call this approach Multi-Fidelity Learning with Elliptic Input Features, or MF-LEIF. The algorithms for the training process and prediction are provided in Algorithms \ref{alg:ml-training} and \ref{alg:prediction} in Appendix \ref{sec:appendix-alg}, respectively.

Our ML model will be constructed with a fully-connected neural network with parameters $\theta$. The training of the model is conducted via the optimization of $\theta$ towards the minimization of a specified loss function, $\lossl$, for our data-driven ML regression problem. We consider the $L^{2}$ mean squared error loss function defined as:

\begin{equation}
    \lossl = \lossldata = \frac{1}{m}\sum_{i}^{m}{\norm{\vecqhf_{i}-\vecqml_{i}}^{2}} \equiv
    \frac{1}{m}\sum_{i}^{m}{\norm{\delta \vecqhf_{i}-\delta \vecqml_{i}}^{2}}, \label{eqn:data_driven_loss_function}
\end{equation}

\noindent
where $\delta \vecqhf = \vecqhf - \vecqlf$, and $m$ refers to the training data size, or number of sampled data points distributed across the training window. We also consider Sobolev training~\cite{Czarnecki2017}, which incorporates spatial gradient data into the loss function:
\begin{align}
    \lossl &= \lossldata + \losslgrad , \\
    \losslgrad &= \frac{1}{m}\sum_{i}^{m}{\norm{\nabla \delta \vecqhf_{i}-\nabla \delta \vecqml_{i}}^{2}}, \label{eqn:data_driven_loss_function_grad}
\end{align}
where $\nabla \delta \vecqhf = \nabla \vecqhf - \nabla \vecqlf$.

With the presence of sparsely sampled regions, it is important to note that the value of the loss function may not provide a suitable assessment of the flow field prediction accuracy away from training points. For example, the presence of under-sampled regions in the loss function may result in overfitting of the ML model, leading to large prediction errors when the model is evaluated away from training data points. To monitor and prevent overfitting, many ML works split their training data into training, testing, and validation datasets. Instead, we perform testing of the prediction based on error norms integrated over the training window using a fine, uniform Cartesian grid, onto which the high-fidelity solution is interpolated. These fine-grid integrated error norms are used to report the field prediction errors in this work.

Further details of the input feature and training data preparation procedures are described in the following subsections. While in the present work we demonstrate only incompressible Navier-Stokes equations, the proposed method is applicable across a range of flow conditions. Future work will address the application of our method to other steady fluid flow regimes, including compressible and turbulent flows.

\subsection{Elliptic Input Features}\label{sec:eif}
Use of EIFs is a key component of the proposed approach toward a generalizable ML model. While a standard approach to constructing ML models for flow predictions often include Cartesian coordinates as input features, this ties the model to a fixed coordinate system in an Eulerian frame and could hinder the generalizability of the model to different design geometries. To avoid this drawback, we use solutions to BVPs for elliptic partial differential equations (PDEs), such as Laplace's equation, to provide input features on the fluid domain. The EIFs are the solution fields of these elliptic BVPs or additional quantities derived from the solution. The algorithm used to find EIFs is provided in Algorithm \ref{alg:eifs}, Appendix \ref{sec:appendix-alg}. To make predictions at a new design point, we simply evaluate the EIFs by solving the corresponding elliptic BVPs and feed them into the trained ML model. This model construction takes advantage of the relative ease of solving the elliptic BVPs and the favorable properties of the solution fields.

While different elliptic BVPs could be used to obtain input features, we will use the following potential flow BVPs on the fluid domain $\Omega$ with wall boundary $\partial \Omega_{W}$. For the potential function $\Phi$, we have 

\begin{align}\label{eq:eif-phi}
    \left\{ \begin{array}{l}
            \nabla^{2}\Phi = 0 \mbox{ in } \Omega \\
            \mathbf{n} \cdot \nabla \Phi = 0 \mbox{ on } \partial \Omega_{W} \, \mbox{(slip wall BC)} \\
            \mbox{ + appropriate farfield BCs,}
    \end{array} \right.
\end{align}
and for the streamfunction $\Psi$, we have
\begin{align}\label{eq:eif-psi}
    \left\{ \begin{array}{l}
            \nabla^{2}\Psi = 0 \mbox{ in } \Omega \\
            \Psi = 0 \mbox{ on } \partial \Omega_{W}  \\
            \mbox{ + appropriate farfield BCs.}
    \end{array} \right.
\end{align}

\noindent
Details of potential flow models can be found in many standard references \cite{Prandtl1957,Moran1984,Anderson2011}.

The EIFs derived from potential flow solution quantities are the potential function $\Phi$, the streamline function $\Psi$, and the velocity magnitude $U=\abs{\nabla \Phi}=\abs{\nabla \Psi}$. The features $\Phi$ and $\Psi$ allow the identification of geometrically similar points across design space, while the feature $U$ provides sensitivity to change in the boundary geometry and is included among the EIFs in order to distinguish similarly located points across different geometries that may experience different physics (e.g., attached versus separated flow near the trailing edge of an airfoil).

\begin{figure}[h!]
  \centering
    \begin{subfigure}{0.23\textwidth}
        \includegraphics[width=\linewidth,trim={.8in .8in .8in 0.8in},clip]{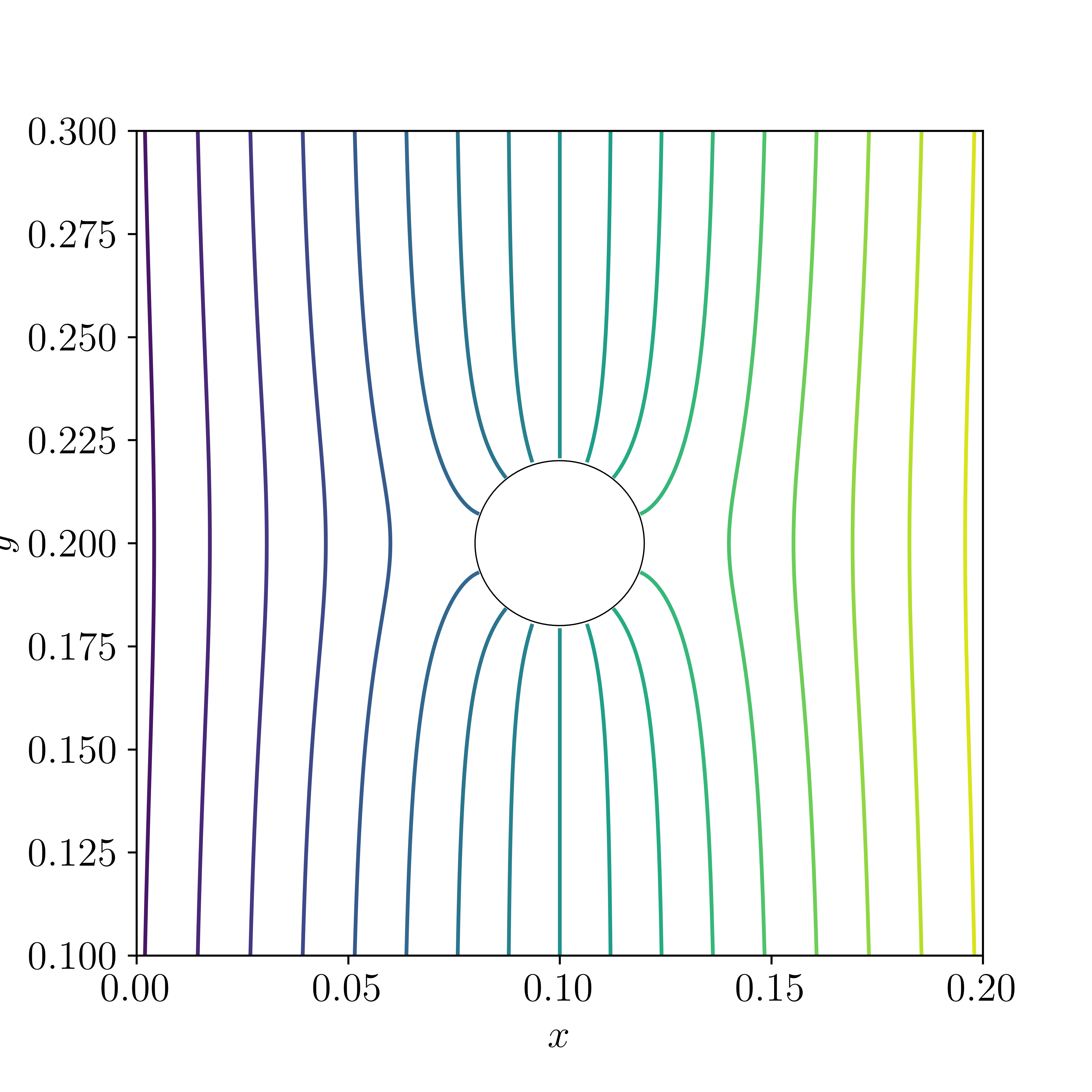}
        \caption{$\Phi$ - cylinder}
    \end{subfigure}
   \begin{subfigure}{0.23\textwidth}
        \includegraphics[width=\linewidth,trim={.8in .8in .8in 0.8in},clip]{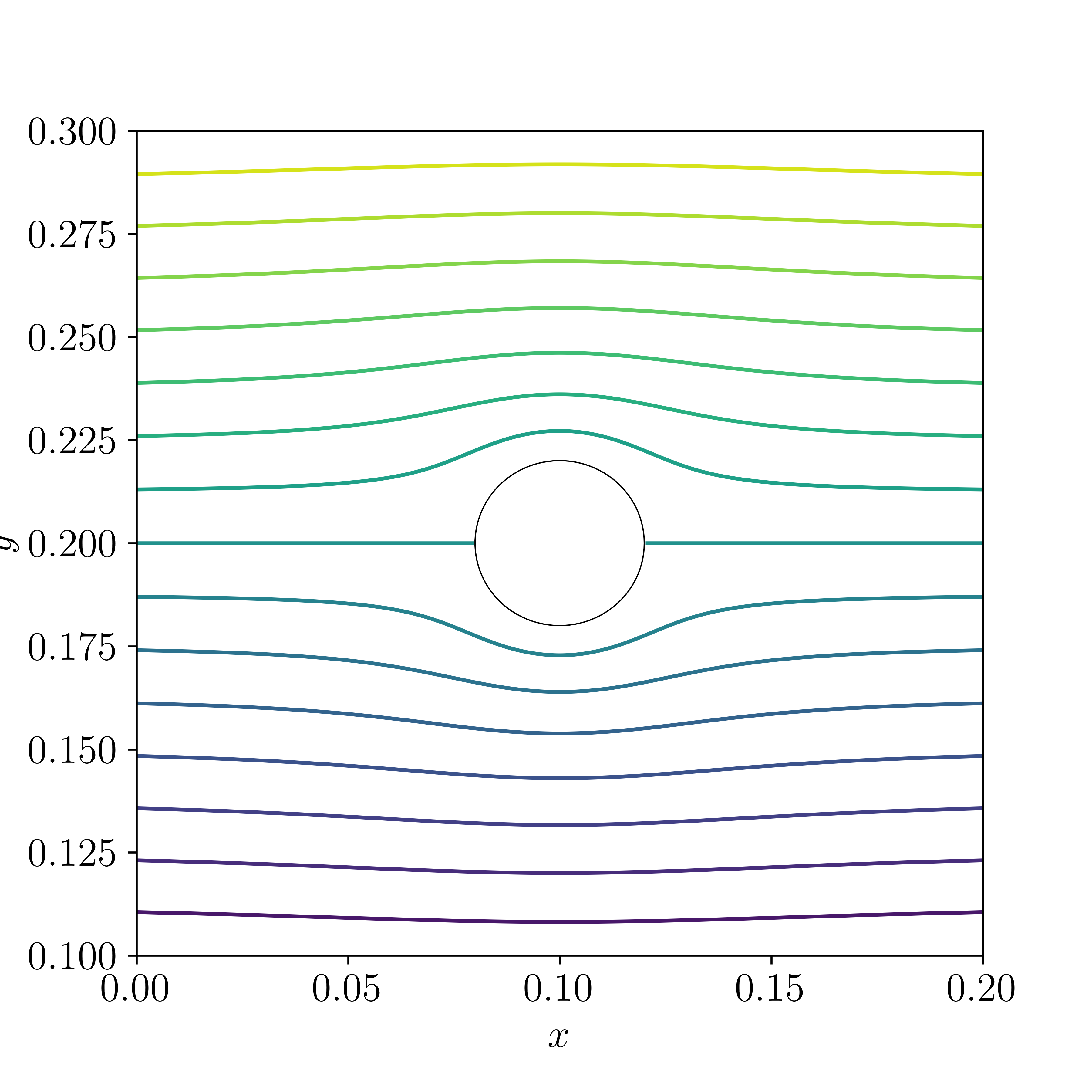}
        \caption{$\Psi$ - cylinder}
    \end{subfigure}
    \begin{subfigure}{0.23\textwidth}
        \includegraphics[width=\linewidth,trim={.8in .8in .8in 0.8in},clip]{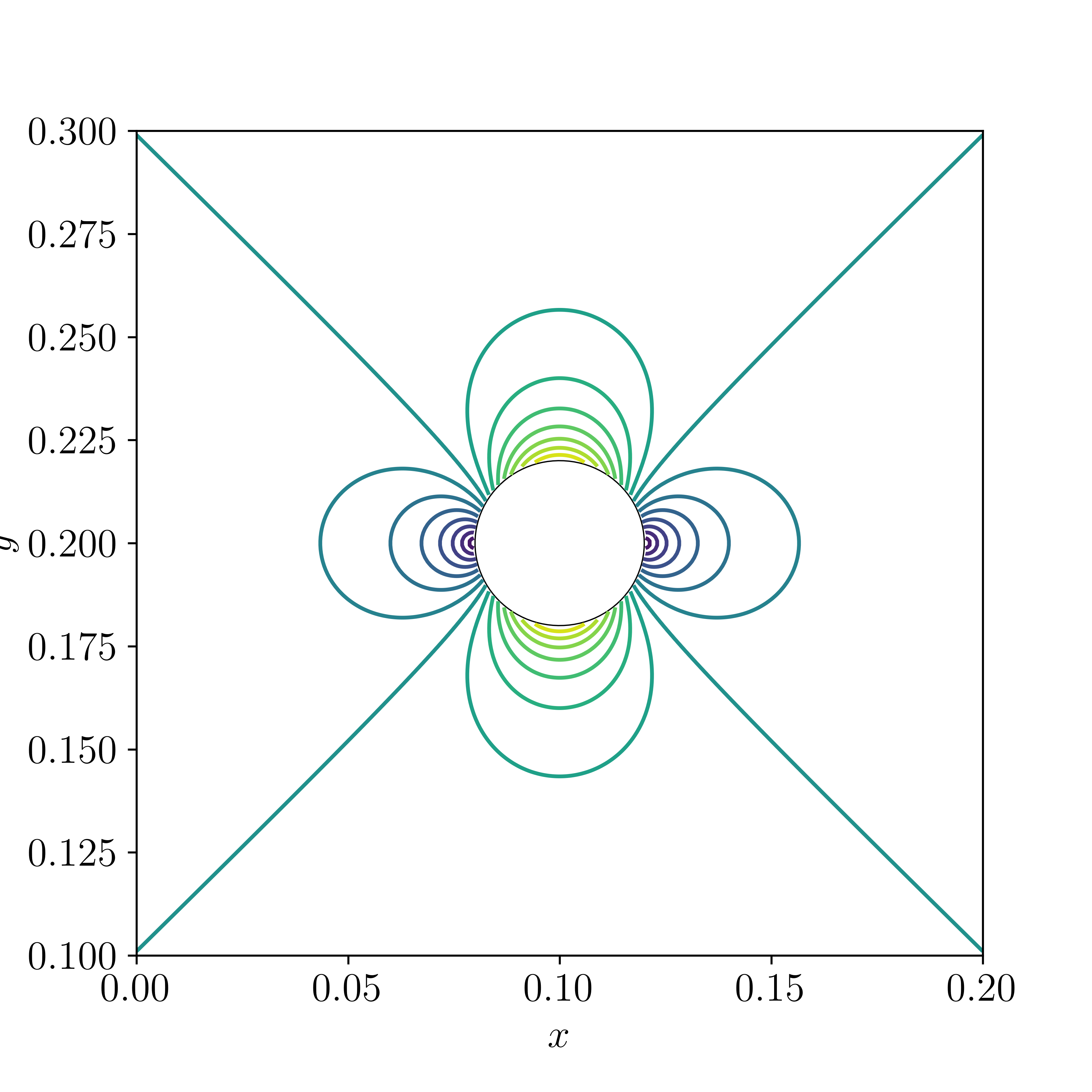}
        \caption{$U$ - cylinder}
    \end{subfigure}
    \\
    \begin{subfigure}{0.23\textwidth}
        \includegraphics[width=\linewidth,trim={.8in .8in .8in 0.8in},clip]{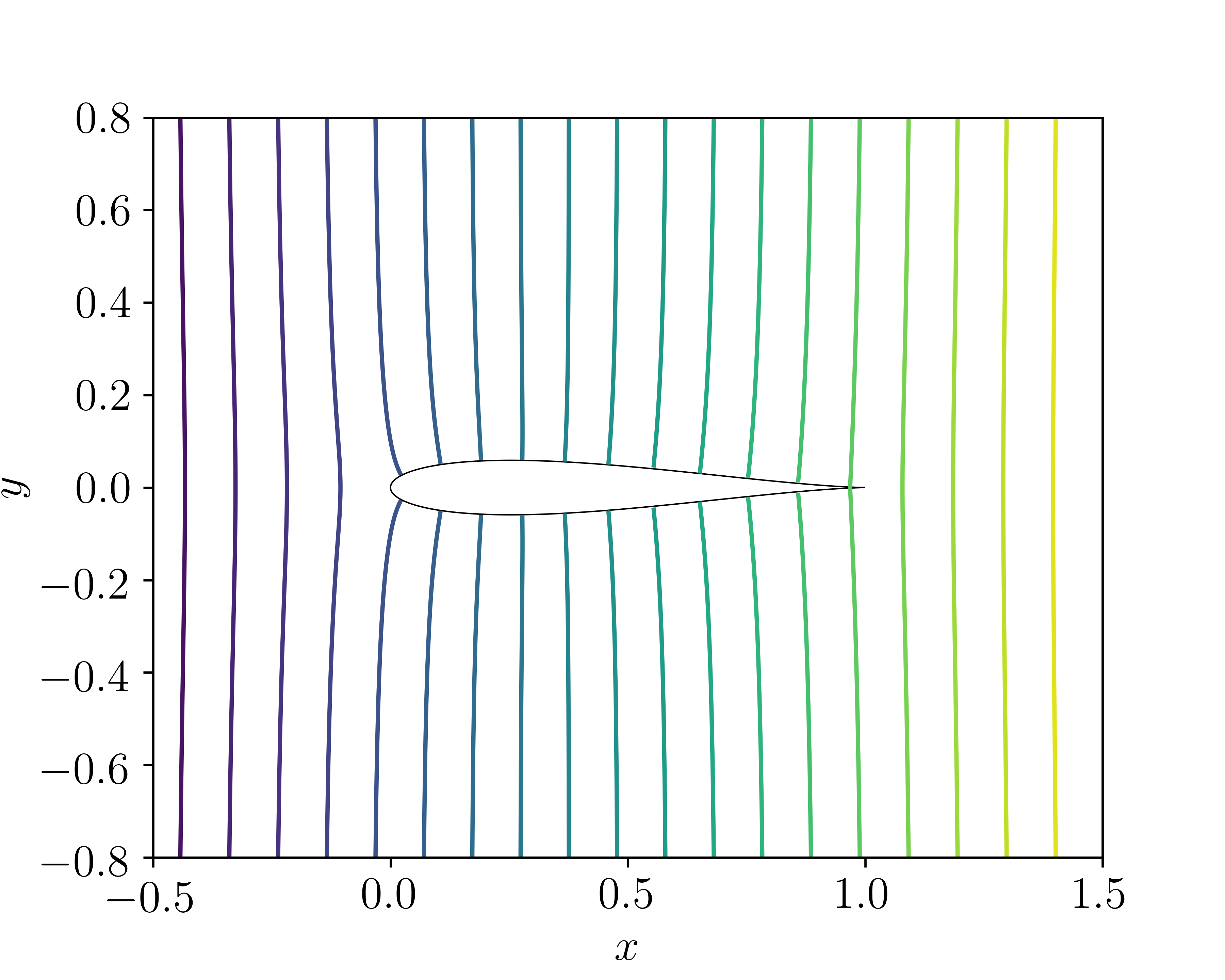}
        \caption{$\Phi$ - airfoil, AoA=0\dg}
    \end{subfigure}
   \begin{subfigure}{0.23\textwidth}
        \includegraphics[width=\linewidth,trim={.8in .8in .8in 0.8in},clip]{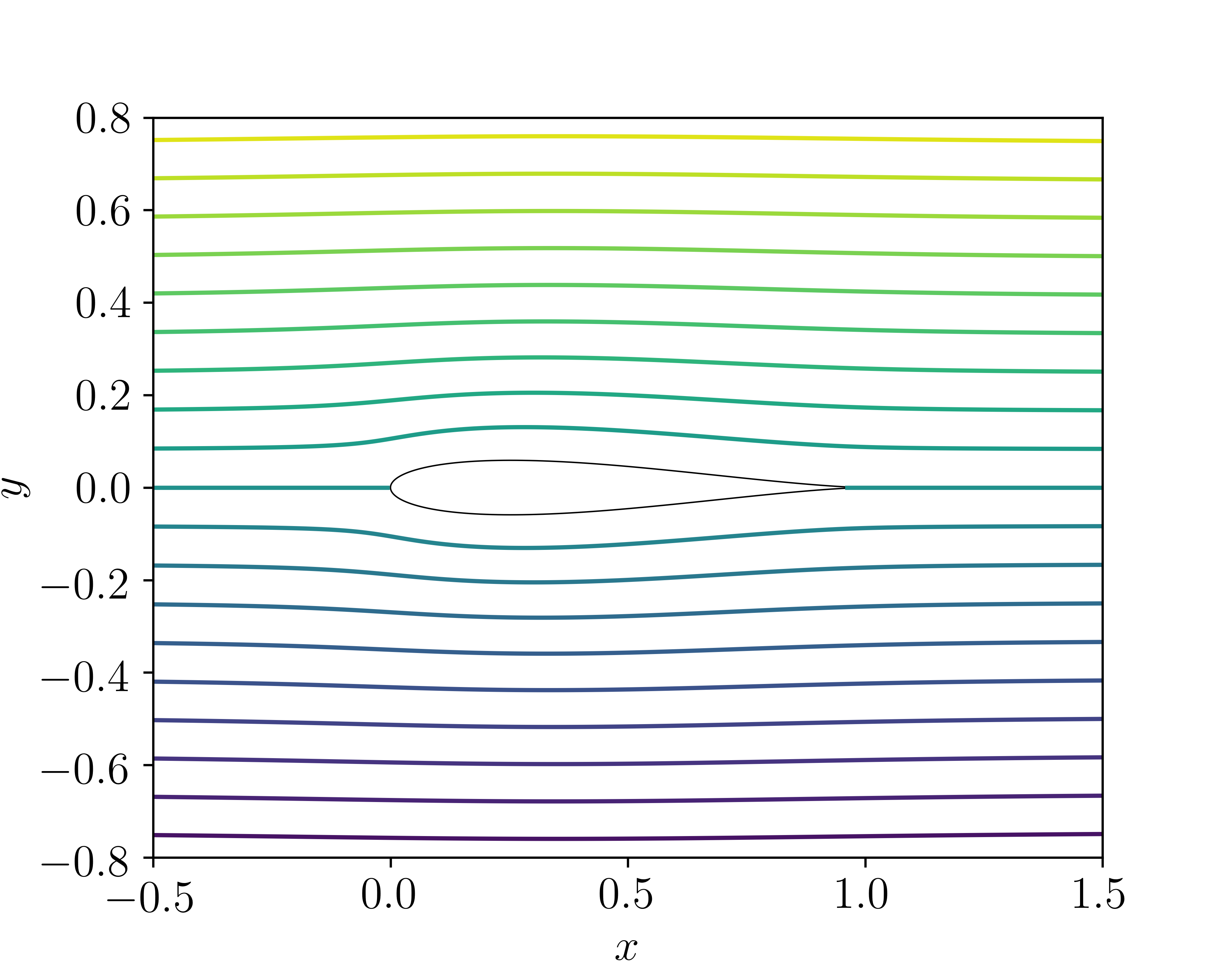}
        \caption{$\Psi$ - airfoil, AoA=0\dg}
    \end{subfigure}
    \begin{subfigure}{0.23\textwidth}
        \includegraphics[width=\linewidth,trim={.8in .8in .8in 0.8in},clip]{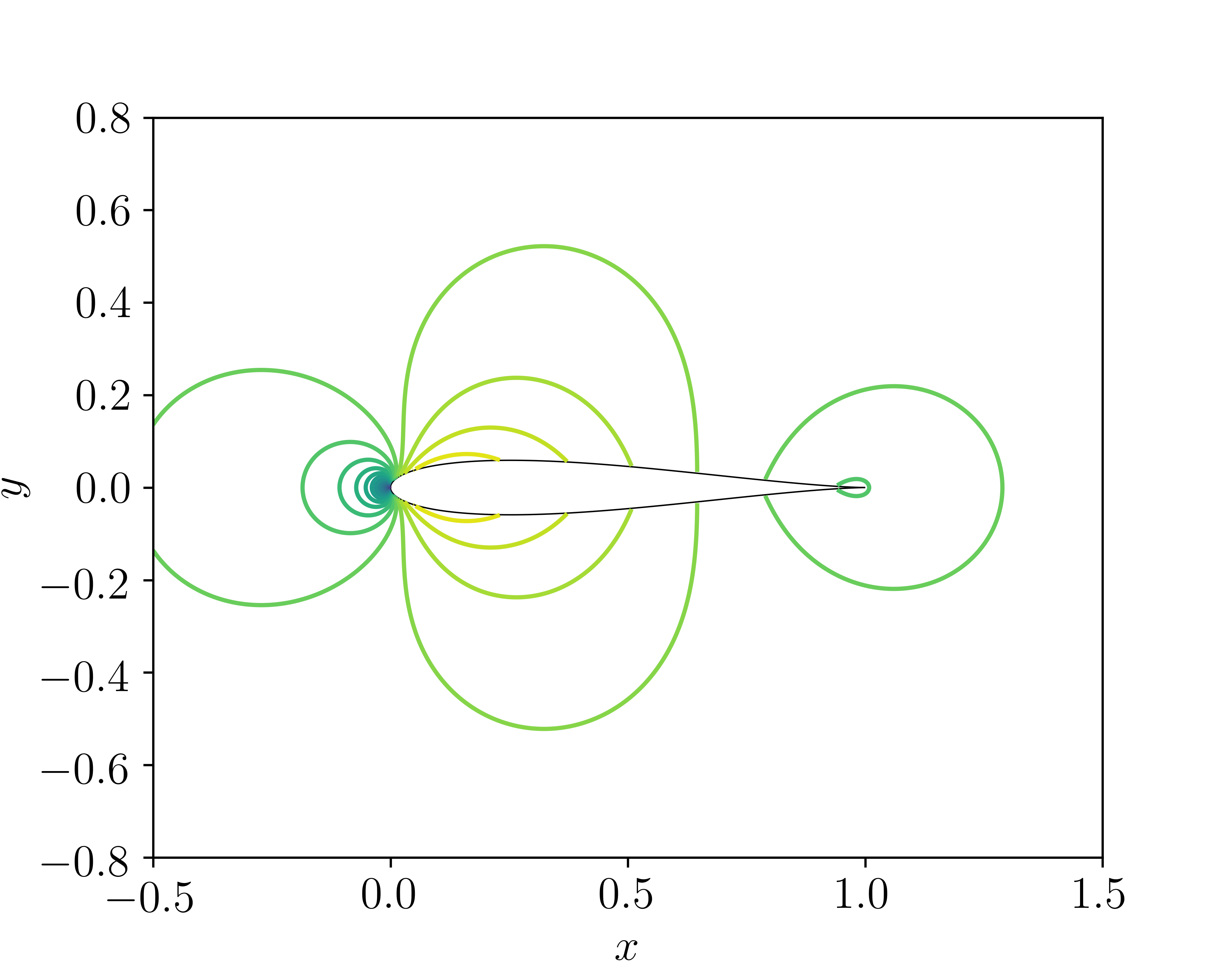}
        \caption{$U$ - airfoil, AoA=0\dg}
    \end{subfigure}
    \\
     \begin{subfigure}{0.23\textwidth}
        \includegraphics[width=\linewidth,trim={.8in .8in .8in 0.8in},clip]{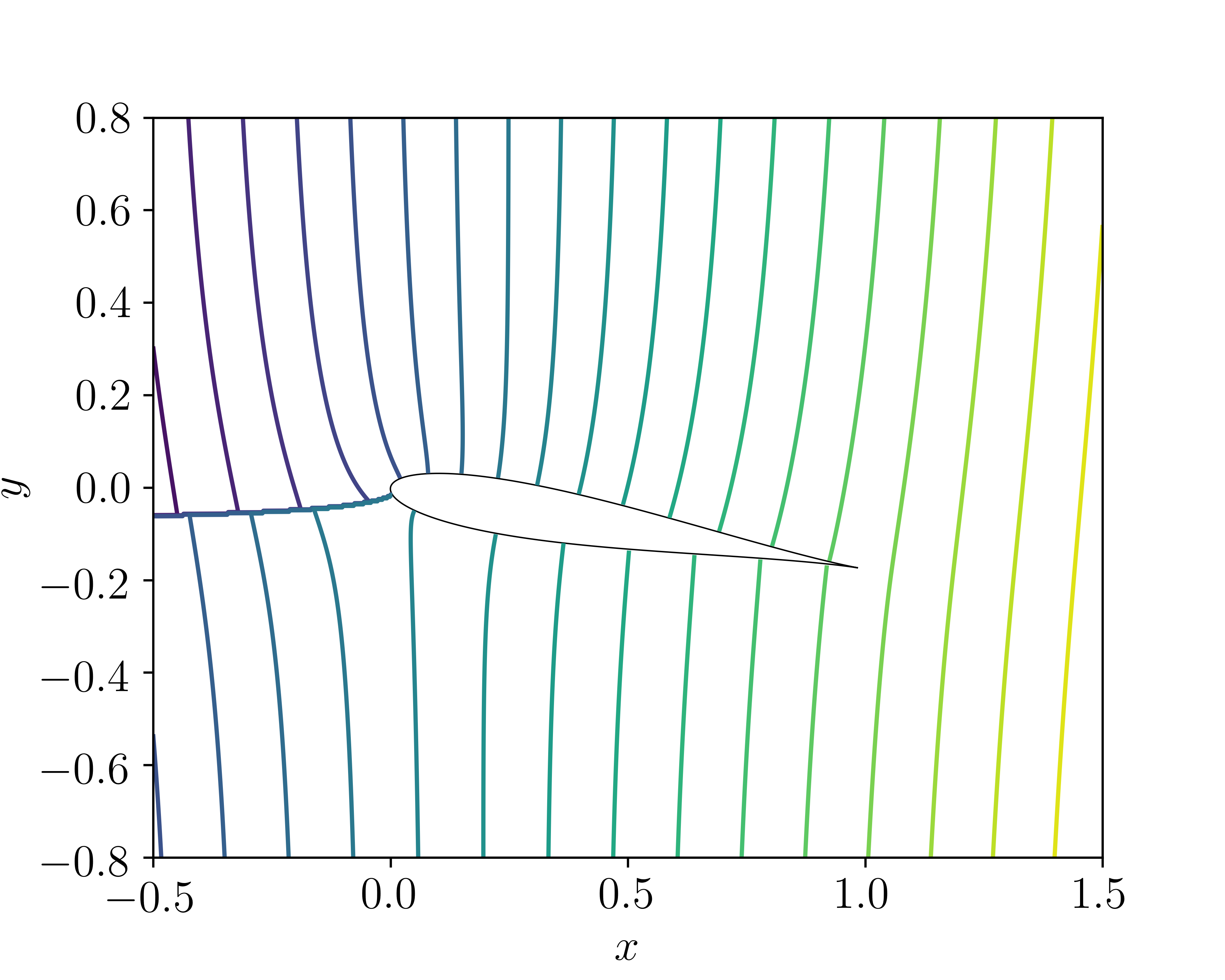}
         \caption{$\Phi$ - airfoil, AoA=10\dg}
    \end{subfigure}
   \begin{subfigure}{0.23\textwidth}
        \includegraphics[width=\linewidth,trim={.8in .8in .8in 0.8in},clip]{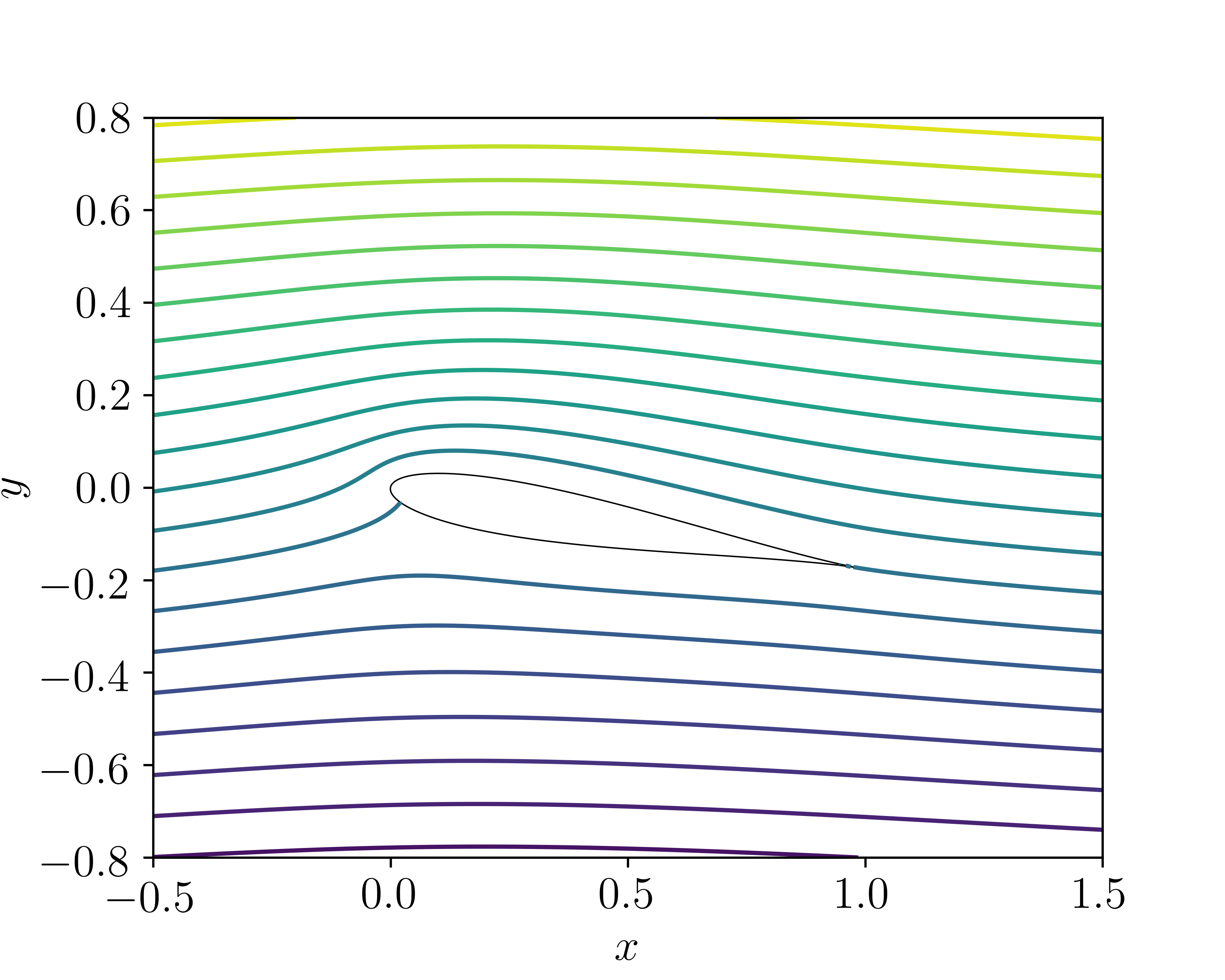}
        \caption{$\Psi$ - airfoil, AoA=10\dg}
    \end{subfigure}
    \begin{subfigure}{0.23\textwidth}
        \includegraphics[width=\linewidth,trim={.8in .8in .8in 0.8in},clip]{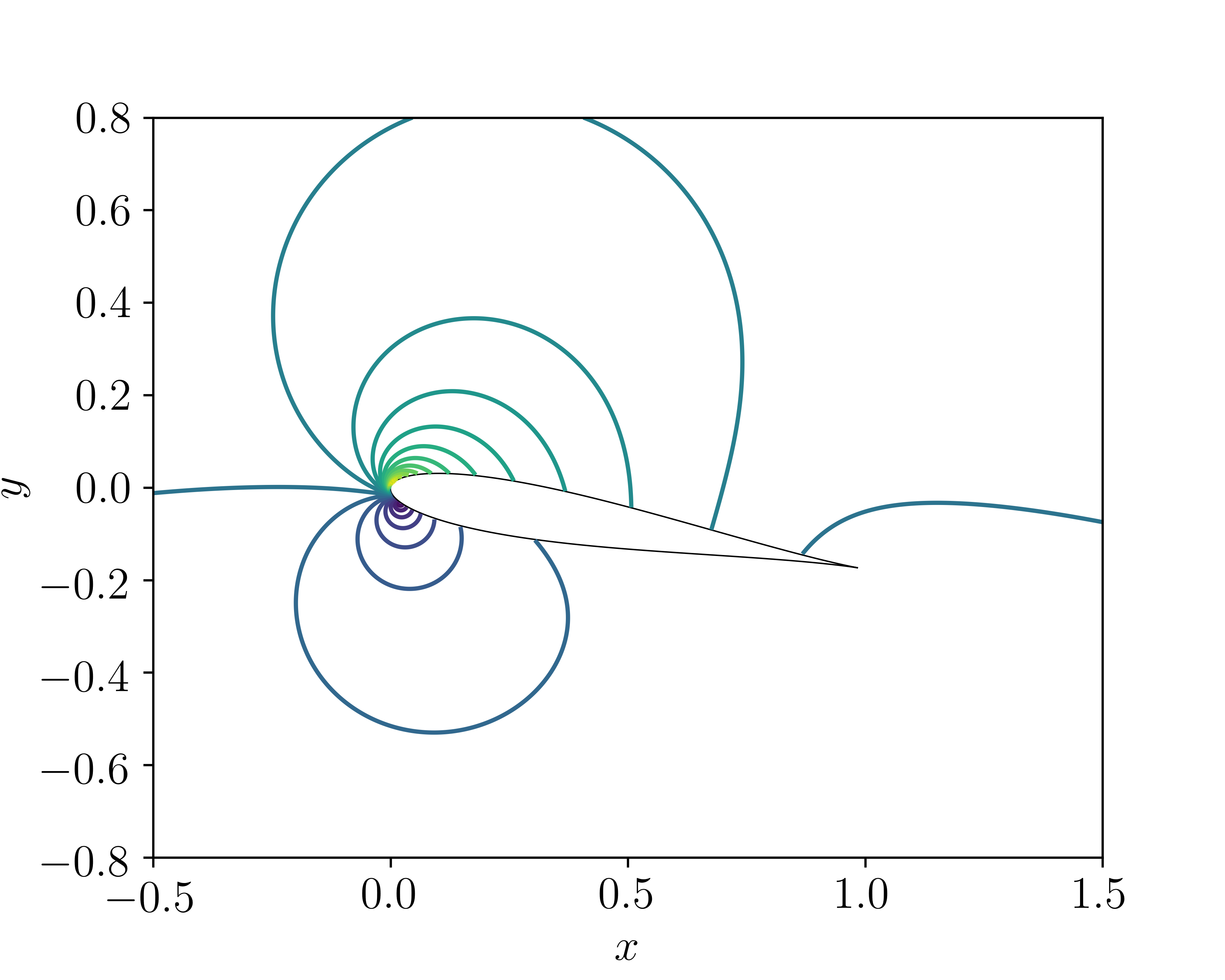}
        \caption{$U$ - airfoil, AoA=10\dg}
    \end{subfigure}
 
 \caption{Examples of elliptic input features for canonical problems.}
  \label{fig:elliptic-input}
\end{figure}

The evaluation of EIFs must be performed accurately and efficiently, because they directly impact the ML model construction and thus the prediction accuracy and efficiency. For complex geometries, standard numerical methods such as finite element or boundary element methods may be employed to compute the EIFs. For simple geometries, such as circular cylinders and Joukowski airfoils, analytical expressions for $\Phi$, $\Psi$, and $U$ can be obtained through conformal mapping. Contours of the potential $\Phi$, stream function $\Psi$ and the velocity magnitude $U$ for the circular cylinder case are illustrated in Fig. \ref{fig:elliptic-input} (a-c). Potential flow features of a Joukowski airfoil can be obtained in a similar manner, and the examples illustrated for zero and 10\dg AoAs shown in Fig. \ref{fig:elliptic-input} (d-i) demonstrate the smooth transformation of potential features across the problem parameter. Derivations of the potential solutions using conformal mapping are summarized in Appendix \ref{sec:appendix-PF} for completeness.

Given that the spatial gradients of the EIFs, $\nabla \Phi$, $\nabla \Psi$, and $\nabla U$, are provided, spatial gradients of $\vecqml$ can be computed using automatic differentiation of ML model outputs with respect to inputs along with the chain rule:
\begin{equation}
    \begin{array}{ll}
        \nabla \vecqml &= \nabla \vecqlf + \nabla \deltavecqml \\
                        &= \nabla \vecqlf + \frac{\partial \deltavecqml}{\partial \Phi}\nabla \Phi + \frac{\partial \deltavecqml}{\partial \Psi}\nabla \Psi  + \frac{\partial \deltavecqml}{\partial U}\nabla U,
    \end{array} \label{eqn:spatial_gradient_chain_rule}
\end{equation}
which enables the computation of the gradient loss function term in Eq.~(\ref{eqn:data_driven_loss_function_grad}).

\subsection{Training Window and Partition-of-Unity (POFU) Extension}\label{sec:pofu}

\begin{figure}[h!]
  \centering
    \begin{subfigure}{0.8\textwidth}
        \centering
        \includegraphics[width=\linewidth]{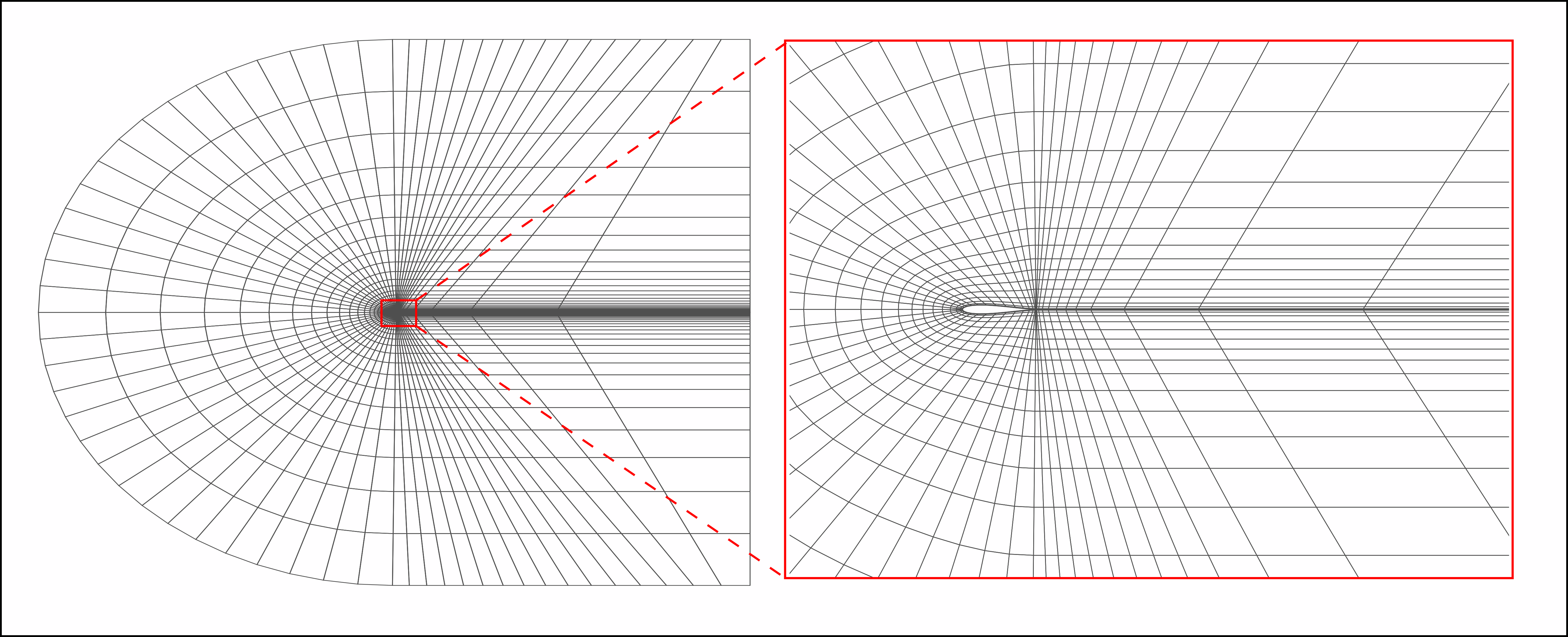}
        \caption{Training window}
    \end{subfigure}
    \\
   \begin{subfigure}{0.4\textwidth}
        \centering
        \includegraphics[width=\linewidth,trim={2.6in 0.5in 0.5in 0.5in},clip]{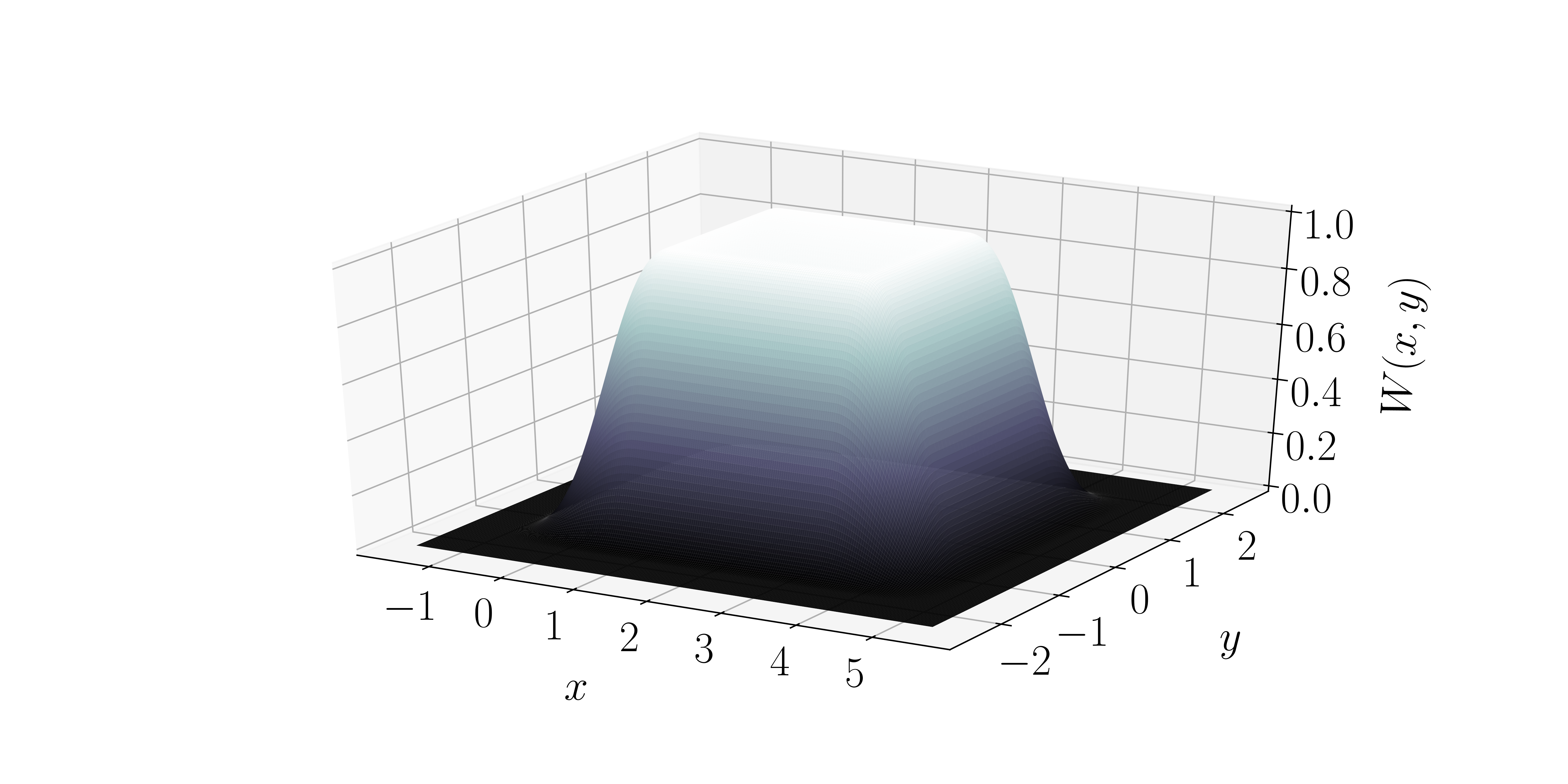}
        \caption{$W\left(x,y\right)$}
   \end{subfigure}
   \begin{subfigure}{0.4\textwidth}
        \centering
        \includegraphics[width=\linewidth,trim={2.6in 0.5in 0.5in 0.5in},clip]{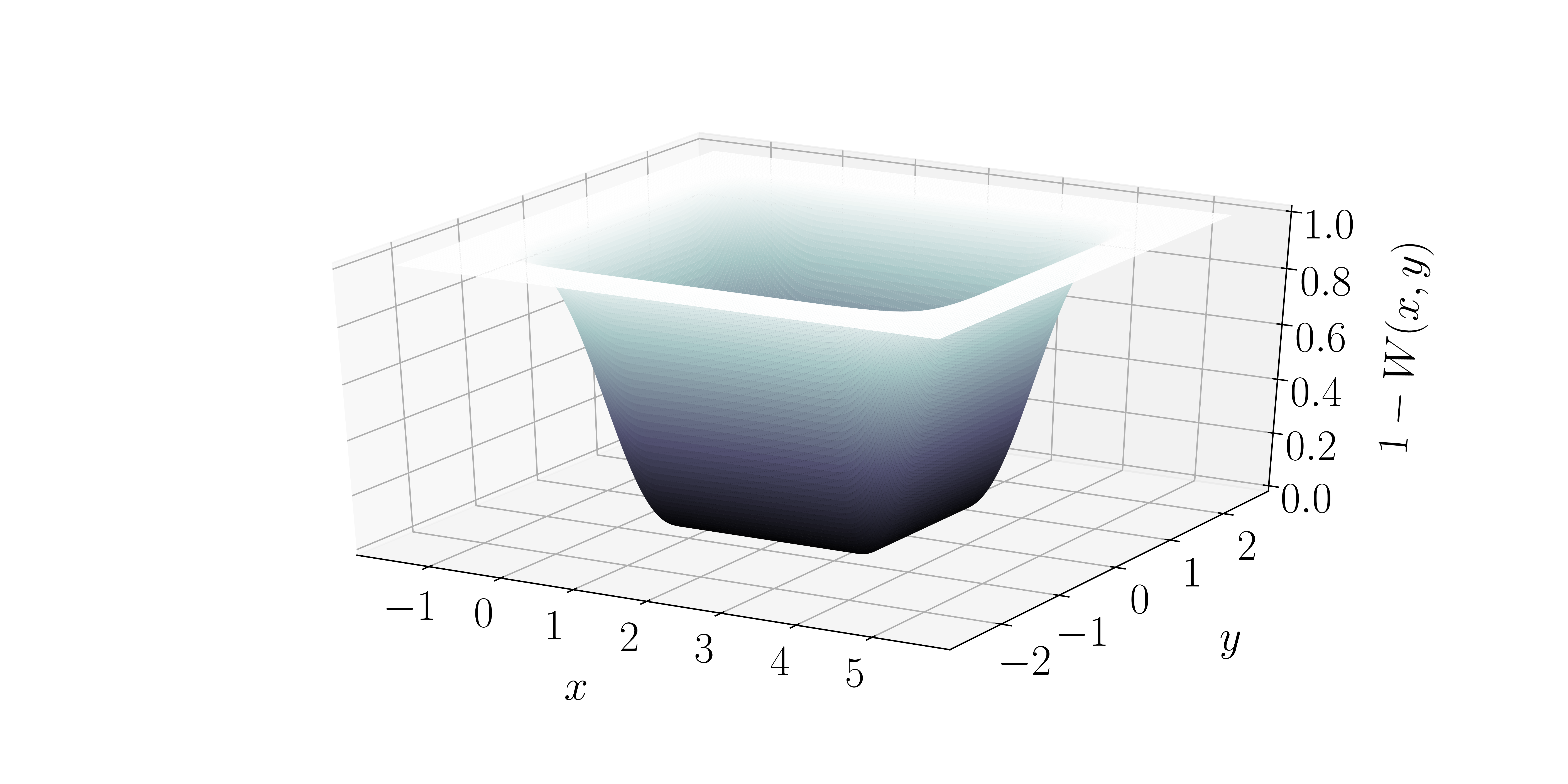}
        \caption{$1-W\left(x,y\right)$}
   \end{subfigure}
 \caption{ML model training window and POFU window functions.}
  \label{fig:pofu}
\end{figure}

While typical CFD simulations of external flows may have meshes or grids that extend 20-100 times the characteristic body length away from the body, most of the salient flow features occur in the near-body region. We will adopt a strategy in which the ML model is trained only on a near-body rectangular training window (see Fig. \ref{fig:pofu} (a) for example), while the model prediction is smoothly extended to freestream conditions outside of the training window using Partition-of-Unity (POFU) extension. The algorithm to prepare POFU extension of the ML model is provided in Algorithm \ref{alg:pofu}, Appendix \ref{sec:appendix-alg}. This strategy concentrates the representational power of the ML model in the key near-body region while permitting us to make physically reasonable predictions for initialization of CFD simulations on meshes of arbitrary topology and extent. The POFU extension allows our initialization to be consistent with freestream farfield boundary conditions and no-slip wall boundary conditions (up to the accuracy of the ML model), which may be beneficial for the stability of the initialization. When multiple reference problems are used for training, the same training window is applied in each case. We rely in the present work on the user to specify a reasonable training window. A trade-off presents itself in that an excessively large training window will not effectively concentrate the representational power of the ML model, while an excessively small training window will clip key flow features and hinder the performance of the CFD initialization.

The POFU extension is constructed with a window function $W\left(x,y\right)$ that smoothly transitions from the value 0 outside of the training window to the value 1 inside through a transition region at the edge of the training window. The POFU extension of the ML model prediction, $\vecqml$, can then be evaluated as:
\begin{equation}
\vecqpofu\left(x,y\right) = W\left(x,y\right)\vecqml\left(\Phi , \Psi, U \right) \left(x,y\right)+\left(1-W\left(x,y\right)\right)\vecqfs, \label{eqn:pofu_extension}
\end{equation}

\noindent
where $\bm{q}_{FS}$ refers to the freestream conditions.

While there are many possibilities in the construction of window functions, in the present work we will proceed as follows. Let the left, right, bottom, and top boundary locations of the rectangular training window be $x=x_{0}$, $x=x_{1}$, $y=y_{0}$, and $y=y_{1}$, respectively. Let $s_{x}$ and $s_{y}$ be the widths of the transition regions in the $x$- and $y$-directions, respectively. The window function is then defined as a product of 1D transition functions:
\begin{equation}\label{eq:window-func}
    W\left(x,y\right) = T\left(\left(x-x_{0}\right)/s_{x}\right)
    T\left(\left(x_{1}-x\right)/s_{x}\right)
    T\left(\left(y-y_{0}\right)/s_{y}\right)
    T\left(\left(y_{1}-y\right)/s_{y}\right),
\end{equation}
where the transition function $T\left(r\right)$ is defined as a twice-continuously differentiable piecewise polynomial given by 
\begin{equation}\label{eq:transition-func}
  T\left(r\right)  =
  \begin{cases}
    0 & \text{if $r < 0$} \\
    6 r^{5} -15 r^{4} + 1- r^{3} & \text{if $0 \le r \le 1$} \\
    1 & \text{if $1 < r$}.
  \end{cases}
\end{equation}

\noindent
An example of a resulting window function $W(x,y)$ and its counterpart $1-W(x,y)$ are shown in Fig. \ref{fig:pofu} (b,c).

\subsection{Training Data Sampling}\label{sec:sampling-methods}
The input features and associated output values are evaluated at sampled points within the training window in each reference problem to generate training data. In this subsection, we discuss methods for training data sampling: mesh node sampling, uniform sampling, random sampling, and a proposed quadtree adaptive sampling method.

As our training data is derived from a mesh-based simulation, the use of mesh node locations (or a random subset thereof) suggests itself. While the node locations in a well-designed mesh used to generate the reference high-fidelity solution provide some information on the distribution of solution field complexity, these distributions do not necessarily coincide with the sampling point distribution required in the ML training. For instance, a fine mesh in the boundary layer and a rapid element size growth away from a body can be expected in a conventional fluid simulation mesh, resulting in a large point-point distance away from the body. However, our experience shows that a finer data sampling resolution may be required during the ML training in the wake region further downstream where a large deviation in the field quantities from the baseline is observed. Hence, we are motivated to seek alternative data sampling methods more suitable for ML training data preparation.

Alternative training data locations may be selected and standard interpolation methods may be used to evaluate the high-fidelity solution data at these locations. Numerous ML works have used uniform (see Algorithm \ref{alg:uniform}, Appendix \ref{sec:appendix-alg}) or random sampling (see Algorithm \ref{alg:random}, Appendix \ref{sec:appendix-alg}) to select training data positions. Simple sampling methods such as uniform and random distributions may be effective enough when data size used is large enough to resolve all the flow features, however, they may require excessive data sizes to achieve resolution of small-scale features such as boundary layers. An ad hoc process may be required to determine sufficient data size. 

\begin{figure}[h!]
  \centering
        \includegraphics[width=2.8 in]{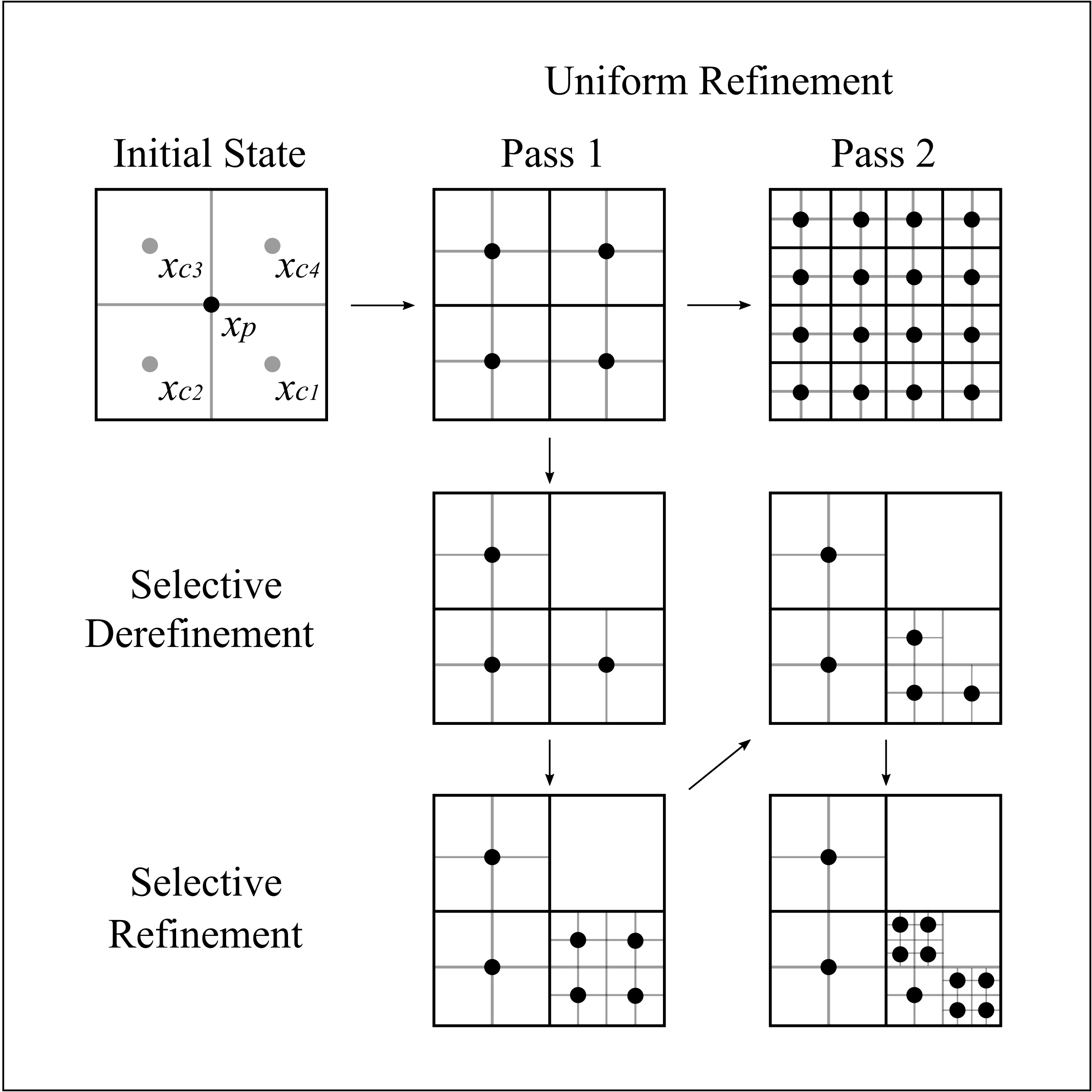}
    \caption{Illustration of a quadtree derefinement and refinement process (Algorithm \ref{alg:qdtree}, Appendix \ref{sec:appendix-alg}).}
  \label{fig:qdtree-process2}
\end{figure}

We propose an adaptive quadtree sampling method to achieve effective resolution of flow features in our training data distribution.
A quadtree is a tree data structure with four child nodes at each parent node. This structure is utilized here to subdivide the fluid domain recursively and selectively to generate sampling points distributed across the domain. Figure \ref{fig:qdtree-process2} illustrates a general quadtree refinement process, and the algorithm is provided in Algorithm \ref{alg:qdtree}, Appendix \ref{sec:appendix-alg}. The problem domain is recursively divided into quadrants with new child nodes placed at the center of each quadrant in each pass until a desired refinement level is reached. Selective refinement applies the partitioning process on selected nodes that satisfy specified criteria; derefinement may be applied to nodes that fall under certain conditions to remove subdivisions. In our work, the norm of difference in the high-fidelity field quantities between parent and child nodes is used as the refinement and derefinement metric, expressed as:

\begin{equation}\label{eq:qdtree}
    d_{cn} = \norm{\bm{q}\prt{\bm{x}_{p}}-\bm{q}\prt{\bm{x}_{cn}}},
\end{equation}

\noindent
for child node $cn$. Here $\bm{x}_{p}$ and $\bm{x}_{c}$ refer to the coordinates of parent and child nodes, respectively. The child nodes are ranked according to this metric, and a certain percentage of top ranking nodes are refined, while bottom ranking nodes are derefined.

At each pass during the quadtree sampling process, it must be verified that only valid nodes are included in the training dataset, since some new child nodes may lie inside the body. A body collision test is performed based on Gauss' Lemma, computing a boundary integral over the body's surface to determine whether a given point lies inside the surface \cite{liu2009fast}. Any node found inside the body is marked as invalid and not included in the training data.

\begin{figure}[h!]
  \centering
   \begin{subfigure}{0.3\textwidth}
        \includegraphics[width=\linewidth,trim={.8in .8in .8in 0.8in},clip]{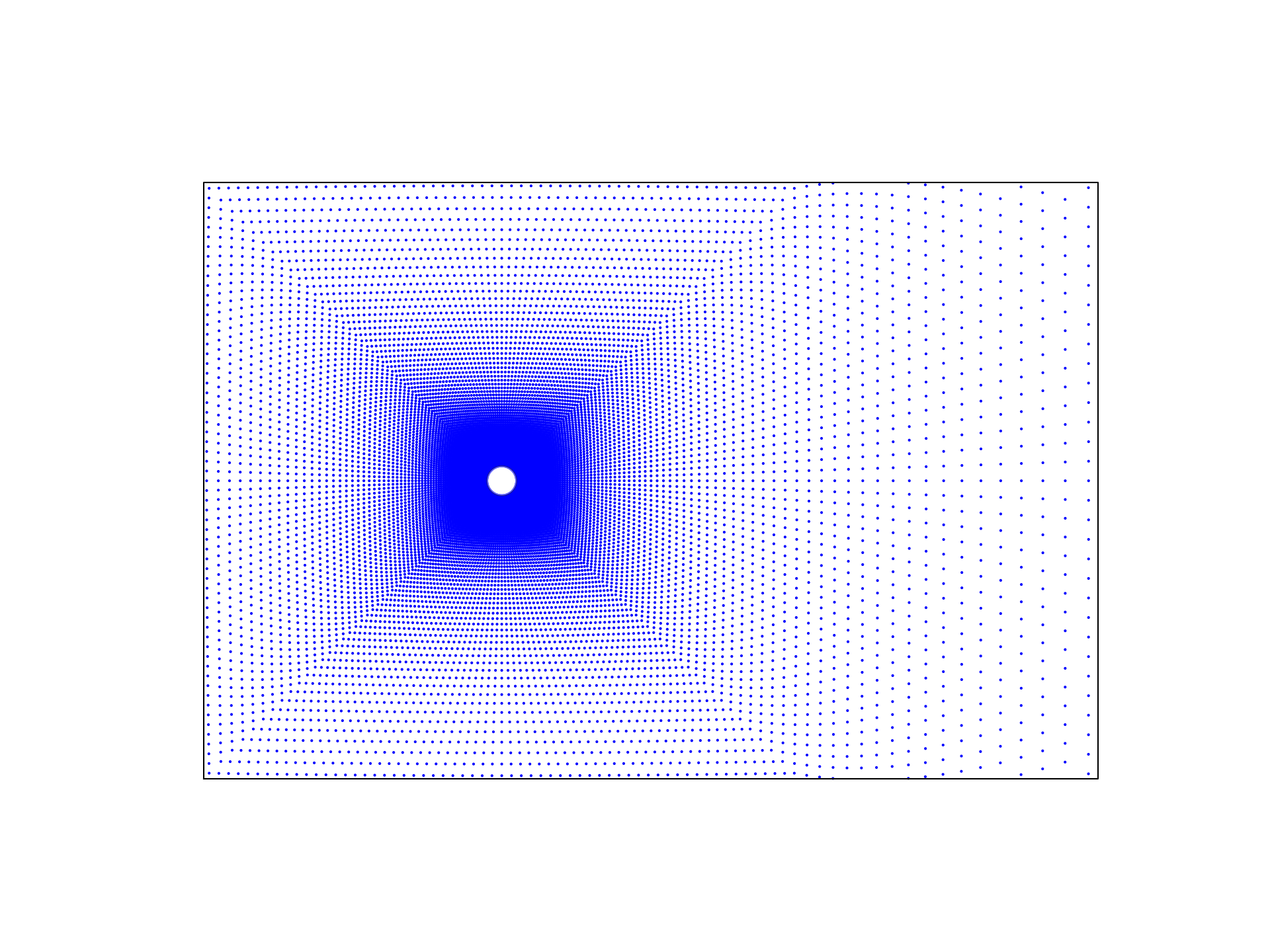}
        \caption{HF simulation mesh grid}
    \end{subfigure}
    \begin{subfigure}{0.3\textwidth}
        \includegraphics[width=\linewidth,trim={.8in .8in .8in 0.8in},clip]{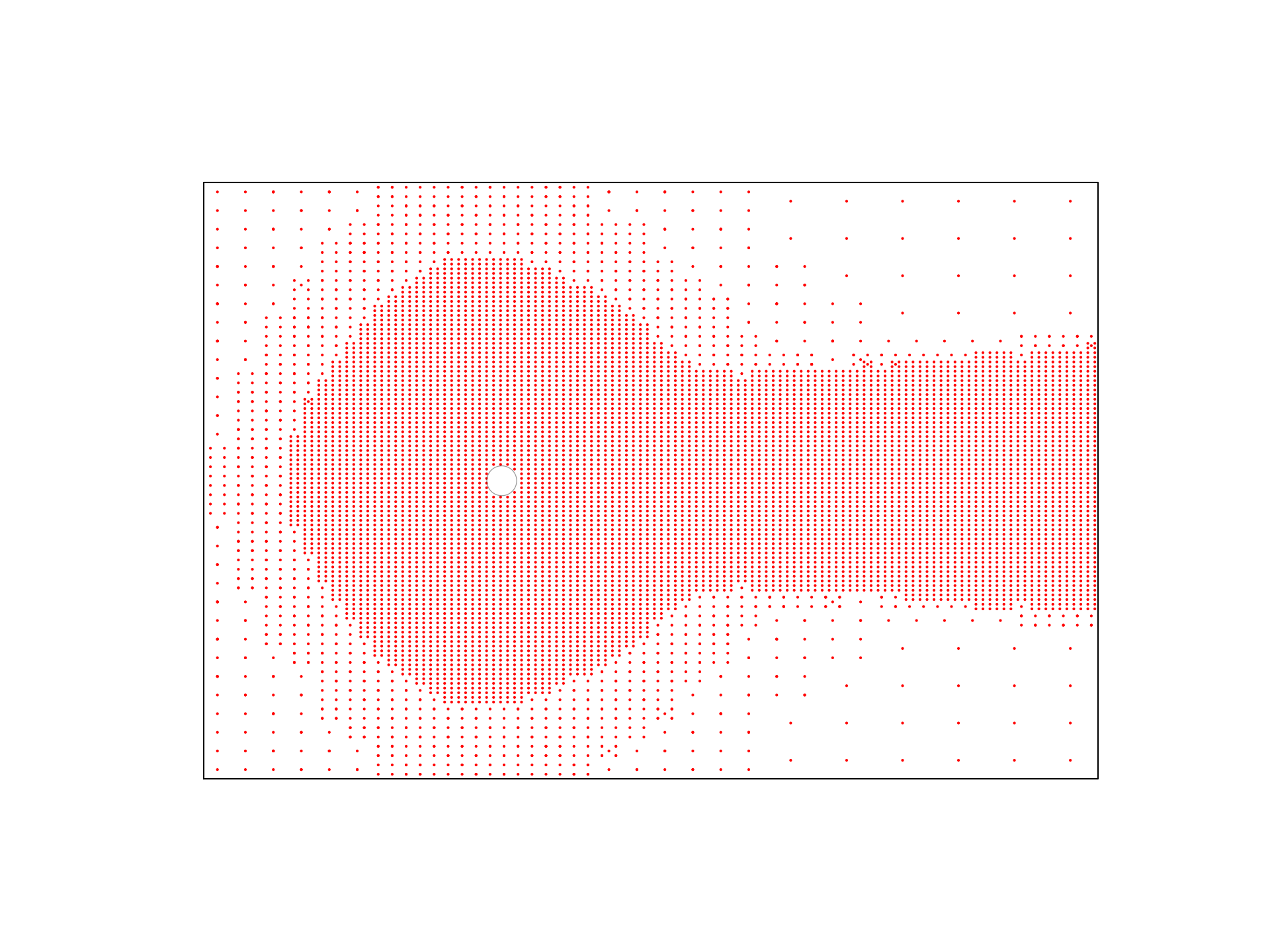}
        \caption{Quadtree refinement pass 3}
    \end{subfigure}
    \begin{subfigure}{0.3\textwidth}
        \includegraphics[width=\linewidth,trim={.8in .8in .8in 0.8in},clip]{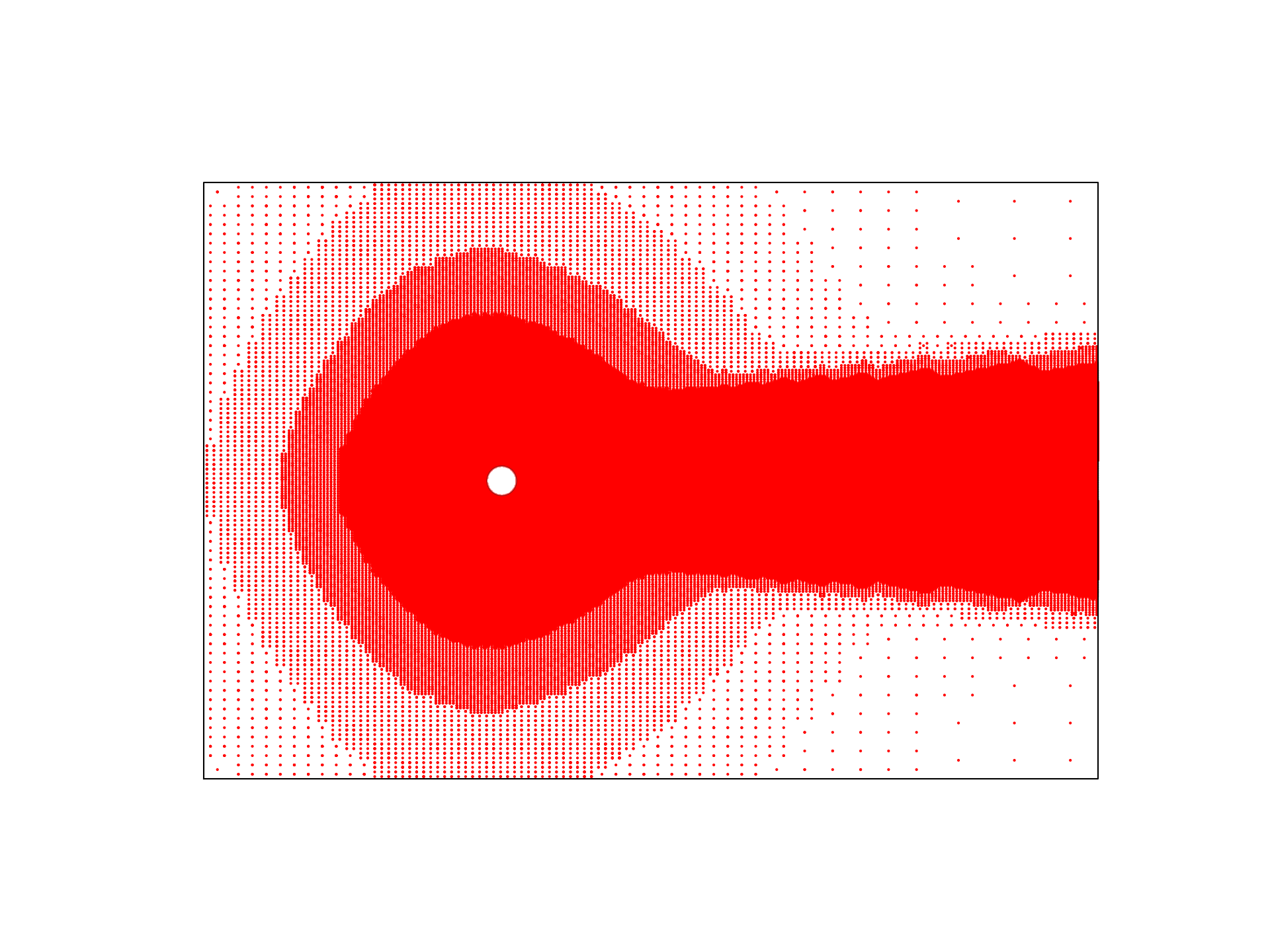}
        \caption{Quadtree refinement pass 6}
    \end{subfigure}

  \caption{Quadtree sampling points for flow around a cylinder.}
  \label{fig:qdtree_sampling}
\end{figure}

Examples of distributed sampling data points for conventional fluid simulation grid points and quadtree sampling are shown for a circular cylinder case in Fig. \ref{fig:qdtree_sampling}. In this example, after 4 uniform refinement passes, top 20\% refinement and bottom 20\% derefinement criteria are applied.

\section{Results}\label{sec:results}
The examples presented here demonstrate the utility of our method for representing steady fluid flows for initialization of high-fidelity simulations. While the data processing and model training procedures outlined earlier are general with respect to the equation set and the solver used for the high-fidelity simulations, the best treatment of the predicted flow fields for CFD initialization is problem and solver dependent. In the following demonstrations, high-fidelity solutions refer to steady-state solutions to incompressible Navier-Stokes equations:

\begin{equation}
    \pd{\bm{u}}{t}+\prt{\bm{u}\cdot\nabla}\bm{u}+\nu_{0}\nabla^{2}\bm{u}=-\frac{1}{\rho_{0}}\nabla p + g, ~~ \nabla \cdot \bm{u} = 0.
\end{equation}

\noindent
The solutions were found using {\it{Parabol}} multi-physics analysis and design optimization tool with extensive capabilities for concurrent shape and topology optimization \cite{makhija_spiral:_2017,makhija_concurrent_2019,makhija_concurrent_2020}. The analysis is carried out using the finite element method in {\it{Parabol}}, and details of the analysis implementation of the two-scale residual-based variational multiscale method for incompressible Navier-Stokes \cite{ahmed_review_2017} can be found in Appendix 1 of \cite{makhija_concurrent_2020}. 

Potential $\Phi$, stream function $\Psi$, and the velocity magnitude $U$ of the potential flow solution were obtained analytically via conformal mapping and used to prepare the EIFs. The potential flow velocity components and pressure $\prt{\ulf, \vlf, \plf}$ were found as:

\begin{equation}\label{eq:u_LF}
    \ulf = \frac{\partial \Phi}{\partial x},
\end{equation}

\begin{equation}\label{eq:v_LF}
    \vlf = \frac{\partial \Phi}{\partial y}, 
\end{equation}

\begin{equation}\label{eq:p_LF}
    \plf = p_{0} + \frac{1}{2}\rho_{0}U_{0}^{2} - \frac{1}{2}\rho_{0}\left(\ulf ^{2} + \vlf^{2}\right),
\end{equation}

\noindent
and used as the baseline solution. Here, zero subscripts refer to the freestream values. ML models were trained to predict the discrepancy between the baseline and high-fidelity solutions. All the flow fields were scaled by non-dimensionalization factors throughout the ML training process ($U_{0}$ for the velocity components and $1/\rho_{0}U_{0}^{2}$ for the pressure). In addition, input and output data scaling using the minmax formula in Eq. (\ref{eq:minmax-data-scaling}) was applied to facilitate the training process further: 

\begin{equation}\label{eq:minmax-data-scaling}
    \delta \bar{q} = \frac{\delta q-\delta q_{min}}{\delta q_{max}-\delta q_{min}}.
\end{equation}

Fully connected neural networks with 10 hidden layers and 100 neurons in each layer with sine activation functions were used. The ML model construction, model parameter differentiation and optimization were all carried out using {\it{TensorFlow version 2}} \cite{tensorflow2015-whitepaper}. L-BFGS optimization tool in the {\it{TensorFlow Probability}} module \cite{lbfgs_minimize} was used to train the ML models, utilizing a {\it{Python}} wrapper by PY \cite{py_optimize_2019} to extract and update the neural network model parameters and gradients. A desktop machine with Intel Core i7-4790K processor at 4GHz and 32GB RAM was used.

\subsection{Circular Cylinder}\label{sec:ex-circle}
The problem of incompressible flow around a circular cylinder at a low Reynolds number was studied in the first example. The flow solver {\it{Parabol}} has been verified for steady and unsteady flow over a cylinder, see \cite{makhija_concurrent_2020} Section 4.1. The problem specifications including the fluid properties and flow conditions used in this example are summarized in Tab. \ref{tab:cylinder-specs}. Radius of the circular cylinder, freestream speed, kinematic viscosity, fluid density and Reynolds number are denoted $a$,  $U_{0}$, $\nu_{0}$, $\rho_{0}$ and \Rey , respectively. The radius of the circular cylinder was considered as a problem parameter in this study, and the range of values used throughout this study is provided in Tab. \ref{tab:cylinder-specs}. Reynolds number changes as the cylinder radius is altered through changes in the characteristic length; the Reynolds number range associated with the radius range is also provided.

\begin{table}[!ht]
\small
\begin{center}
\begin{tabular}{ c|ccccc }
    quantity & cylinder radius & freestream speed & kinematic viscosity & fluid density & Reynolds number \\ 
    \hline
    symbol & $a$ & $U_{0}$ & $\nu_{0}$ & $\rho_{0}$ & \Rey\\
    value & [0.045,0.055] & 0.2 & 0.001 & 1.0 & [18,22]\\
\end{tabular}
\caption{Problem specifications of the cylinder in flow example.}\label{tab:cylinder-specs}
\end{center}
\end{table}

\begin{figure}[!ht]
    \centering
    \includegraphics[width=2.8 in]{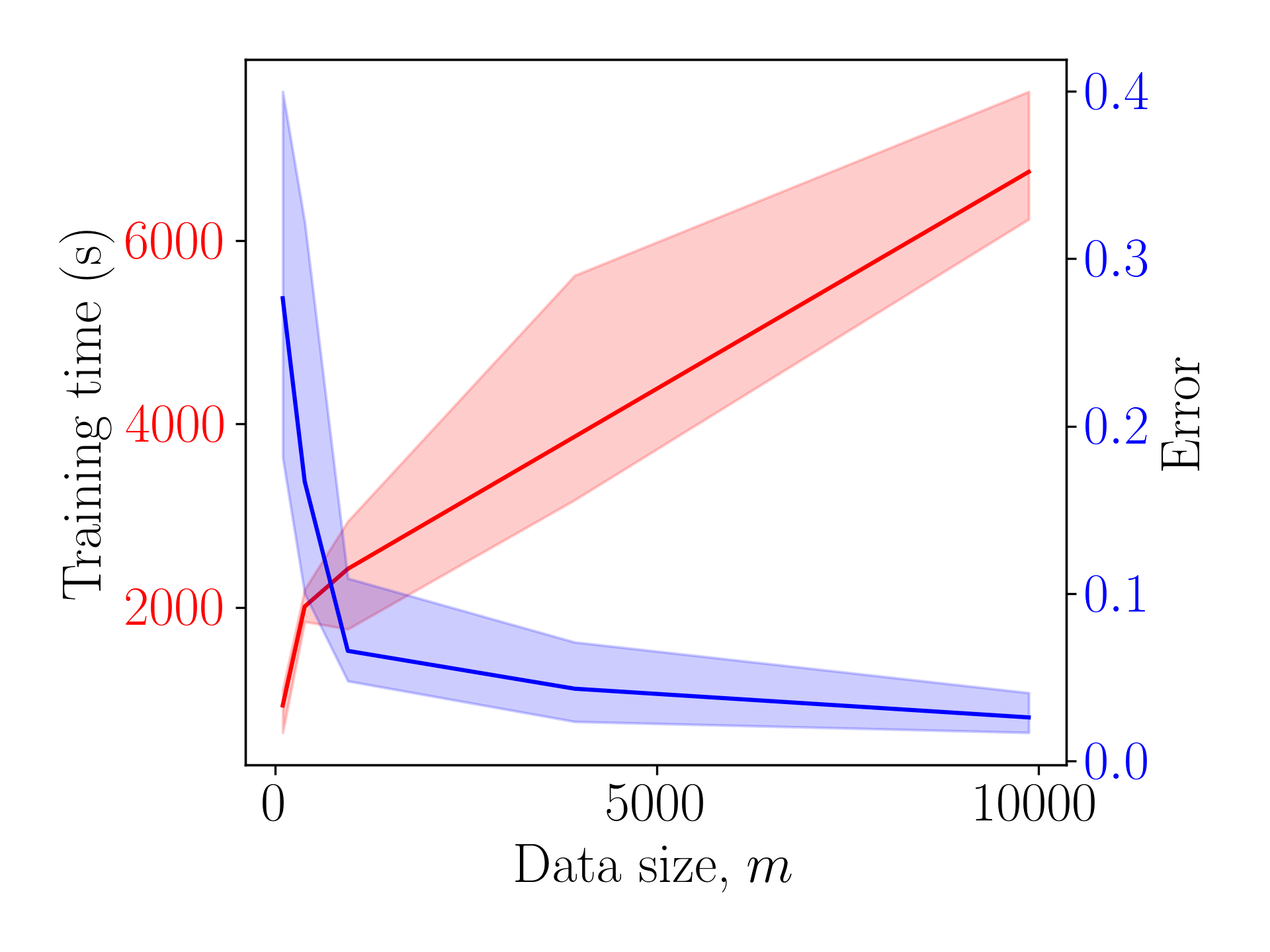}
    \caption{ML model training time scaling over data size $m$ (red). The model training was carried out using uniformly distributed data points within the training window, at $m=$ 96, 384, 949, 3921 and 9870. Sum of integrated $L^{2}$ errors from the three flow fields, $u$, $v$ and $p$ are reported (blue).}
    \label{fig:training-time-scaling}
\end{figure}

First, the ML model training for a reference problem with cylinder radius $a$=0.05 was carried out using various numbers of sampling data points distributed uniformly throughout a training window of $l_{x} \times l_{y}=30a \times 20a$. The training time and errors for the ML training processes at various data sizes are plotted in Fig. \ref{fig:training-time-scaling}. Solid curves and colored shades show averaged values and ranges over 5 ML training runs with different random initial model parameters. The plot shows a general trend of increasing ML training time and decreasing error with data size. Furthermore, the rapid drop in the error near $m$=1000 suggests that a good trade-off between training cost and accuracy occurs at a relatively small data size. When a relatively small sampling locations are used, it becomes important to determine an efficient and effective sampling method. 

\begin{figure}[!ht]
    \centering
    \begin{subfigure}{0.3\textwidth}
        \includegraphics[width=\linewidth,trim={.8in 0.5in .8in 0.5in},clip]{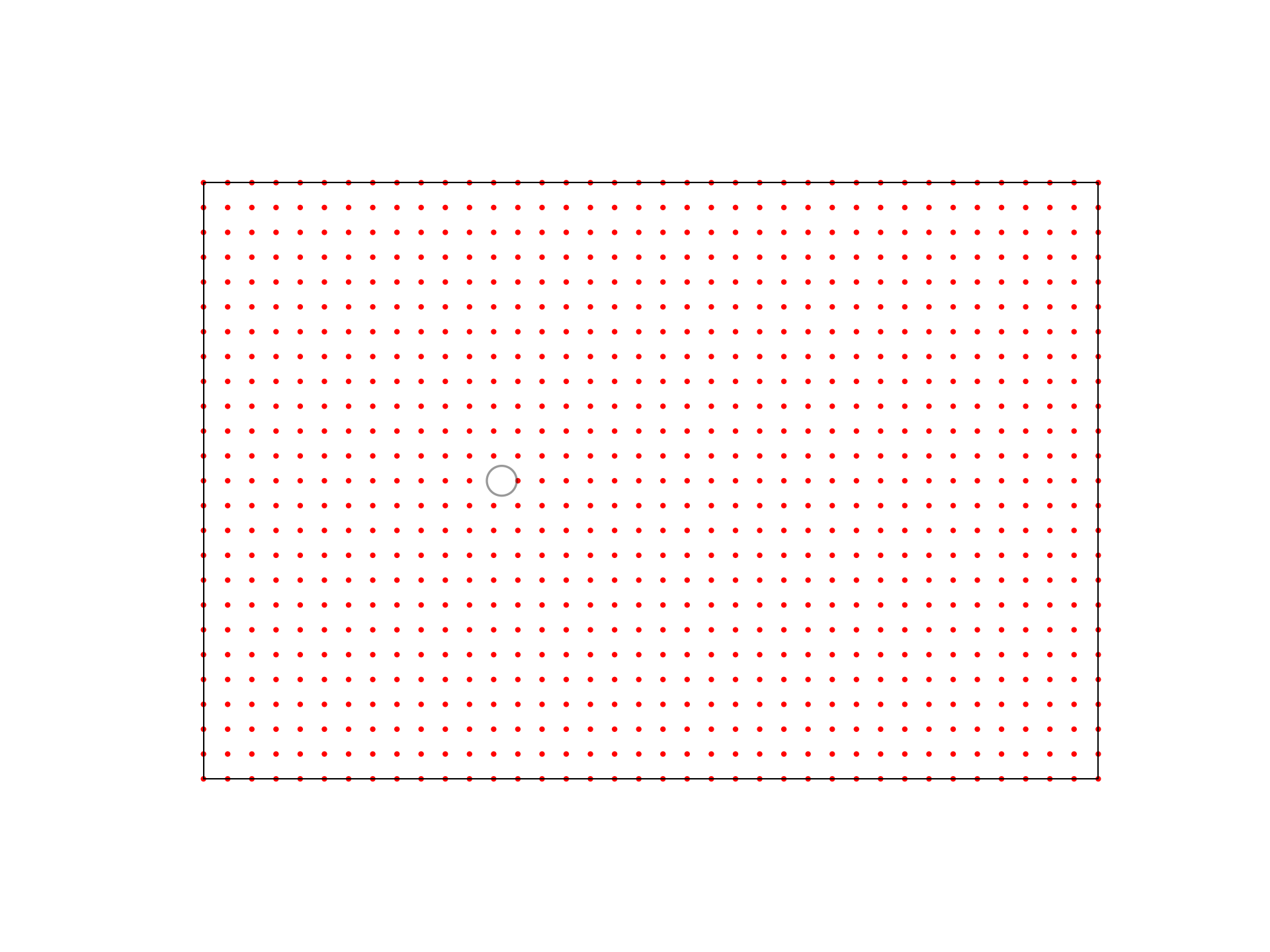}
        \caption{Uniform sampling}
    \end{subfigure}
    \begin{subfigure}{0.3\textwidth}
        \includegraphics[width=\linewidth,trim={.8in 0.5in .8in 0.5in},clip]{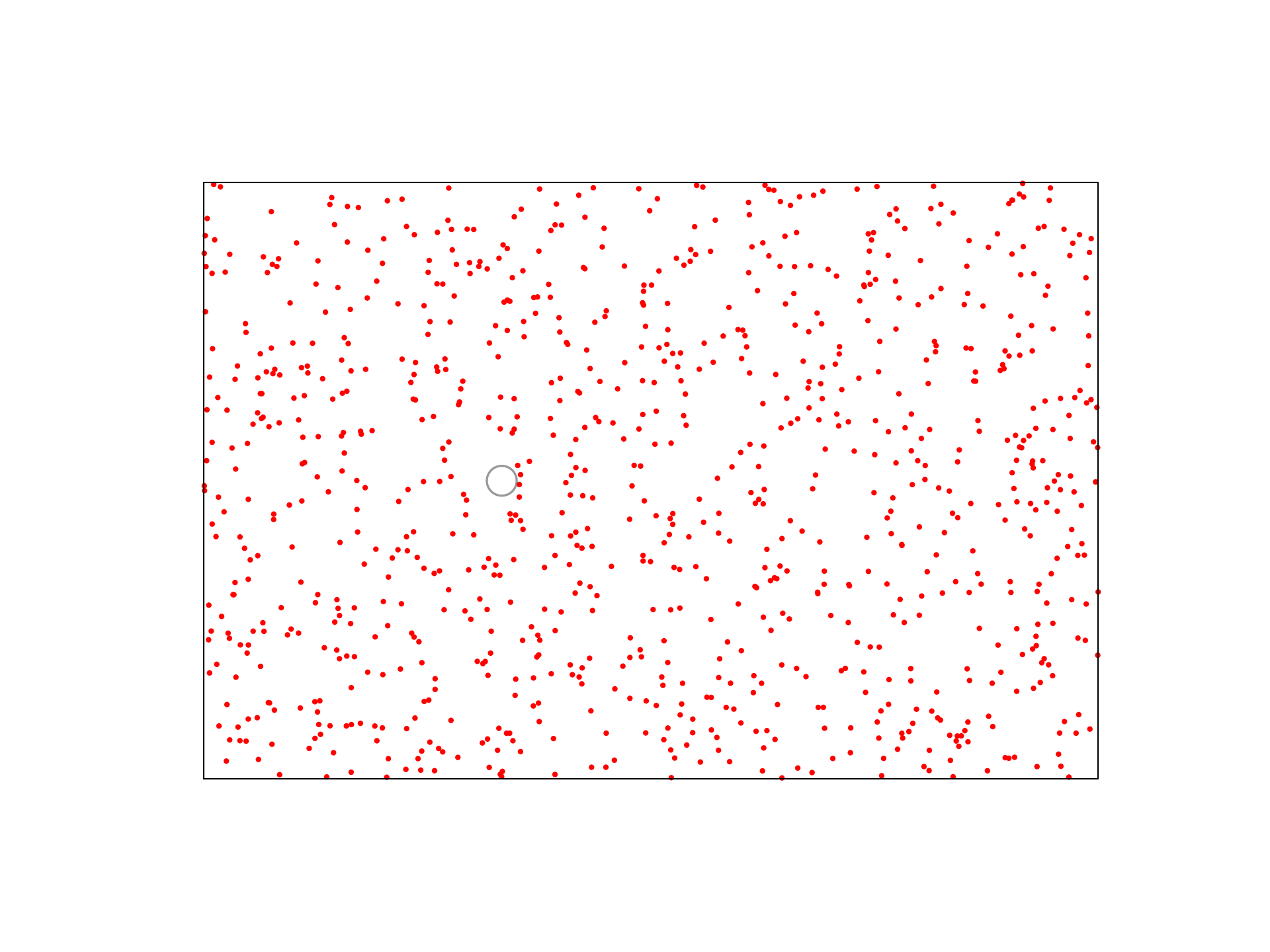}
        \caption{Random sampling}
    \end{subfigure}
\\
    \begin{subfigure}{0.3\textwidth}
        \includegraphics[width=\linewidth,trim={.8in 0.5in .8in 0.5in},clip]{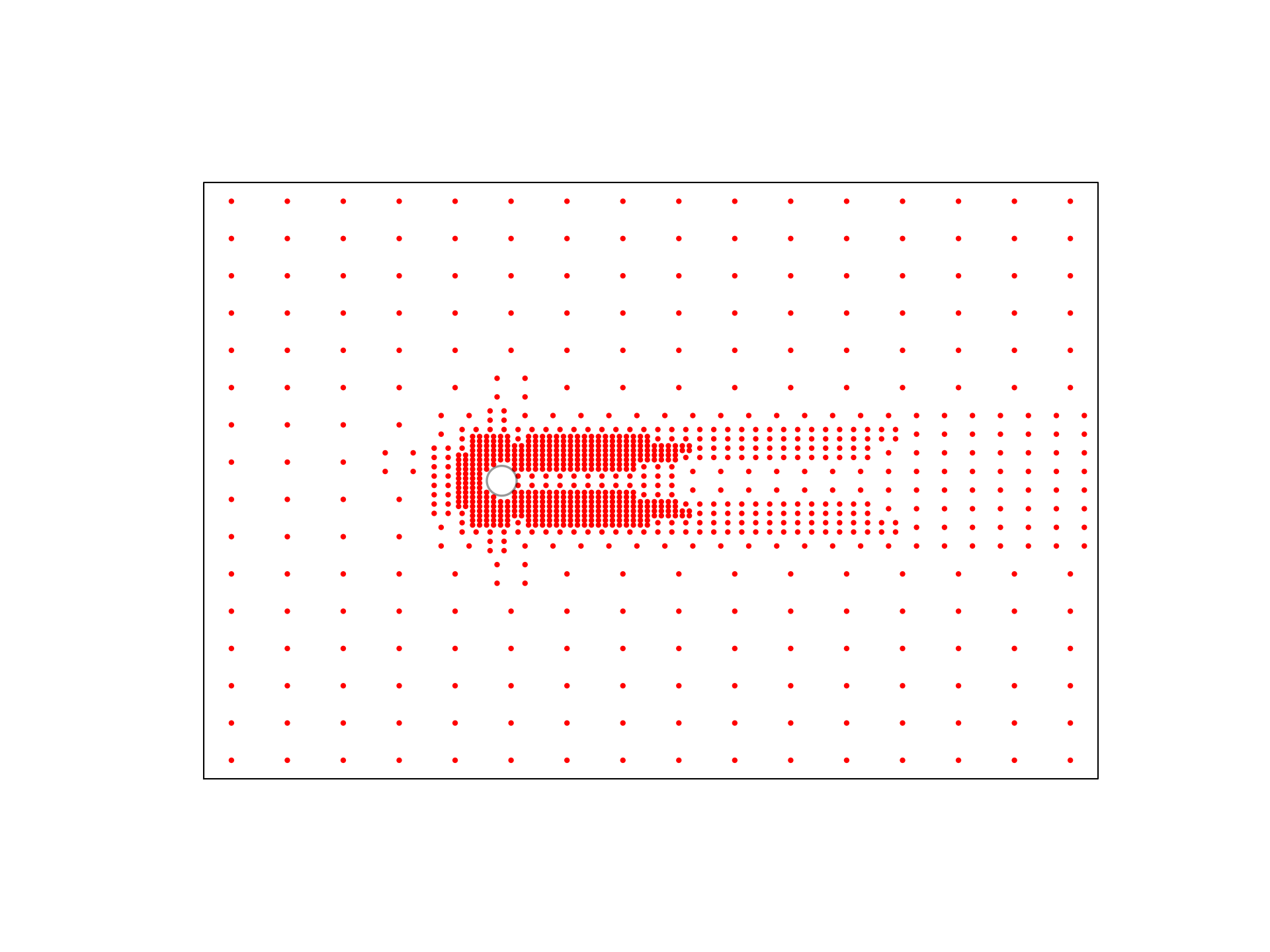}
        \caption{Quadtree sampling}
    \end{subfigure}
    \begin{subfigure}{0.3\textwidth}
        \includegraphics[width=\linewidth,trim={.4in 0.2in .4in 0.2in},clip]{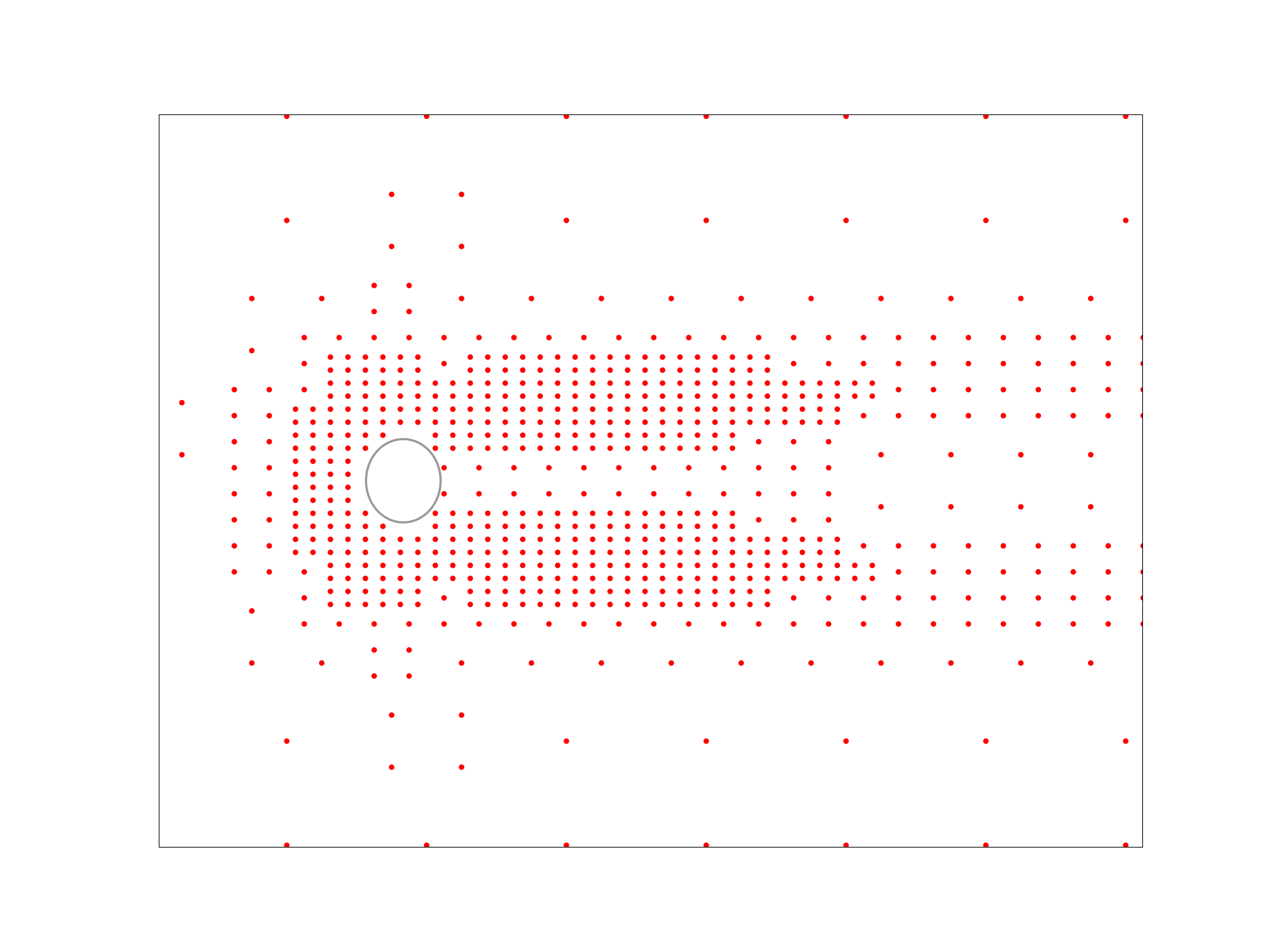}
        \caption{Quadtree sampling, zoomed-in}
    \end{subfigure}
    \caption{Distributions of 949, 950 and 1001 data points for the problem of flow around a circular cylinder using (a) uniform, (b) random, and (c,d) quadtree sampling methods, respectively.}
    \label{fig:cylinder-sampling-methods-comp}
\end{figure}

Next, the three data sampling techniques discussed in Sec. \ref{sec:sampling-methods} were compared. The training data consisting of 949, 950, and 1001 points were prepared using the uniform, random and quadtree sampling methods, respectively (see Fig. \ref{fig:cylinder-sampling-methods-comp} for the data distributions). Data points that fell within the body were removed from the uniformly and randomly distributed data retroactively. The quadtree sampling method eliminated points within the body through the adaptive process, and the refinement process was terminated once the requested number of points was reached. While the uniformly distributed data points cover much of the training window at a reasonable resolution away from the body, there is a rectangular region around the body with no data points. Likewise, presence of a cluster of data points around the body is generally not guaranteed with randomly distributed data, but higher local resolutions are observed at random locations. The quadtree sampling leads to a higher density of points around the body and a sparse data distribution away from the body, as seen in Fig. \ref{fig:cylinder-sampling-methods-comp} (c,d).

\begin{figure}[!ht]
    \centering
   \begin{subfigure}{0.32\textwidth}
        \includegraphics[width=\linewidth]{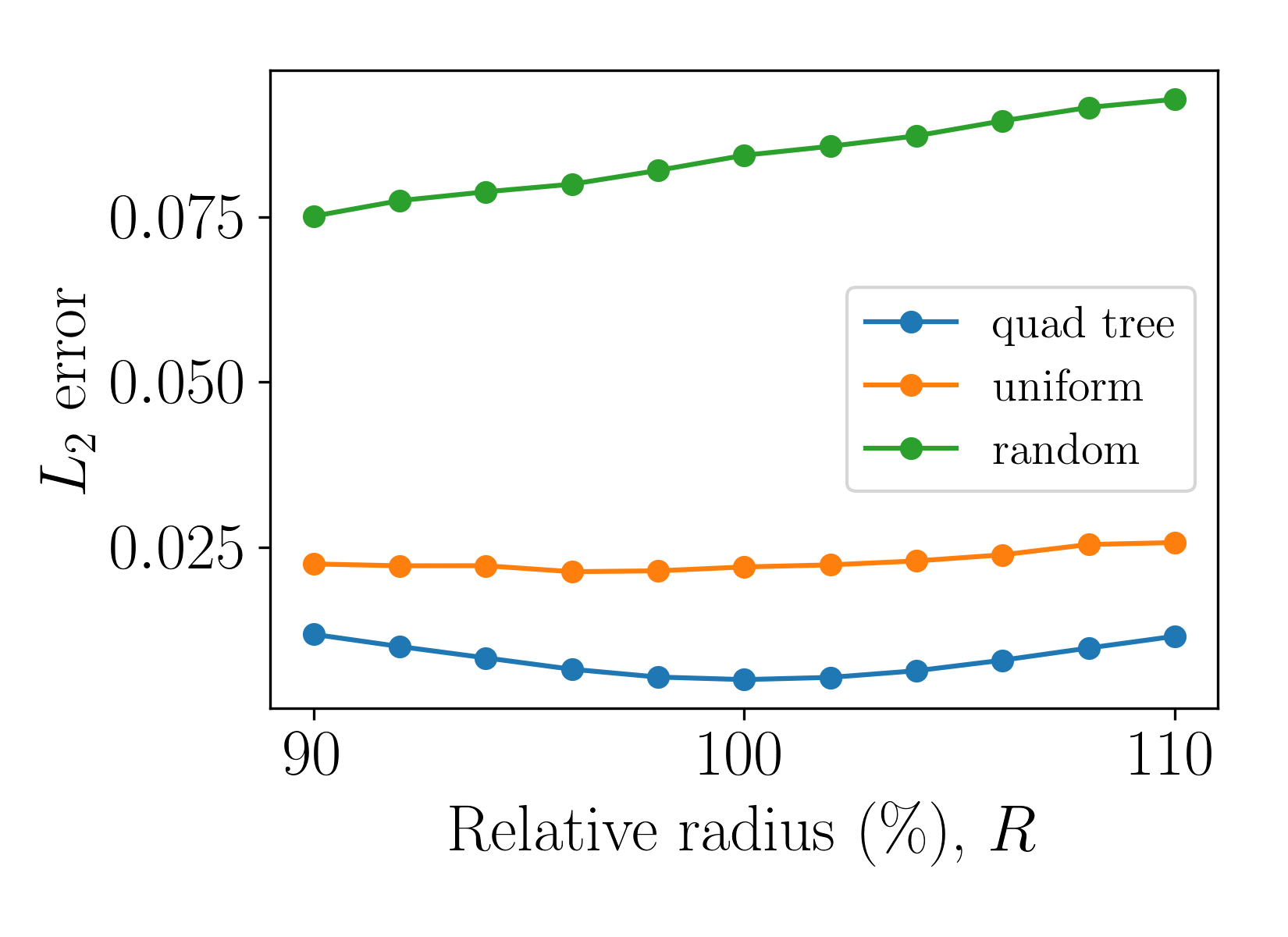}
        \caption{Velocity, $x-$component}
    \end{subfigure}
    \begin{subfigure}{0.32\textwidth}
        \includegraphics[width=\linewidth]{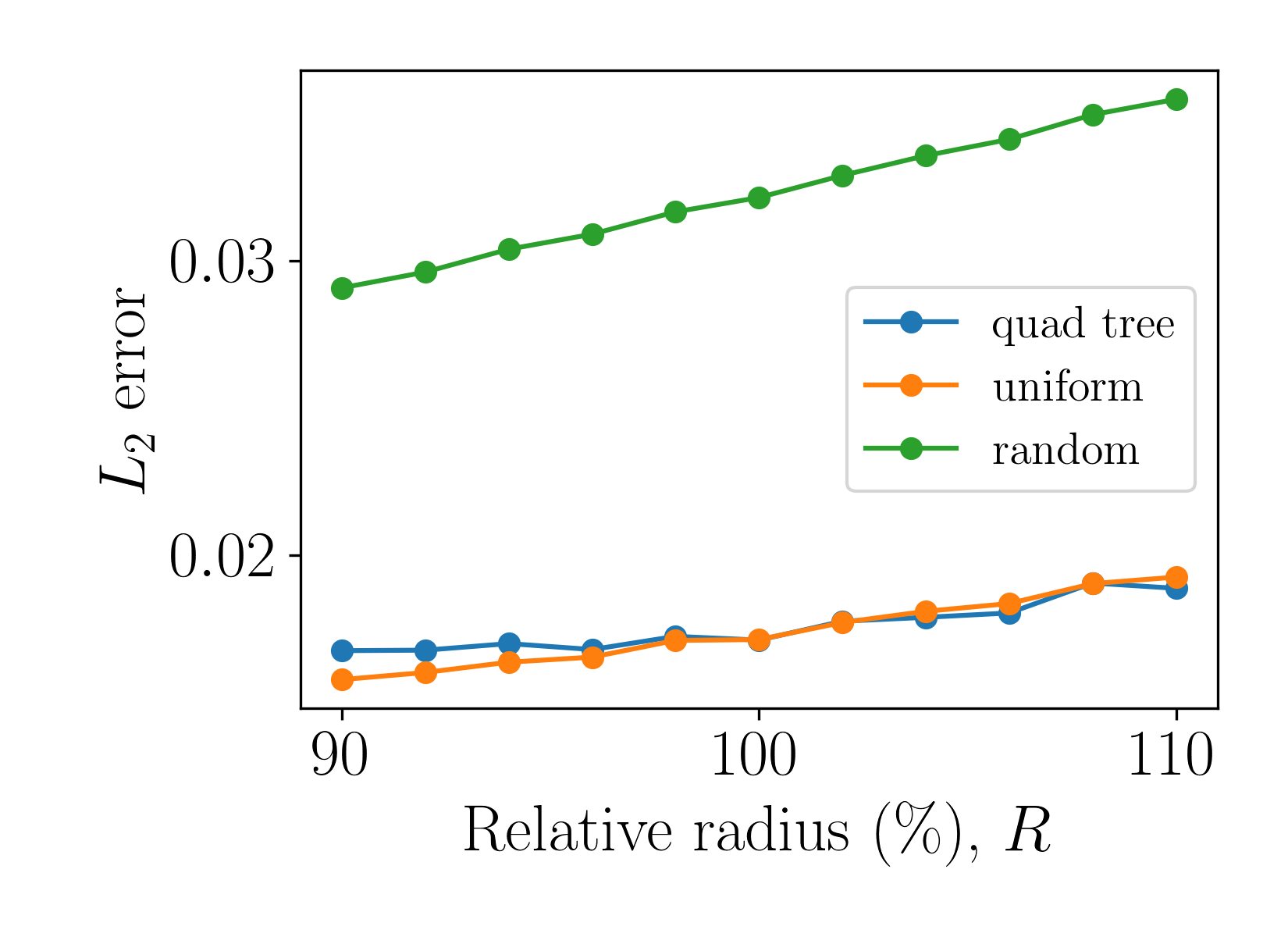}
        \caption{Velocity, $y-$component}
    \end{subfigure}
    \begin{subfigure}{0.32\textwidth}
        \includegraphics[width=\linewidth]{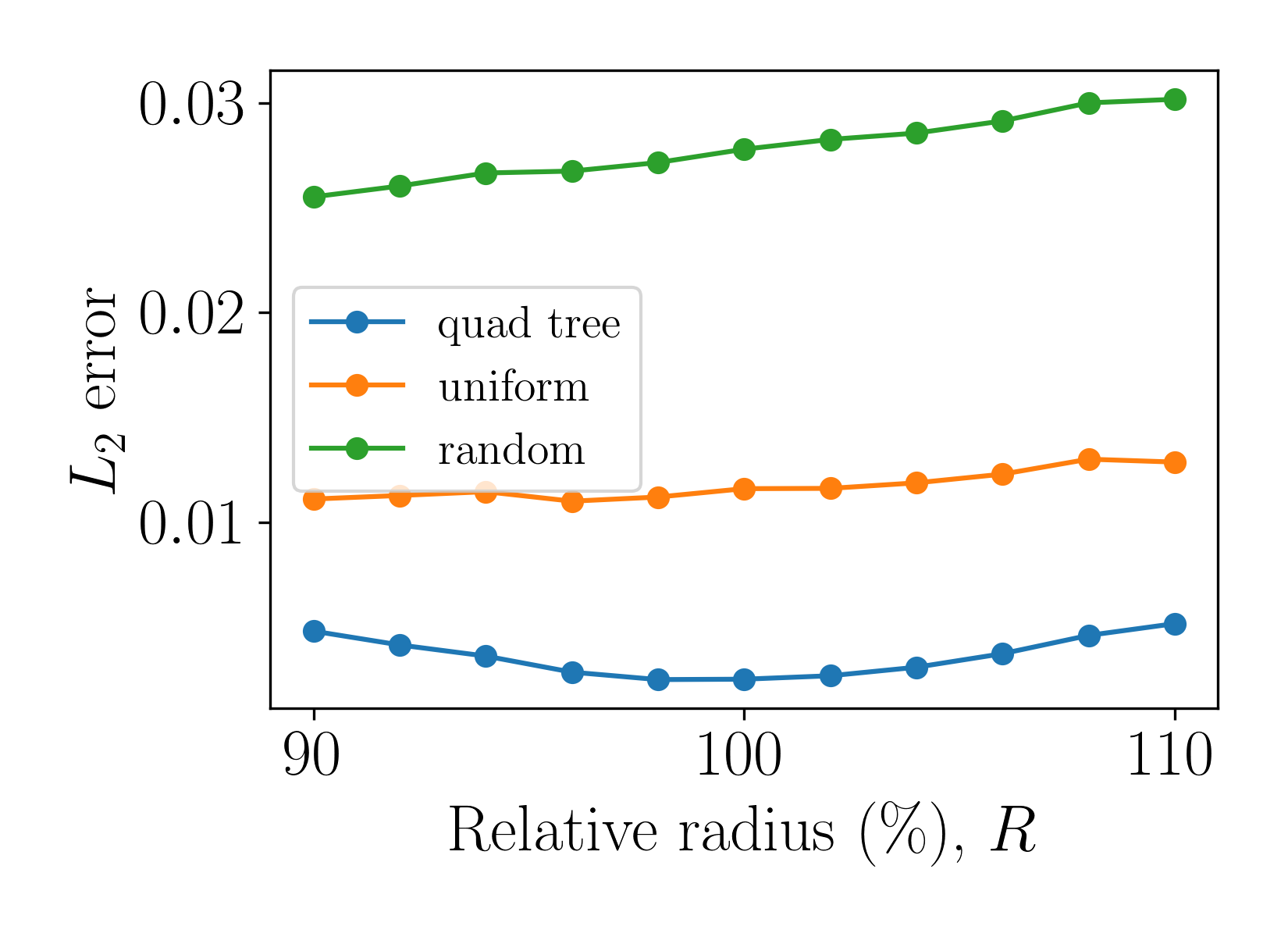}
        \caption{Pressure}
    \end{subfigure}
    \caption{Prediction errors for flow around a circular cylinder using ML models trained at the reference circular radius of $a=0.05$ ($R=$ 100\%) for the (a) $x$ and (b) $y-$components of the velocity and (c) the pressure.}
    \label{fig:cylinder-sampling-error-comp}
\end{figure}

\begin{figure}[!ht]
  \centering
  \begin{subfigure}{0.35\textwidth}
        \includegraphics[width=\linewidth,trim={0 0 1.2in 0},clip]{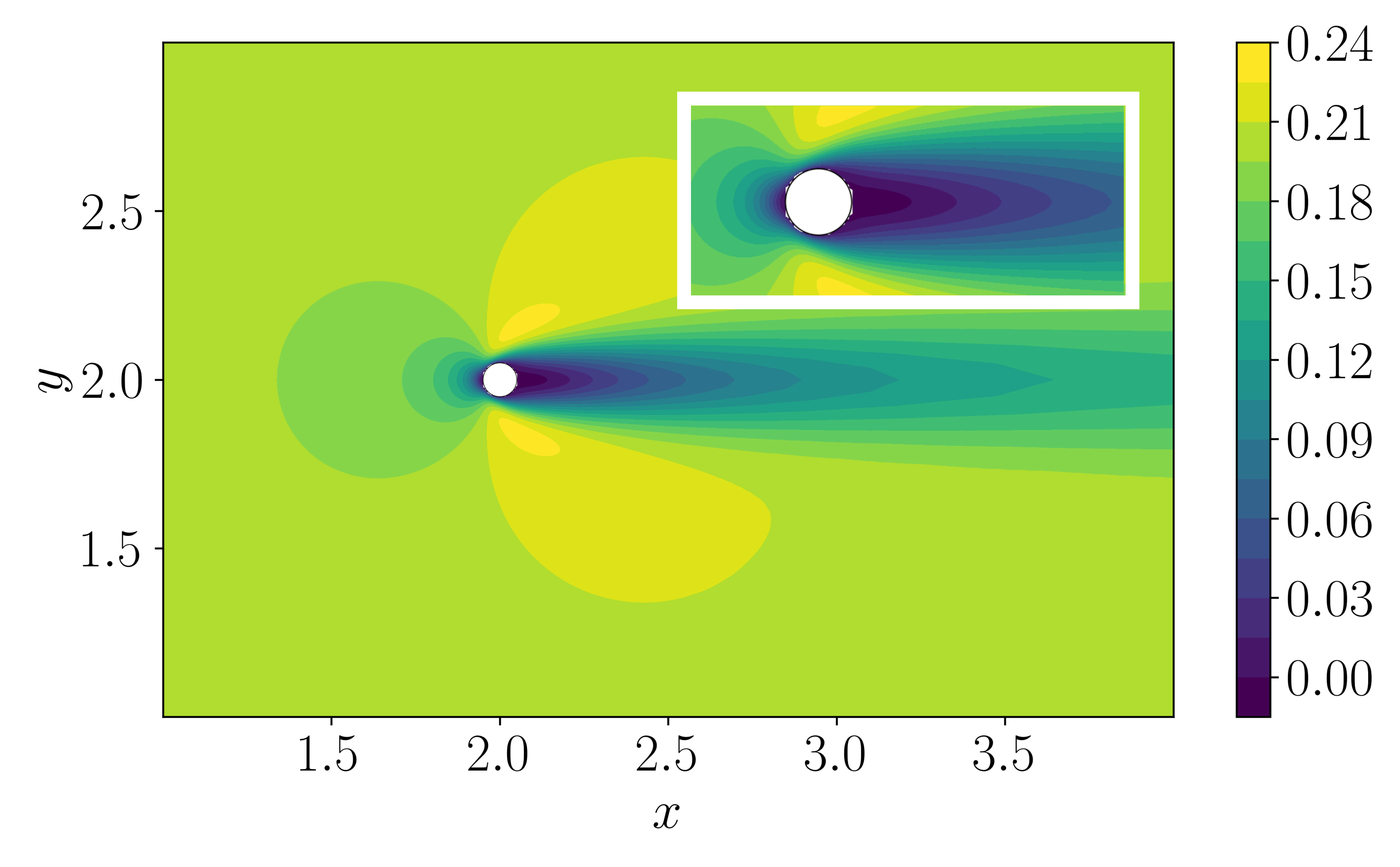}
        \caption{HF solution}
    \end{subfigure}
    \begin{subfigure}{0.35\textwidth}
        \includegraphics[width=\linewidth,trim={0 0 1.2in 0},clip]{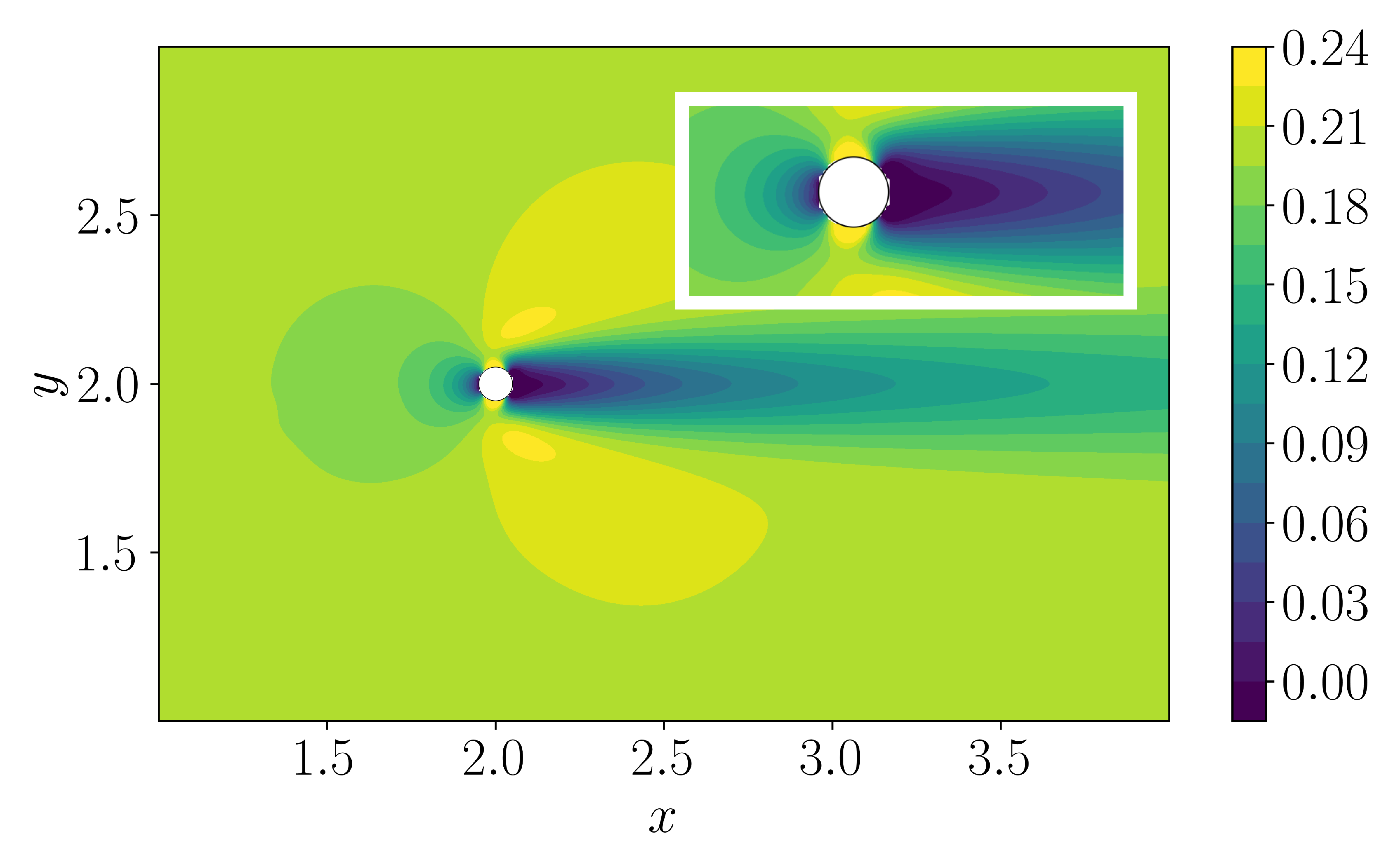}
        \caption{ML, uniform sampling}
    \end{subfigure}
\\
    \begin{subfigure}{0.35\textwidth}
        \includegraphics[width=\linewidth,trim={0 0 1.2in 0},clip]{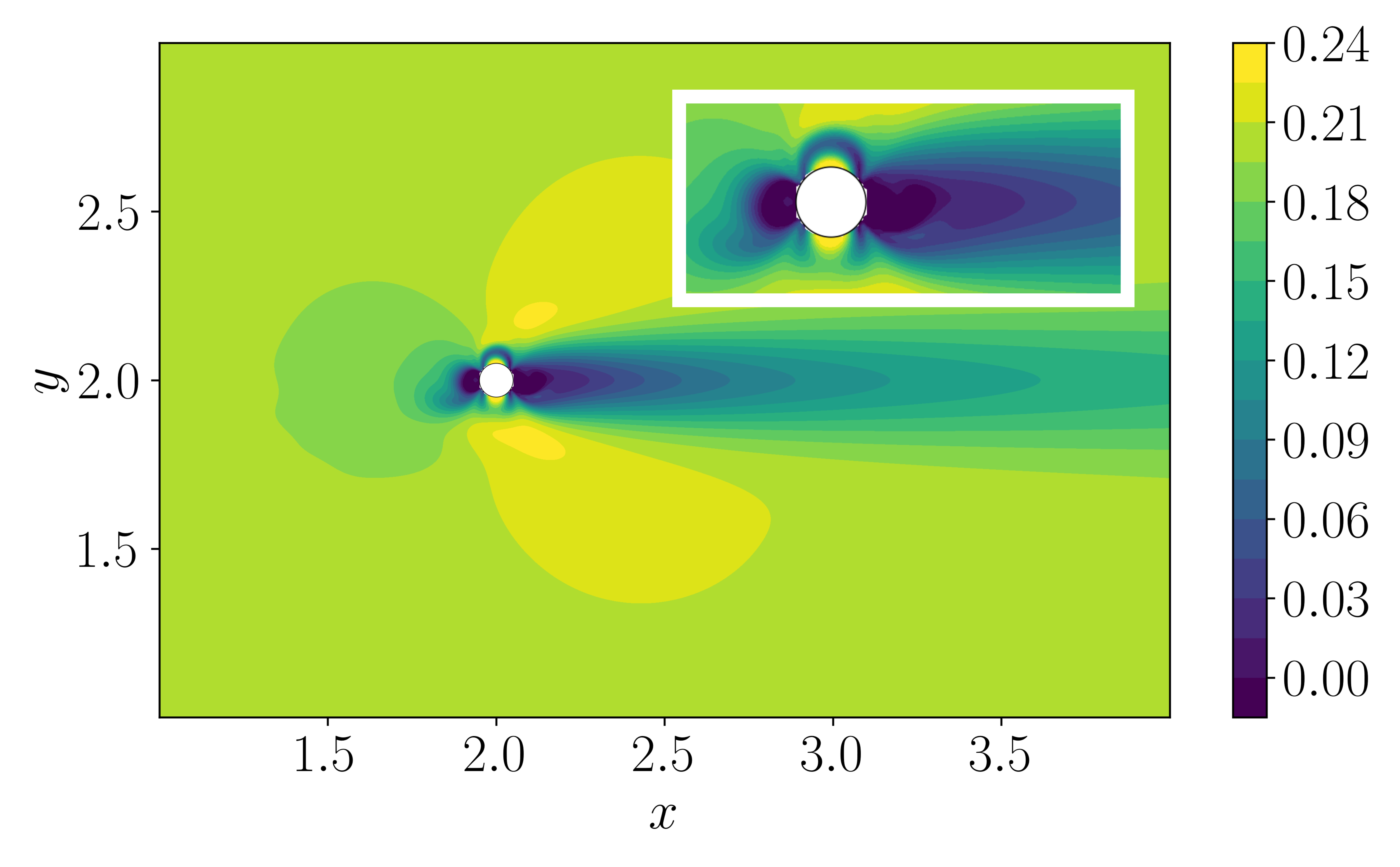}
        \caption{ML, random sampling}
    \end{subfigure}
    \begin{subfigure}{0.35\textwidth}
        \includegraphics[width=\linewidth,trim={0 0 1.2in 0},clip]{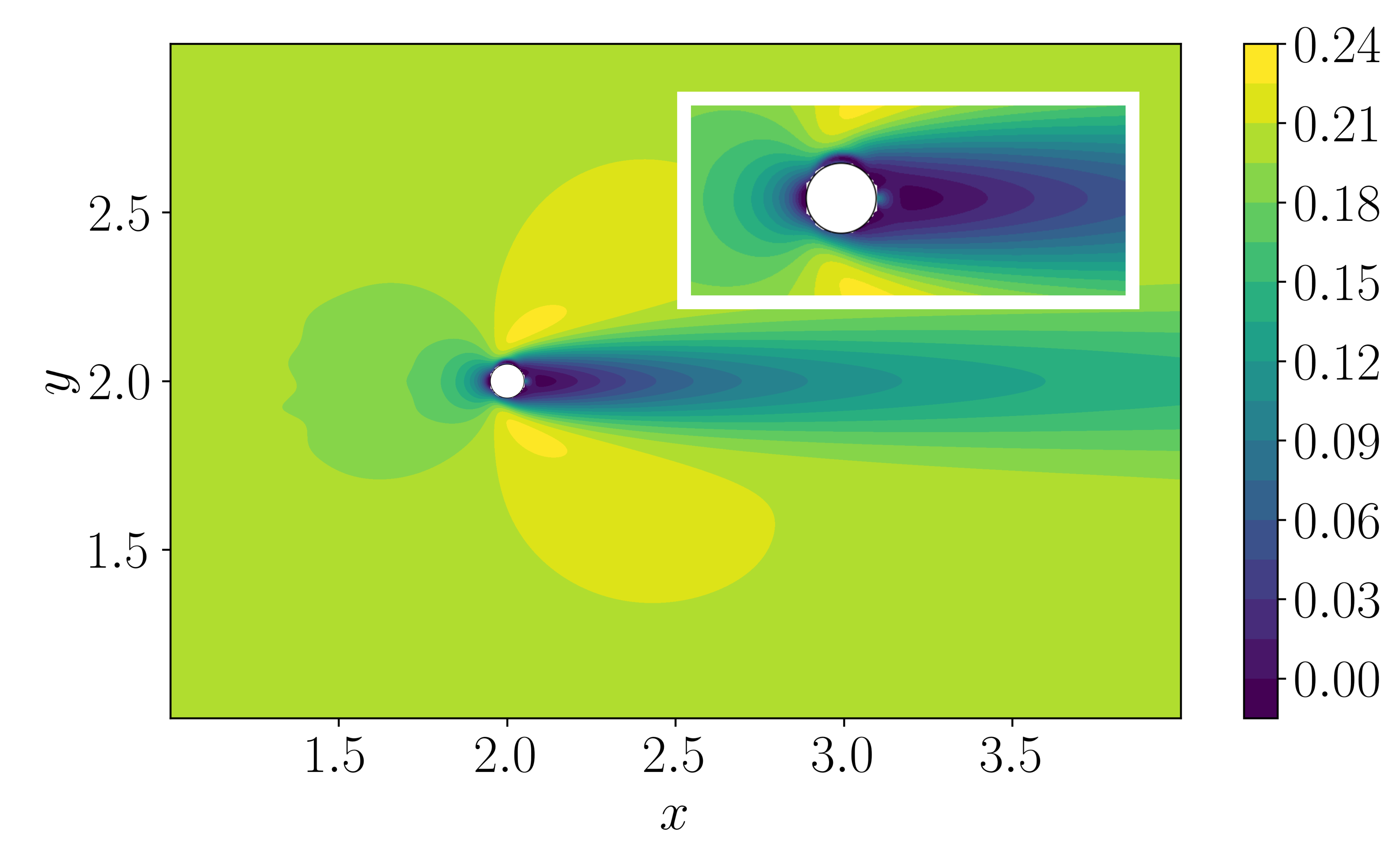}
        \caption{ML, quadtree sampling}
    \end{subfigure}
    \caption{ML-predicted $x-$component of the velocity at the reference circular radius of $R=$100\% (training point) is compared to the HF solution (a). The ML model training was carried out using (b) uniform, (c) random, and (d) quadtree sampling.}
  \label{fig:u-contour-circle}
\end{figure}

To assess the prediction capabilities of the machine-learned model, the flow fields were evaluated for problems at various cylinder radii. The predictions were made by finding the potential flow solutions to prepare the EIFs and the low-fidelity baseline solution, following Algorithm \ref{alg:prediction} in Appendix \ref{sec:appendix-alg}. In order to evaluate the prediction accuracy, high-fidelity solutions were prepared using a mesh of 27,900 points in {\it{Parabol}} for these problems. The integrated errors are reported over the relative percentage circle radius, $R$ (compared to the reference radius of $a=0.05$). 
The plots in Fig. \ref{fig:cylinder-sampling-error-comp} show smaller prediction errors using the quadtree sampling method at the training point at $R=$100\% for all the field quantities. 
The visual inspection of the field contours in Figs. \ref{fig:u-contour-circle} shows the quadtree sampling producing better matched flow features especially near the body.

\begin{figure}[!ht]
  \centering
  \begin{subfigure}{0.3\textwidth}
        \includegraphics[width=\linewidth,trim={0 0 0in 0},clip]{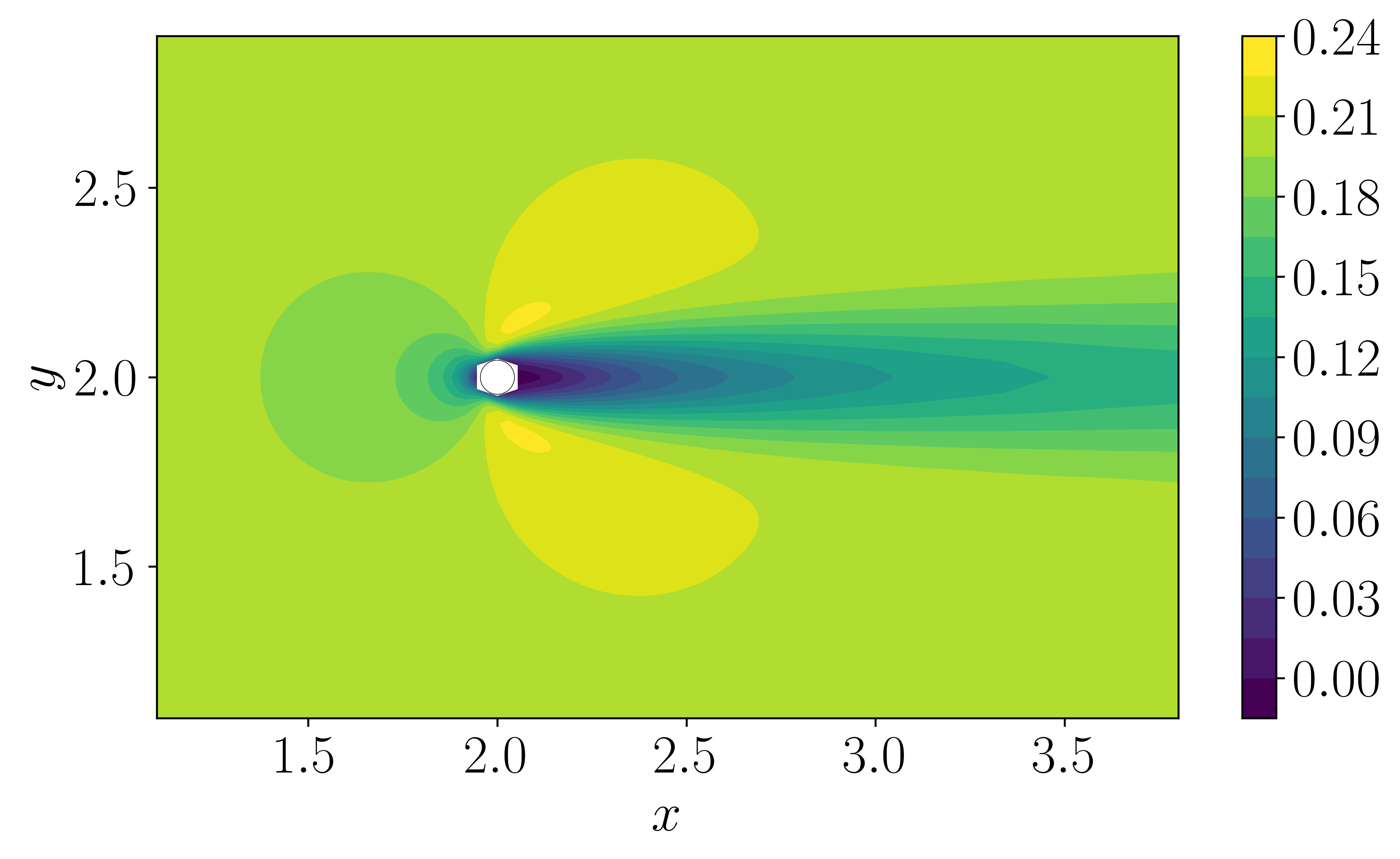}
        \caption{HF solution, $u$}
    \end{subfigure}
    \begin{subfigure}{0.3\textwidth}
        \includegraphics[width=\linewidth,trim={0 0 0in 0},clip]{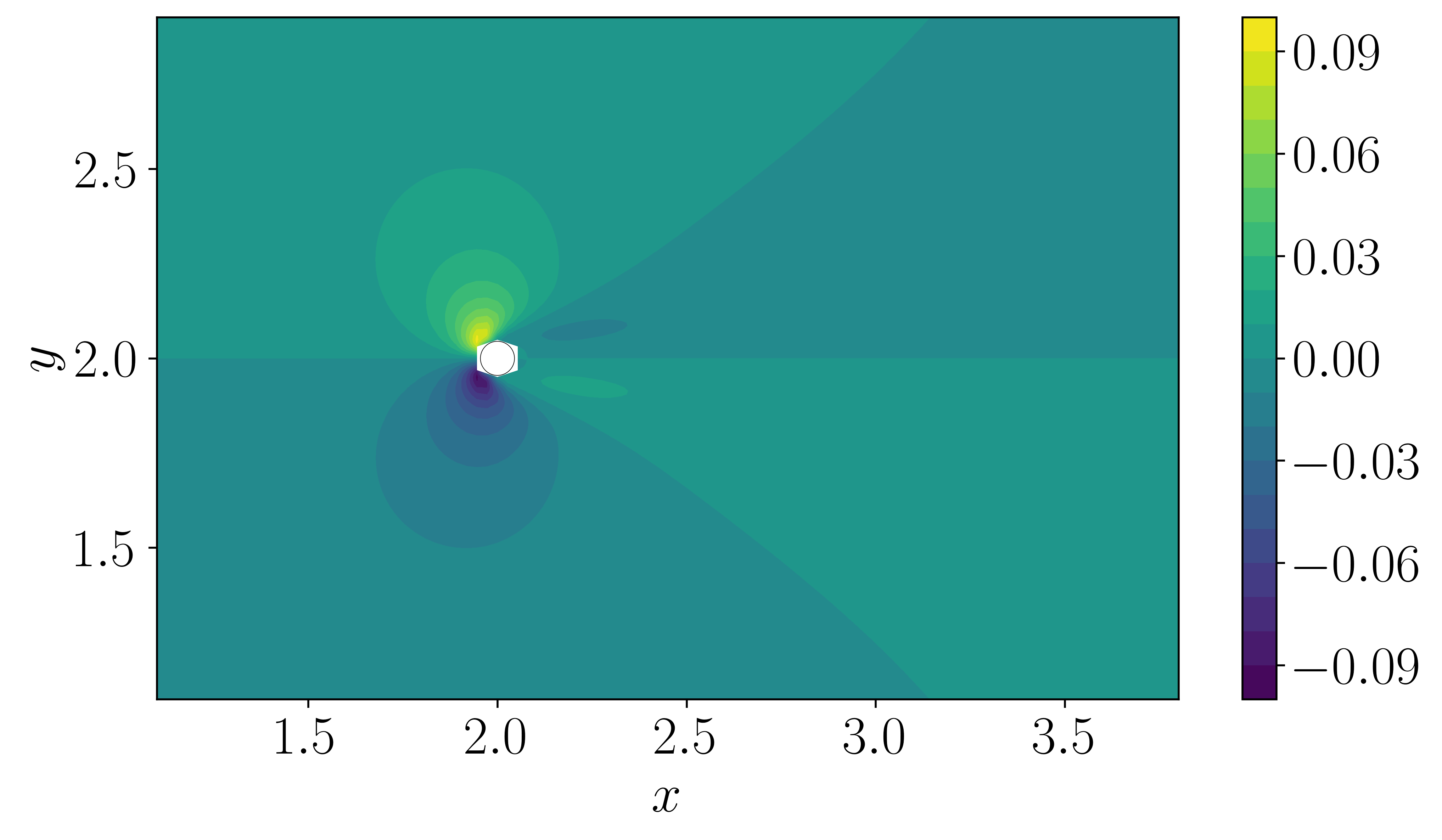}
        \caption{HF solution, $v$}
    \end{subfigure}
    \begin{subfigure}{0.3\textwidth}
        \includegraphics[width=\linewidth,trim={0 0 0in 0},clip]{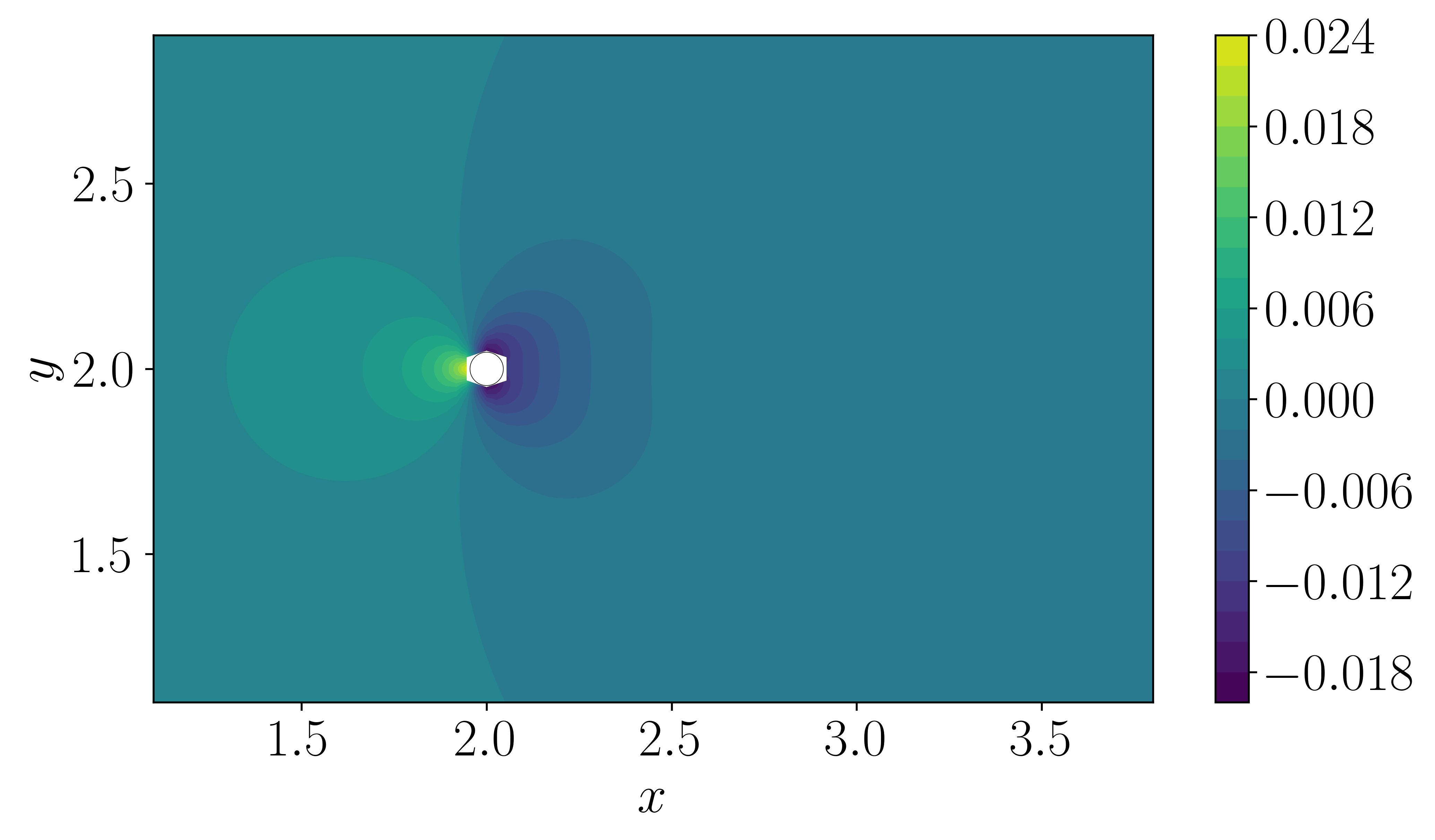}
        \caption{HF solution, $p$}
    \end{subfigure}
\\
  \begin{subfigure}{0.3\textwidth}
        \includegraphics[width=\linewidth,trim={0 0 0in 0},clip]{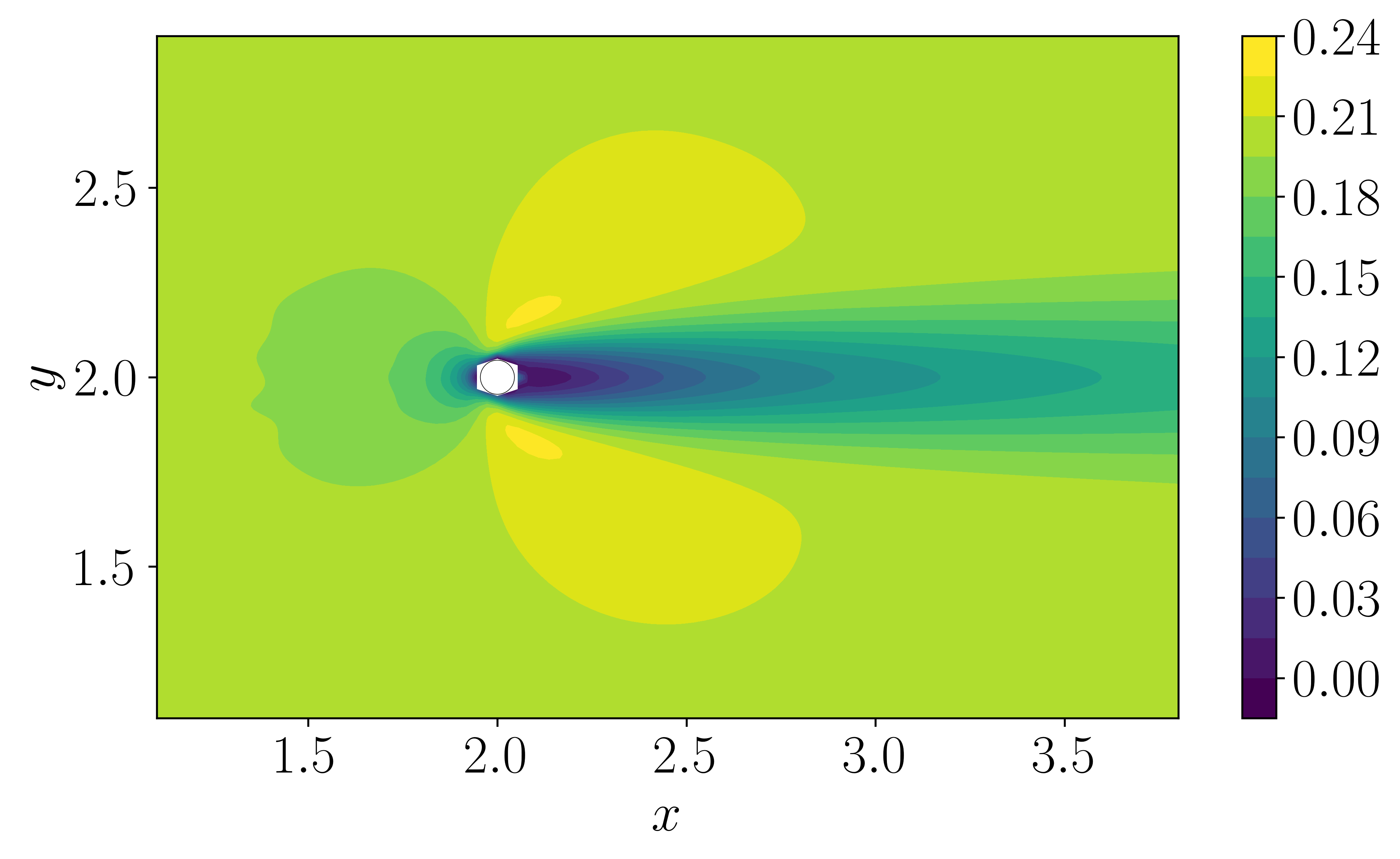}
        \caption{ML prediction, \uml}
    \end{subfigure}
    \begin{subfigure}{0.3\textwidth}
        \includegraphics[width=\linewidth,trim={0 0 0in 0},clip]{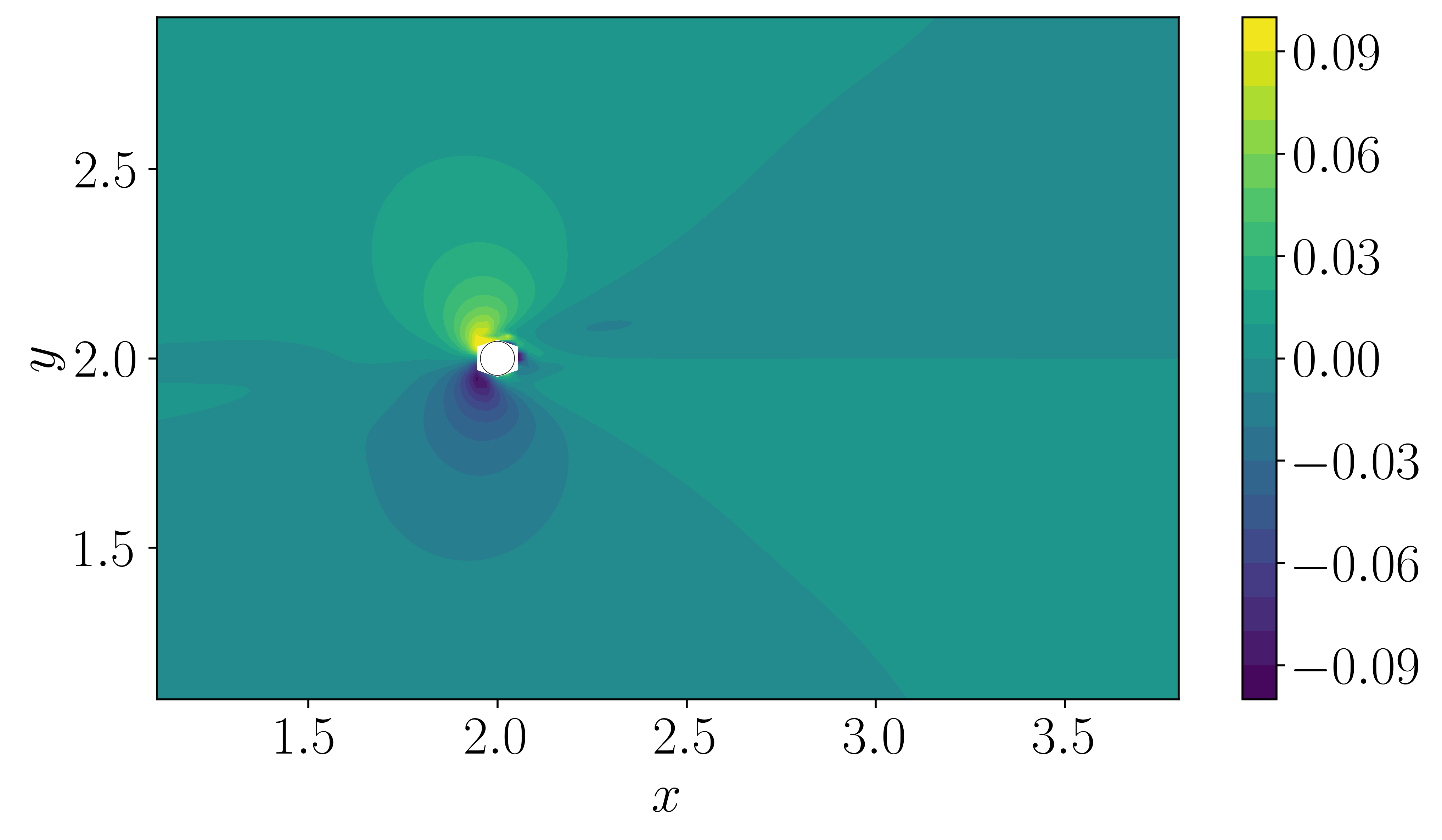}
        \caption{ML prediction, \vml}
    \end{subfigure}
    \begin{subfigure}{0.3\textwidth}
        \includegraphics[width=\linewidth,trim={0 0 0in 0},clip]{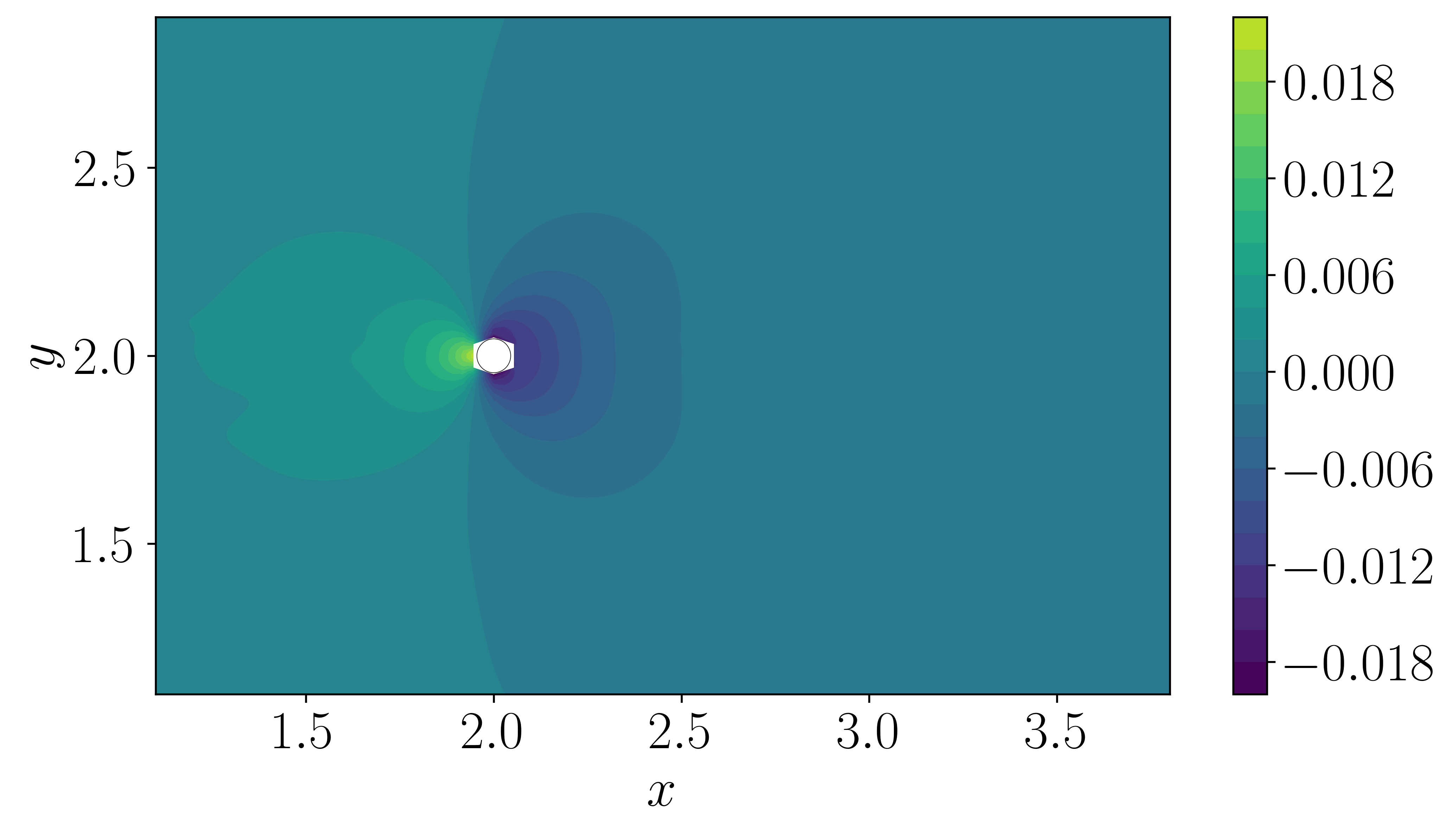}
        \caption{ML prediction, \pml}
    \end{subfigure}
    \caption{The trained ML model was used to predict the flow fields at a smaller circular cylinder at $R=$90\% reference radius (d-f) and compared to the HF solution (a-c).}
  \label{fig:u-predict-90-circle}
\end{figure}

\begin{figure}[!ht]
  \centering
  \begin{subfigure}{0.3\textwidth}
        \includegraphics[width=\linewidth,trim={0 0 0in 0},clip]{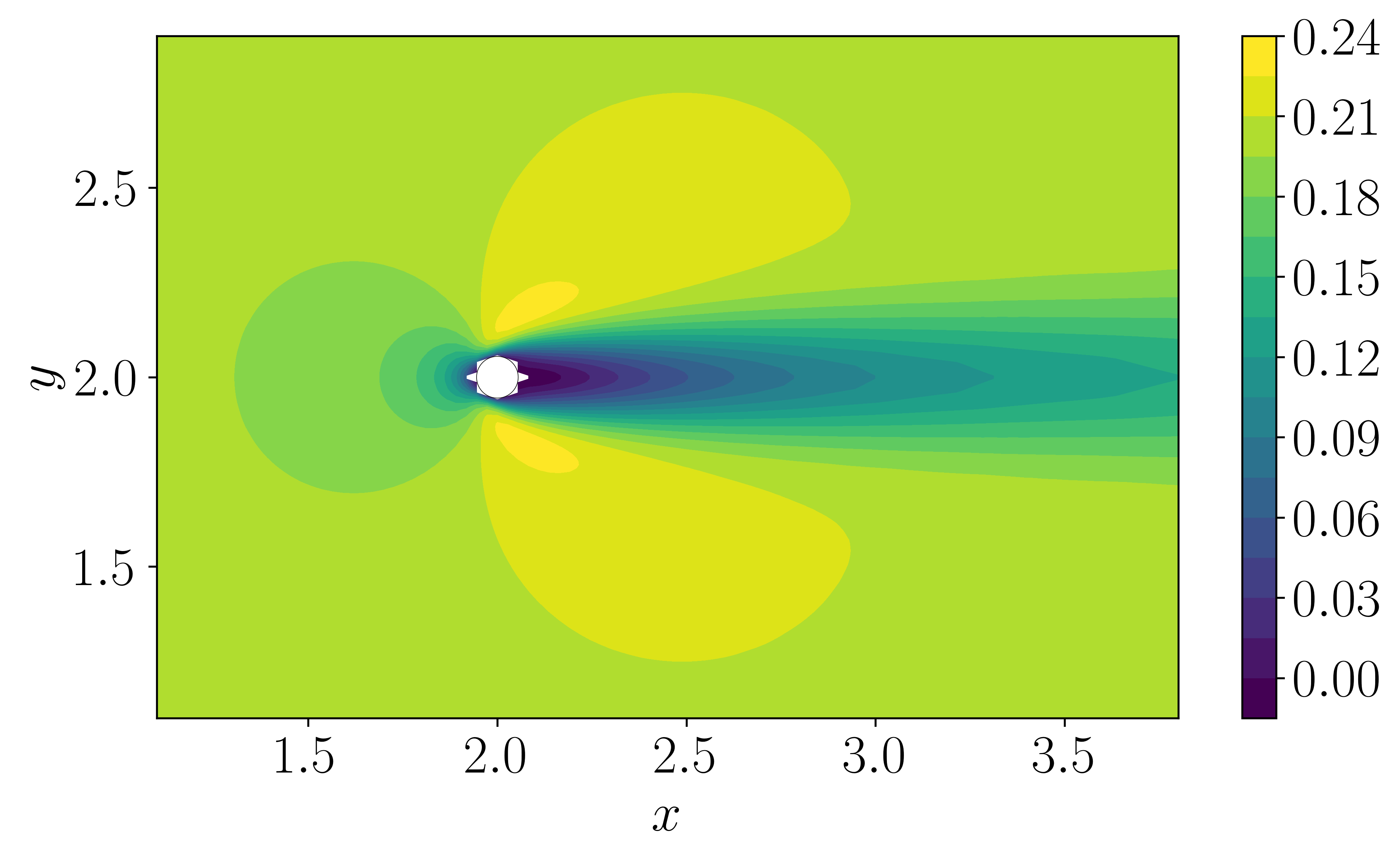}
        \caption{HF solution, $u$}
    \end{subfigure}
    \begin{subfigure}{0.3\textwidth}
        \includegraphics[width=\linewidth,trim={0 0 0in 0},clip]{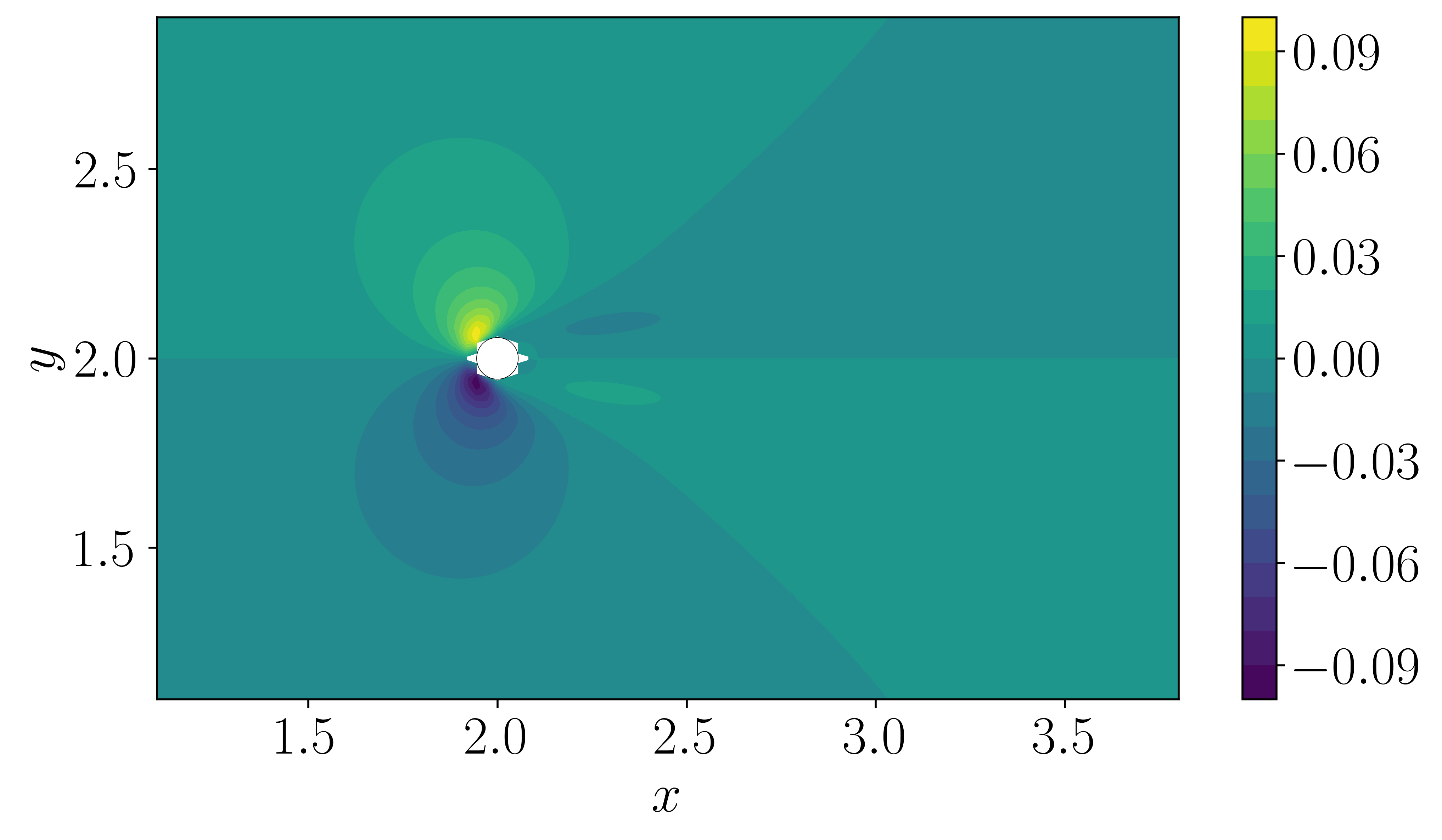}
        \caption{HF solution, $v$}
    \end{subfigure}
    \begin{subfigure}{0.3\textwidth}
        \includegraphics[width=\linewidth,trim={0 0 0in 0},clip]{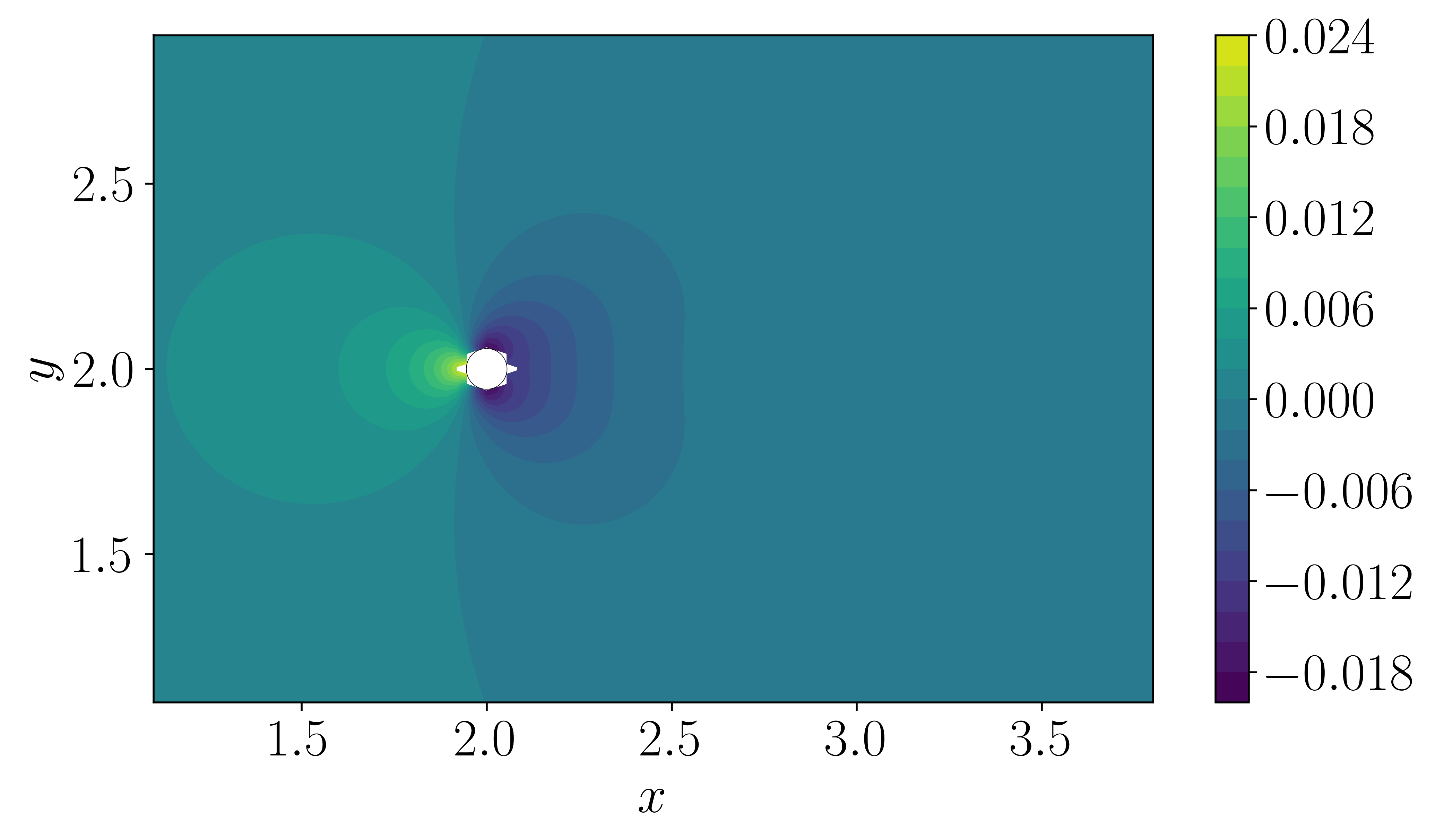}
        \caption{HF solution, $p$}
    \end{subfigure}
\\
  \begin{subfigure}{0.3\textwidth}
        \includegraphics[width=\linewidth,trim={0 0 0in 0},clip]{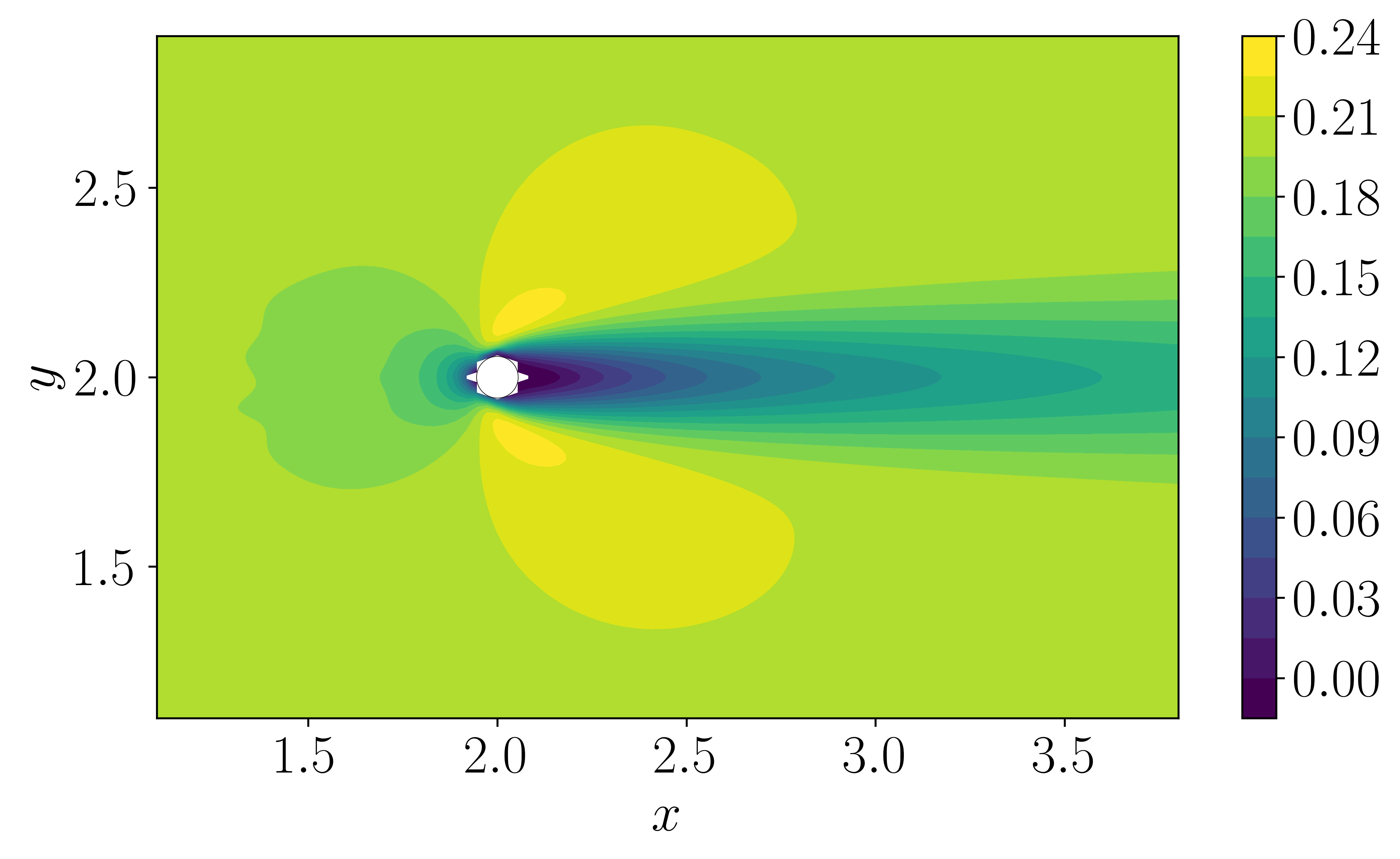}
        \caption{ML prediction, \uml}
    \end{subfigure}
    \begin{subfigure}{0.3\textwidth}
        \includegraphics[width=\linewidth,trim={0 0 0in 0},clip]{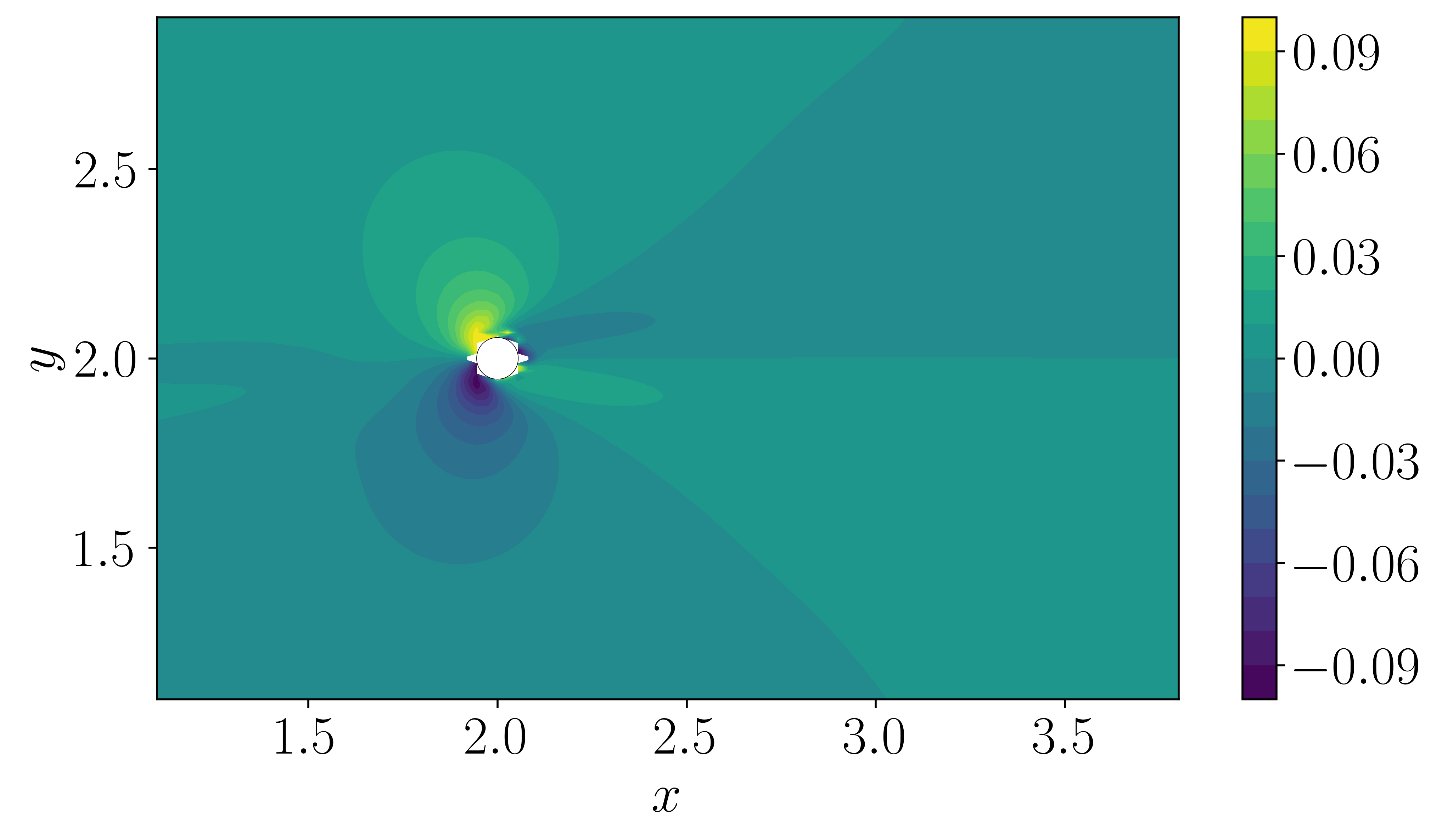}
        \caption{ML prediction, \vml}
    \end{subfigure}
    \begin{subfigure}{0.3\textwidth}
        \includegraphics[width=\linewidth,trim={0 0 0in 0},clip]{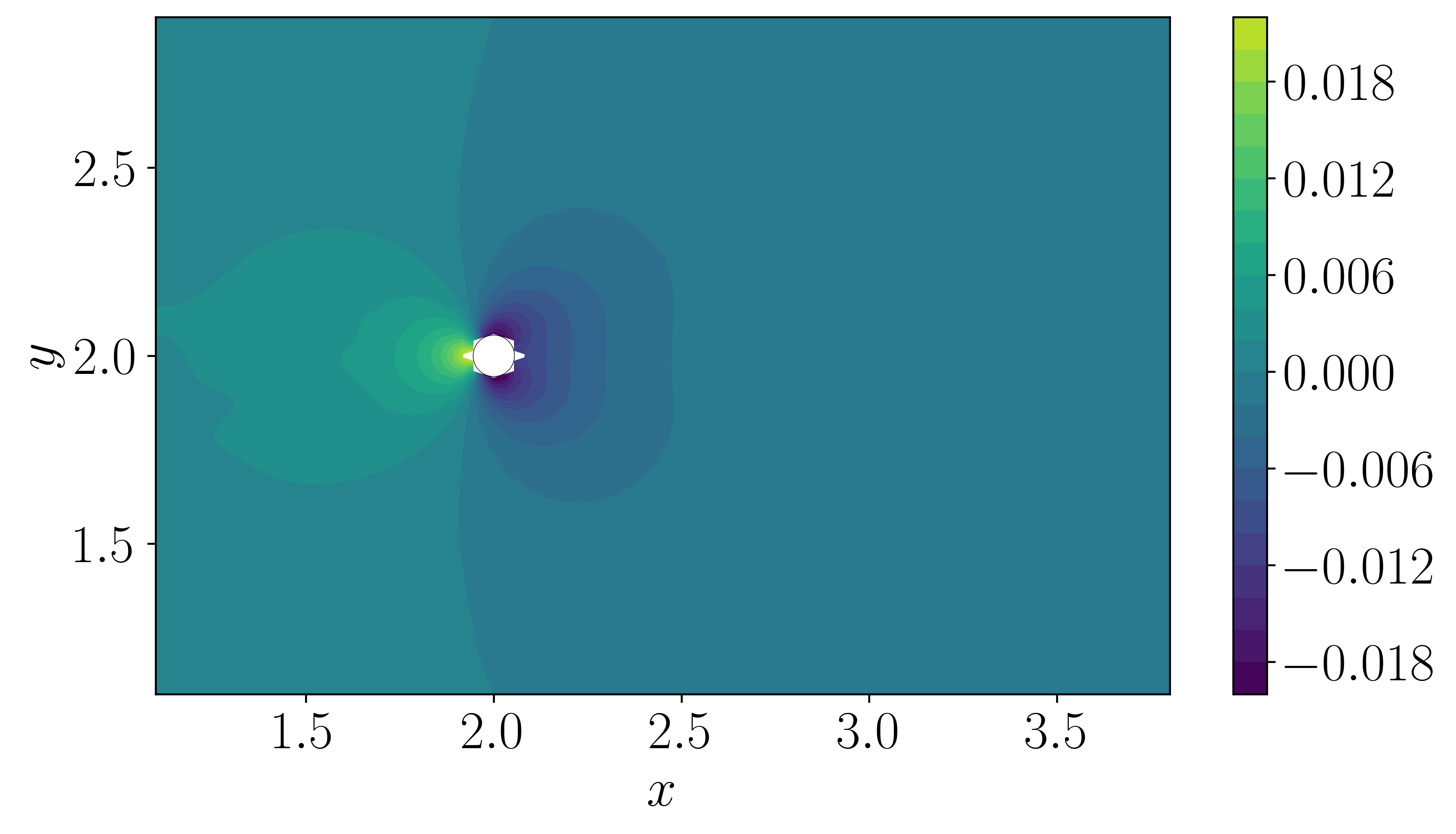}
        \caption{ML prediction, \pml}
    \end{subfigure}
    \caption{The trained ML model was used to predict the flow fields at a larger circular cylinder at $R=$110\% reference radius (d-f) and compared to the HF solution (a-c).}
  \label{fig:u-predict-110-circle}
\end{figure}

The trained ML model using quadtree sampling method was then used to generate contour plots of the flow fields at off-training circular cylinder radii. The comparisons between HF solution and ML prediction for the flow problem with smaller ($R=$90\%) and larger ($R=$110\%) cylinders are shown in Fig. \ref{fig:u-predict-90-circle} and \ref{fig:u-predict-110-circle}, respectively.

\subsection{Joukowski Airfoil}\label{sec:ex-airfoil}
The problem of flow around a Joukowski airfoil for an incompressible fluid at \Rey =1000 was studied in this section. This example demonstrates the use of a trained model to predict solutions at various AoAs. The properties of the fluid and flow conditions used in this example are summarized in Tab. \ref{tab:airfoil-specs}. A unit chord length $c=1.0$ was used. 

\begin{table}[!ht]
\small
\begin{center}
\begin{tabular}{ c|ccccc }
    quantity &  freestream speed & kinematic viscosity & fluid density & Reynolds number & angle of attack\\ 
 \hline
    symbol & $U_{0}$ & $\nu_{0}$ & $\rho_{0}$ & \Rey & $\alpha$ \\
    value & 10.0 & 0.01 & 1.0 & 1000 & [0\dg , 8\dg ]\\
\end{tabular}
\caption{ Problem specifications of the low Reynolds number Joukowski airfoil example.}\label{tab:airfoil-specs}
\end{center}
\end{table}

As before, the high-fidelity solution of the reference problem was prepared using {\it{Parabol}}. The meshes used for the CFD simulations were generated in {\it{Gmsh}} \cite{geuzaine_gmsh_2009} using a {\it{Python}} script provided by the {\it{5th International Workshop on High-Order CFD Methods}} \cite{noauthor_vl1_nodate}. 
Farfield distance from the airfoil was set to 100 times the chord length. Structured meshes with linear quadrilateral elements were adopted for our CFD runs. Meshes at three refinement levels corresponding to 2,128, 8,352, and 33,088 nodes were prepared. A rotation was applied to the mesh to run simulations at off-zero AoAs. High-fidelity runs refer to simulations carried out using the finest mesh.

\begin{figure}[!ht]
    \centering
   \begin{subfigure}{0.32\textwidth}
        \includegraphics[width=\linewidth]{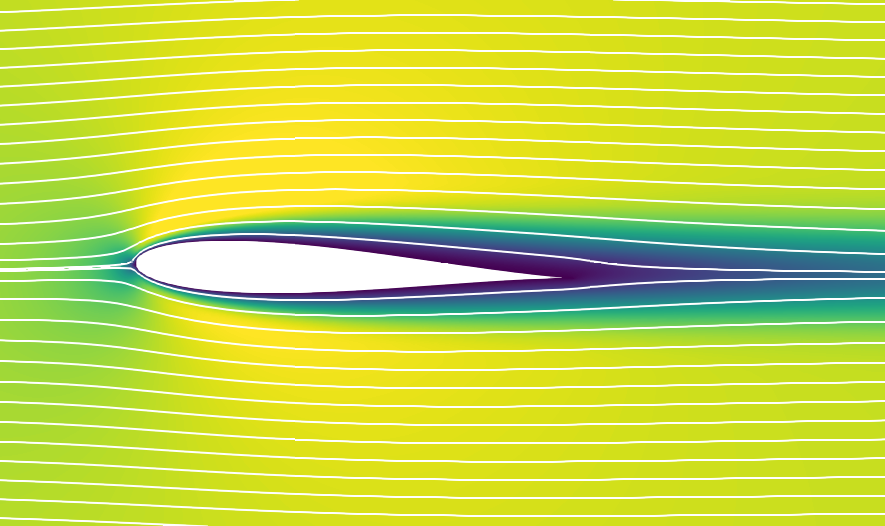}
        \caption{AoA=2\dg }
    \end{subfigure}
    \begin{subfigure}{0.32\textwidth}
        \includegraphics[width=\linewidth]{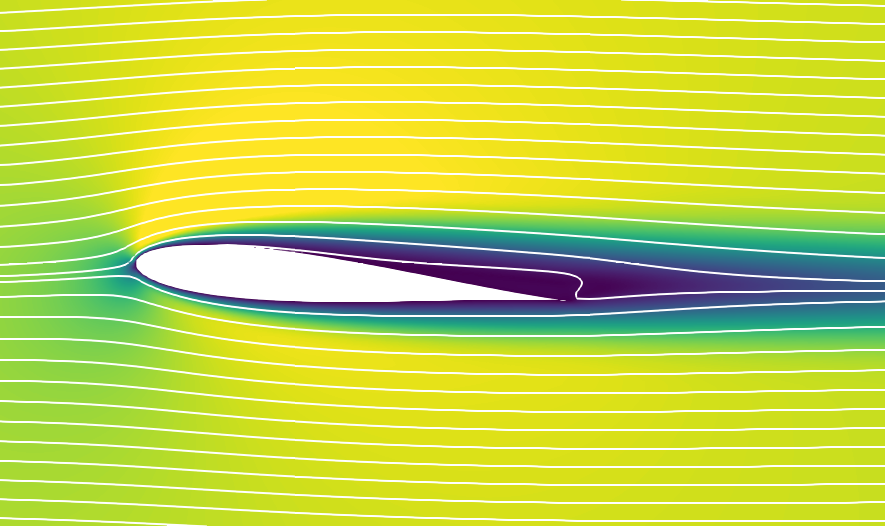}
        \caption{AoA=5\dg }
    \end{subfigure}
    \begin{subfigure}{0.32\textwidth}
        \includegraphics[width=\linewidth]{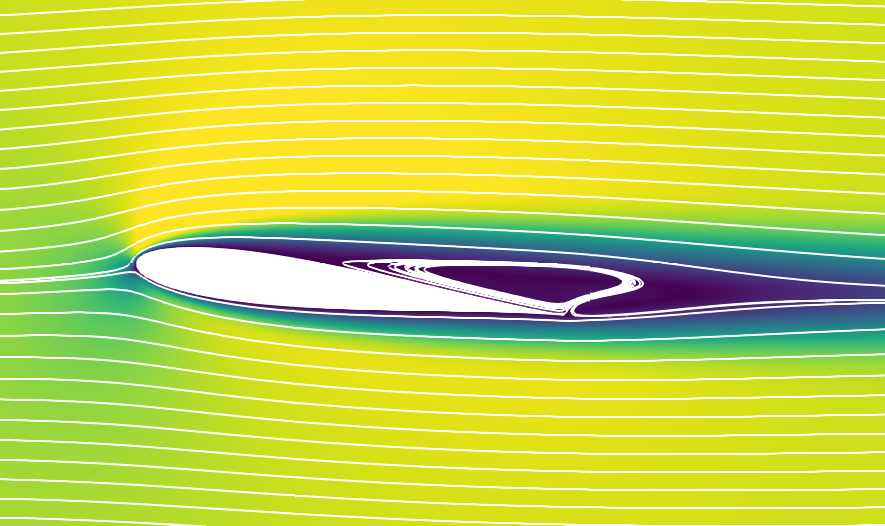}
        \caption{AoA=7\dg }
    \end{subfigure}
    \caption{HF solutions (velocity magnitude and streamlines) for flow around a Joukowski airfoil indicate separation at AoA=5\dg .}
    \label{fig:airfoil-hf}
\end{figure}

The high-fidelity solutions at various AoAs are shown in Fig. \ref{fig:airfoil-hf} along with streamlines. The flow is attached up to AoA=5\dg , at which separation is observed (Fig. \ref{fig:airfoil-hf} (b)). The vortex grows until the flow becomes unsteady beyond AoA=7\dg . The proposed ML model is intended for the prediction of steady state flow fields only, and discussions of the treatment of time dependency in a ML model development are reserved for the future work.

\begin{figure}[!ht]
    \centering
    \begin{subfigure}{0.3\textwidth}
        \includegraphics[width=\linewidth,trim={.8in 0.5in .8in 0.5in},clip]{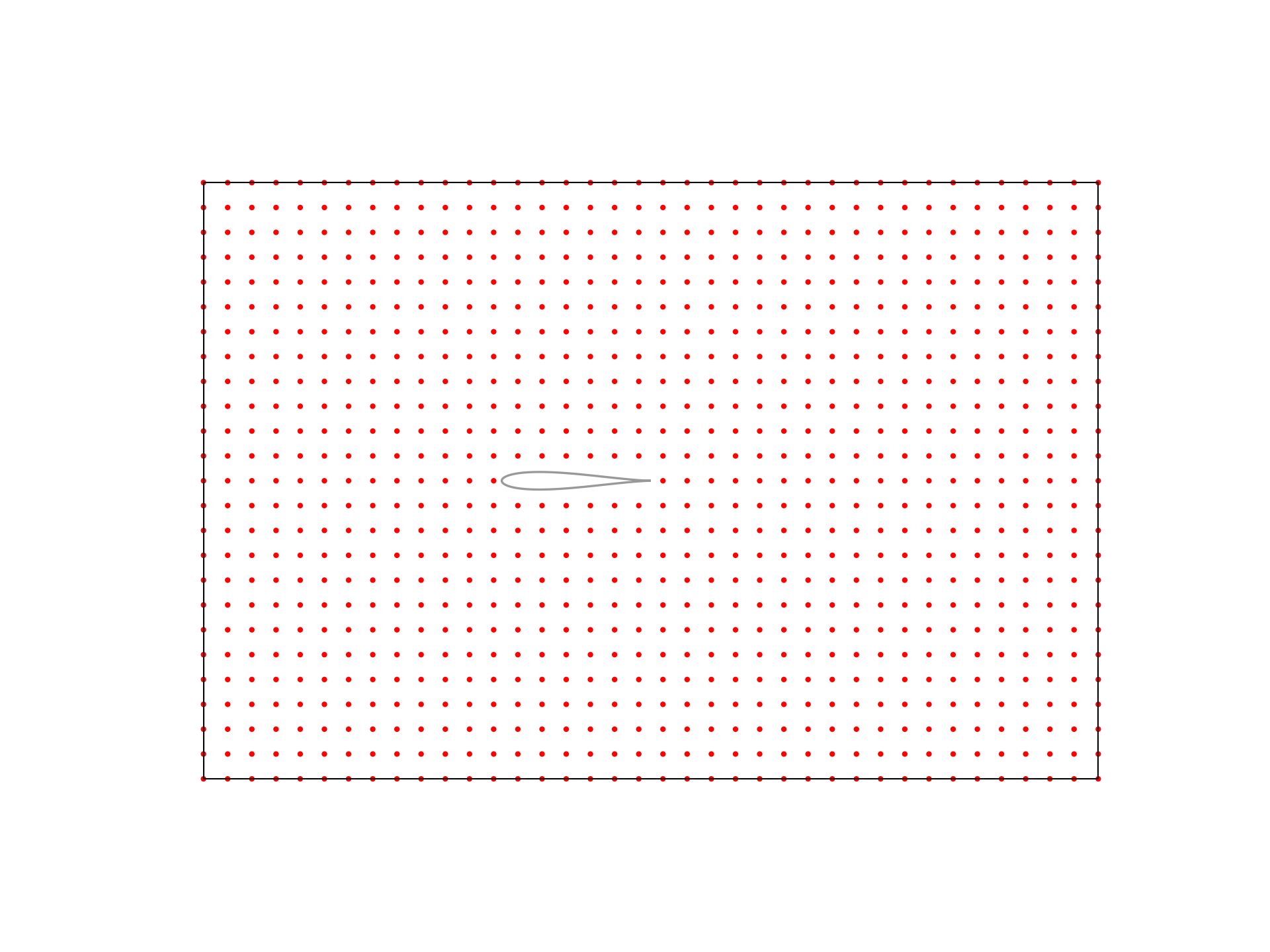}
        \caption{Uniform sampling}
    \end{subfigure}
    \begin{subfigure}{0.3\textwidth}
        \includegraphics[width=\linewidth,trim={.8in 0.5in .8in 0.5in},clip]{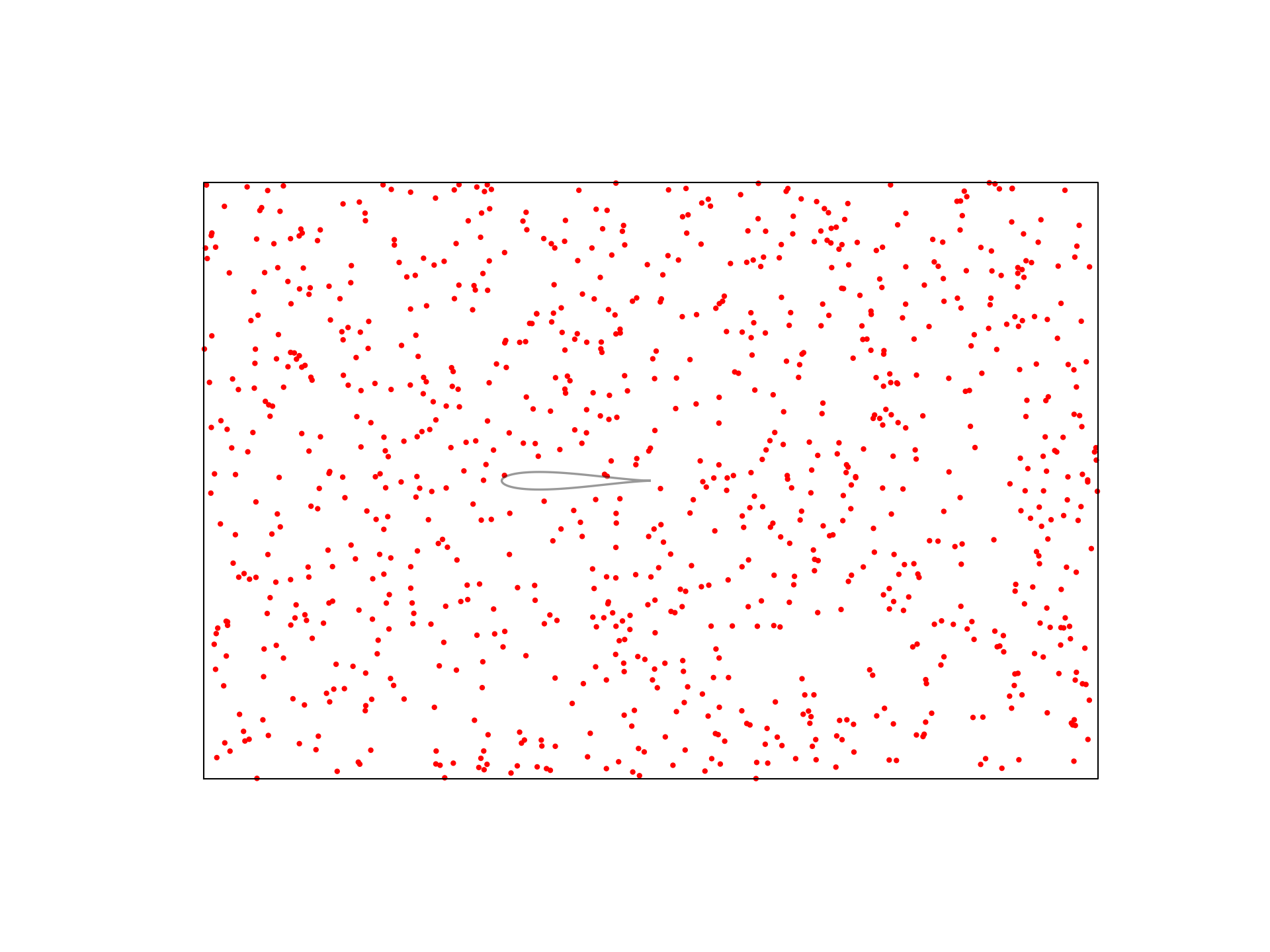}
        \caption{Random sampling}
    \end{subfigure}
    \\
    \begin{subfigure}{0.3\textwidth}
        \includegraphics[width=\linewidth,trim={.8in 0.5in .8in 0.5in},clip]{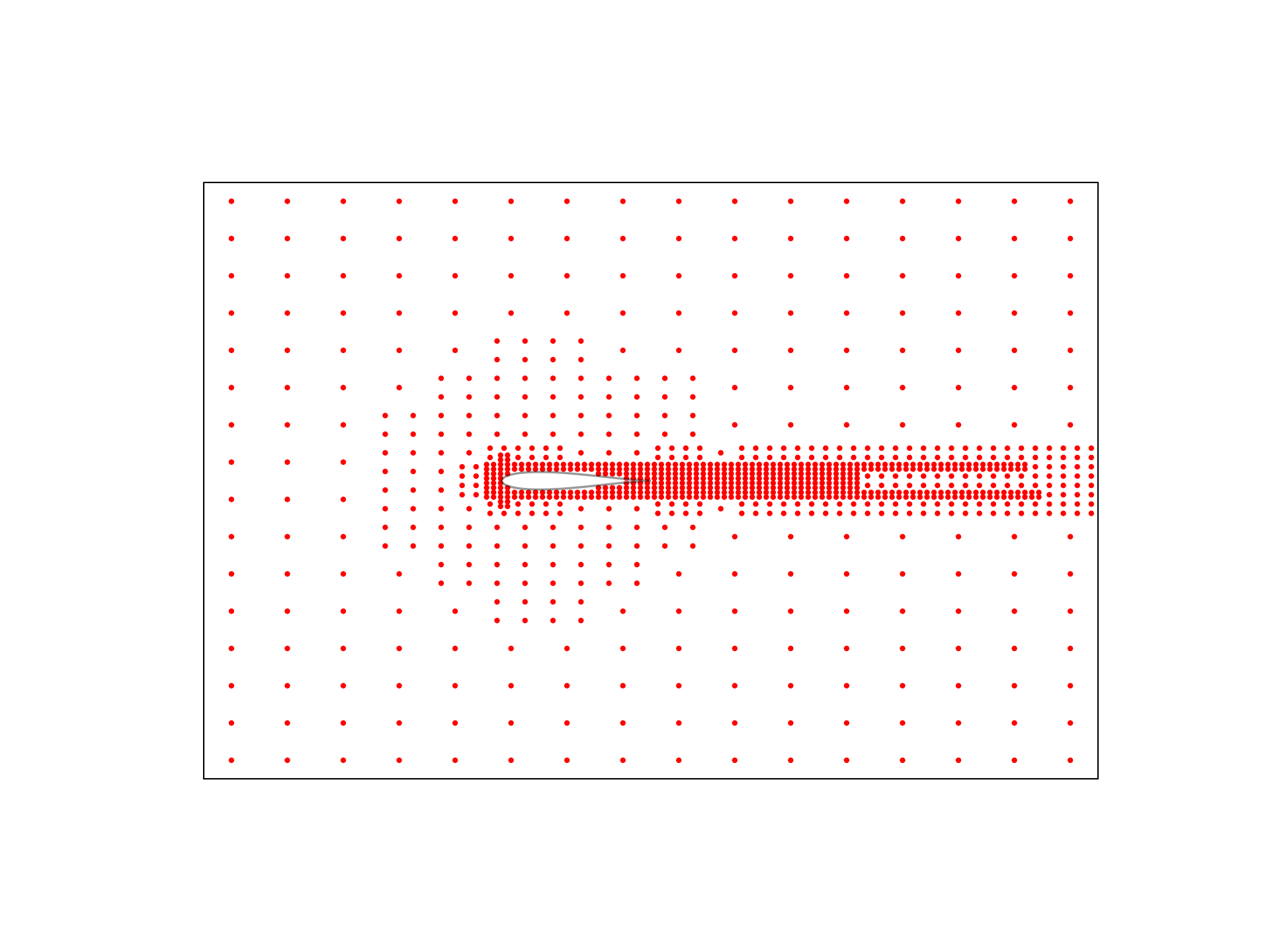}
        \caption{Quadtree sampling}
    \end{subfigure}
    \begin{subfigure}{0.3\textwidth}
        \includegraphics[width=\linewidth,trim={.4in 0.2in .4in 0.2in},clip]{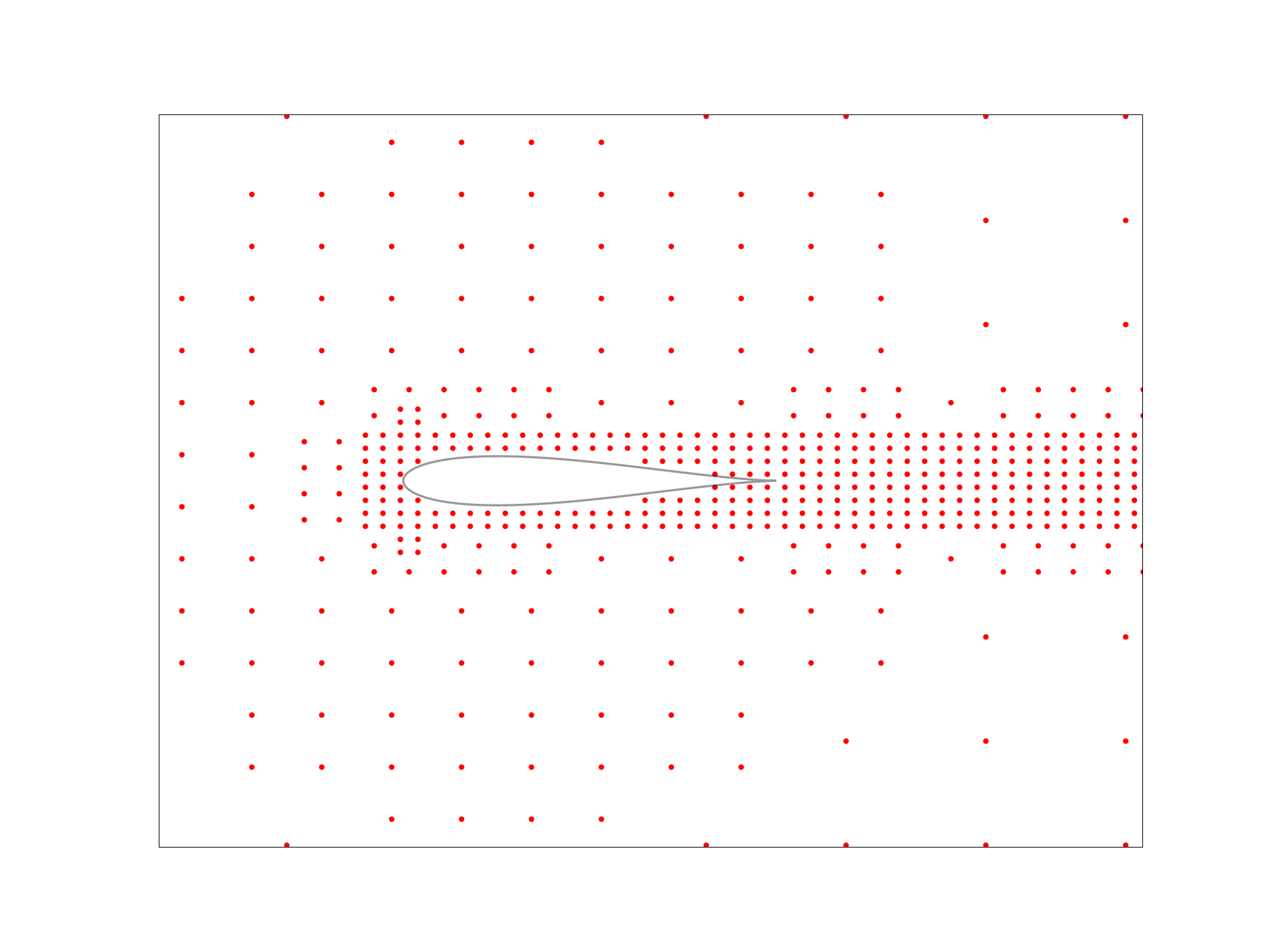}
        \caption{Quadtree sampling, zoomed-in}
    \end{subfigure}
    \caption{Distributions of 944, 950, and 1002 data points for the problem of flow around a Joukowski airfoil using (a) uniform, (b) random, and (c,d) quadtree sampling methods.}
    \label{fig:airfoil-sampling-methods-comp}
\end{figure}

The data sampling methods were compared to determine the efficient data preparation, similarly to the cylinder case. The data distributions of 944, 950 and 1002 points using uniform, random and quadtree sampling methods are shown in Fig. \ref{fig:airfoil-sampling-methods-comp}. A similar observation as in the circular cylinder case is made. A generally denser distribution in the freestream region is seen in uniform and random distributions; clustering of points around the airfoil is observed in the quadtree sampling case. 

\begin{figure}[!ht]
    \centering
   \begin{subfigure}{0.32\textwidth}
        \includegraphics[width=\linewidth]{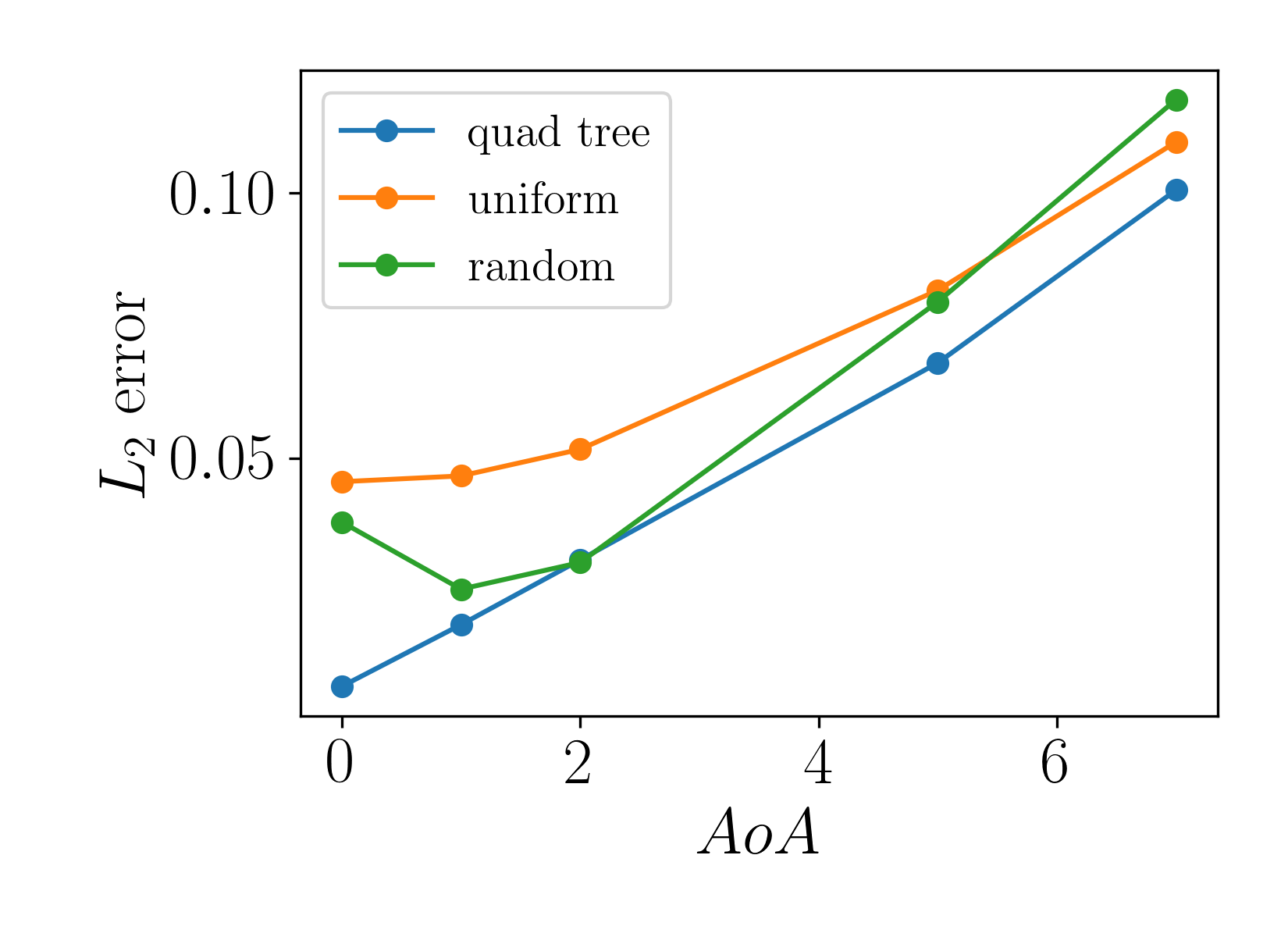}
        \caption{Velocity, x-component}
    \end{subfigure}
    \begin{subfigure}{0.32\textwidth}
        \includegraphics[width=\linewidth]{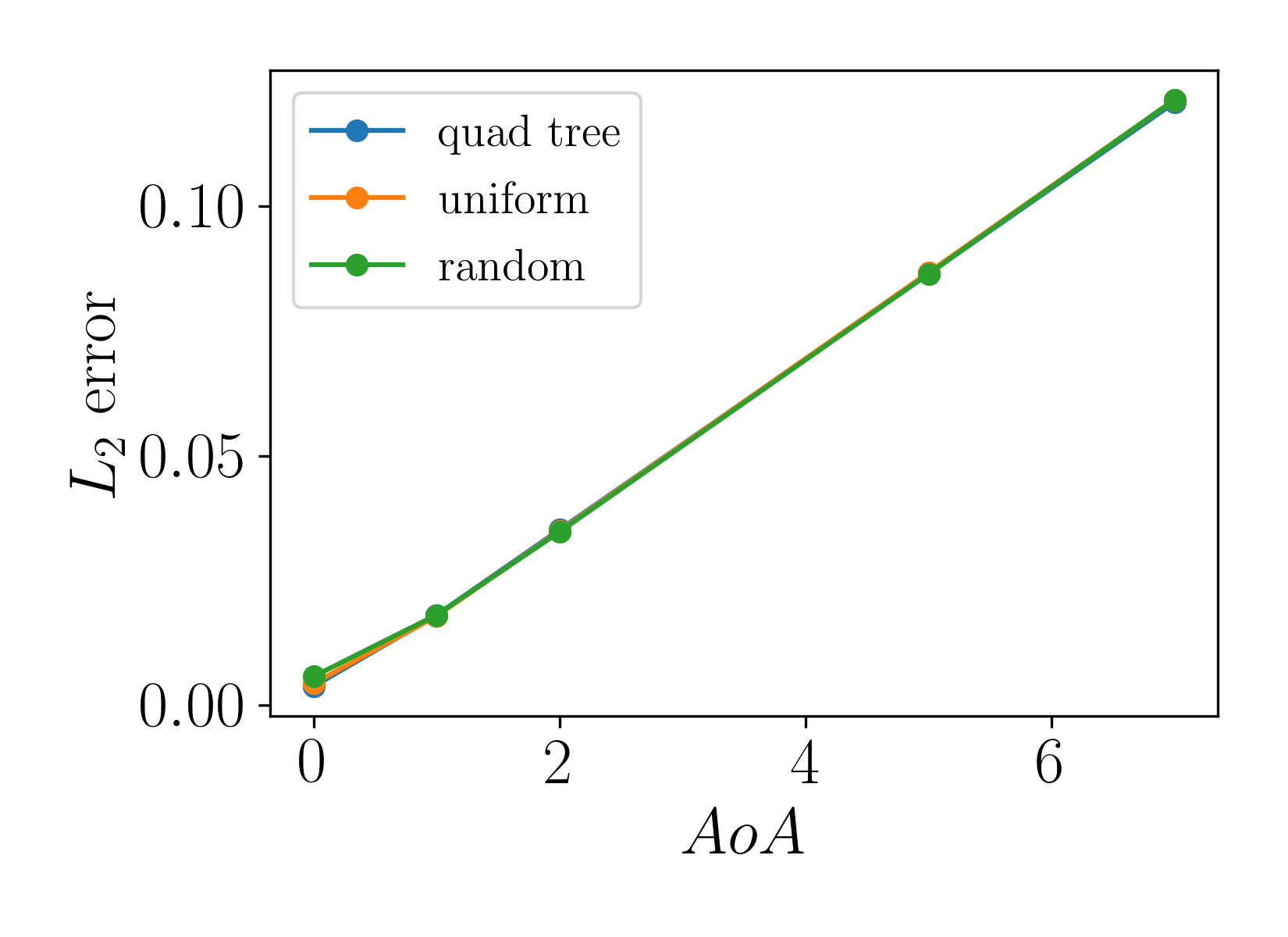}
        \caption{Velocity, y-component}
    \end{subfigure}
    \begin{subfigure}{0.32\textwidth}
        \includegraphics[width=\linewidth]{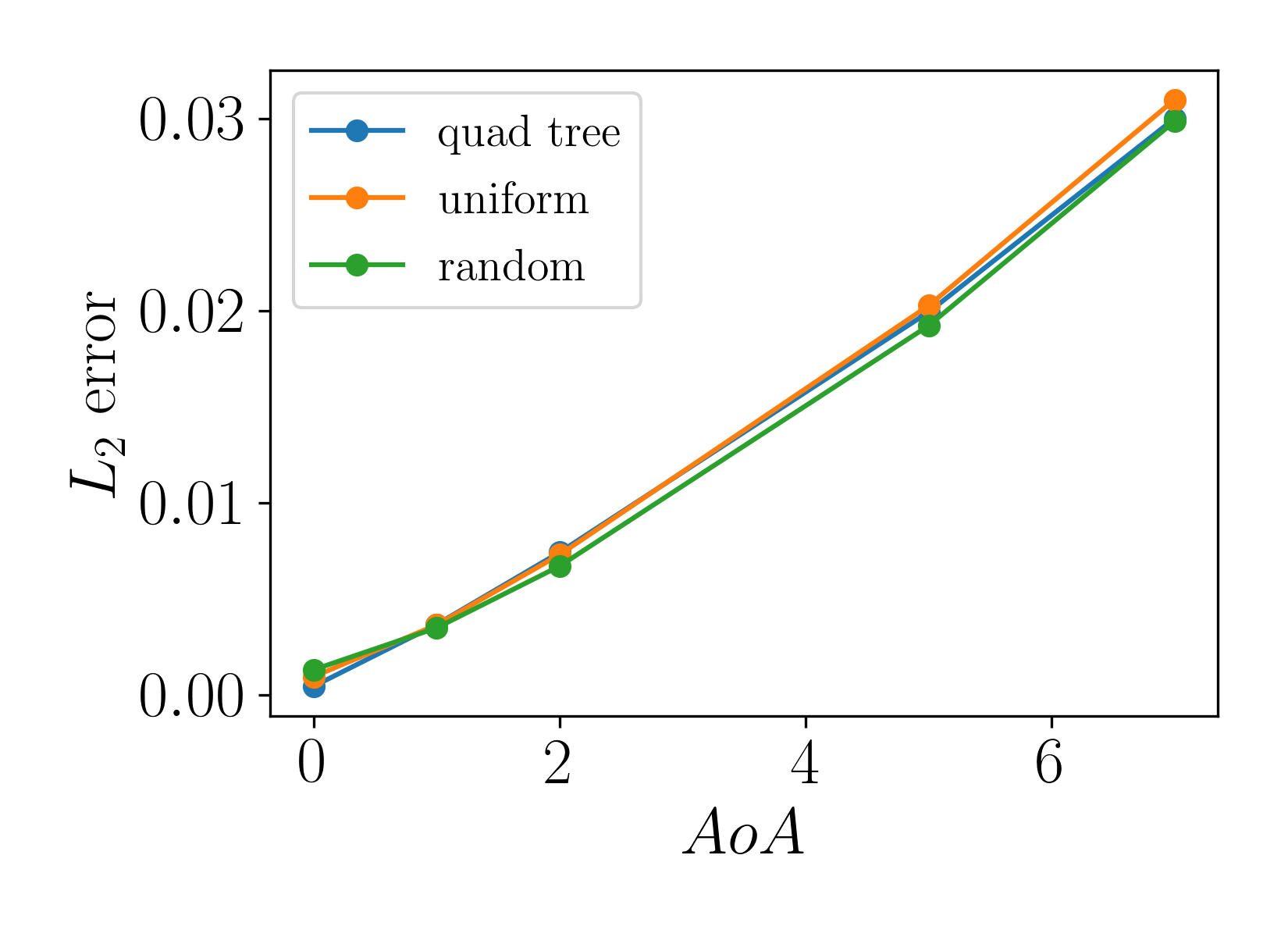}
        \caption{Pressure}
    \end{subfigure}
    \caption{Prediction errors for flow around a Joukowski airfoil using ML models trained at AoA=0\dg . Data points were distributed using (a) uniform, (b) random, and (c) quadtree sampling methods.}
    \label{fig:airfoil-sampling-error-comp}
\end{figure}

Three separate ML models were trained using the three sampling methods at the reference problem at AoA=0\dg . The prediction errors were computed and plotted over a range of AoAs (see Fig. \ref{fig:airfoil-sampling-error-comp}). The errors for the velocity components were similar for all sampling methods across the entire range of AoAs considered, and the errors for the pressure were smaller using the quadtree sampling, especially at the training point at AoA=0\dg .

\begin{figure}[!ht]
  \centering
  \begin{subfigure}{0.35\textwidth}
        \includegraphics[width=\linewidth,trim={0 0 1.2in 0},clip]{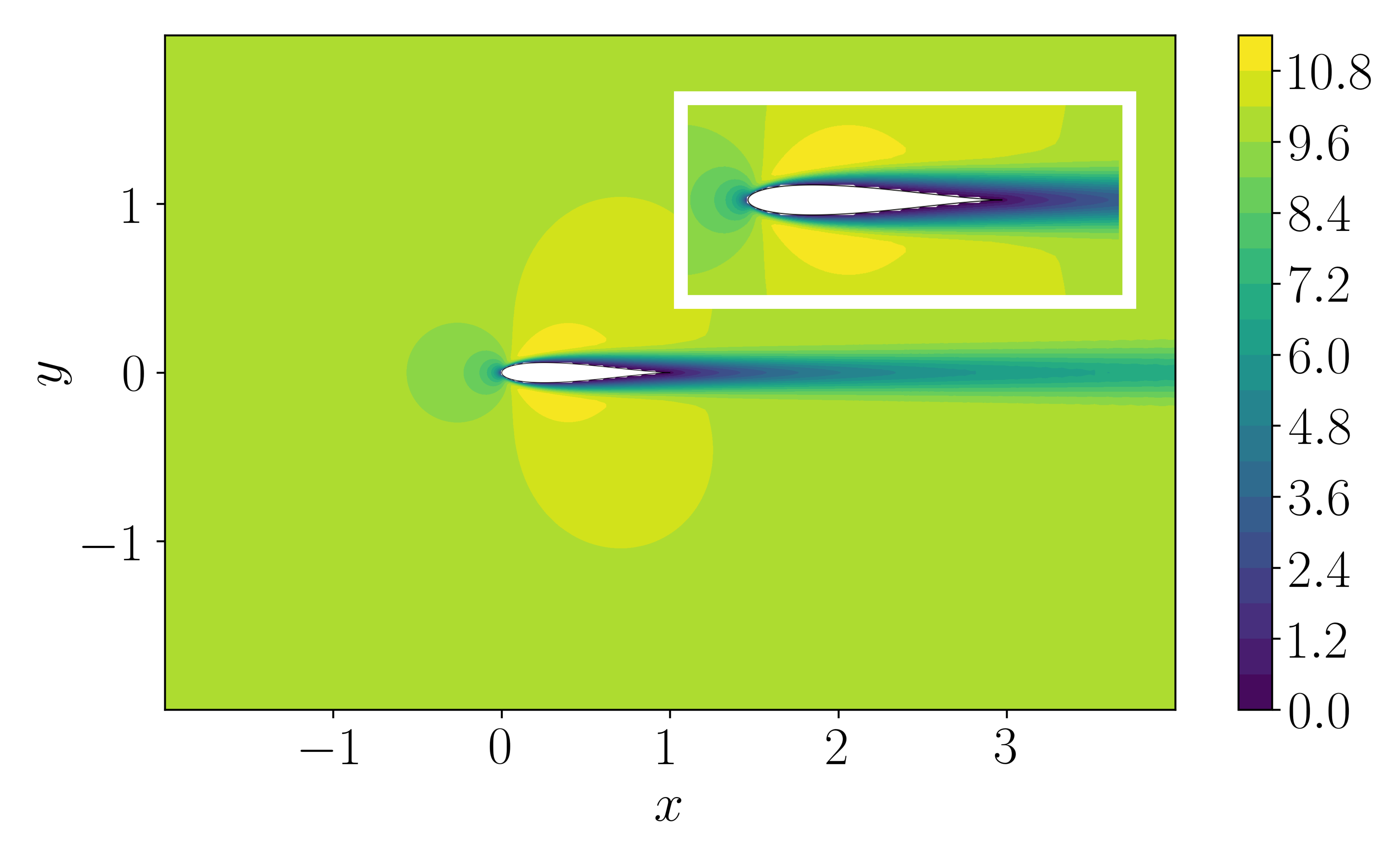}
        \caption{HF solution}
    \end{subfigure}
    \begin{subfigure}{0.35\textwidth}
        \includegraphics[width=\linewidth,trim={0 0 1.2in 0},clip]{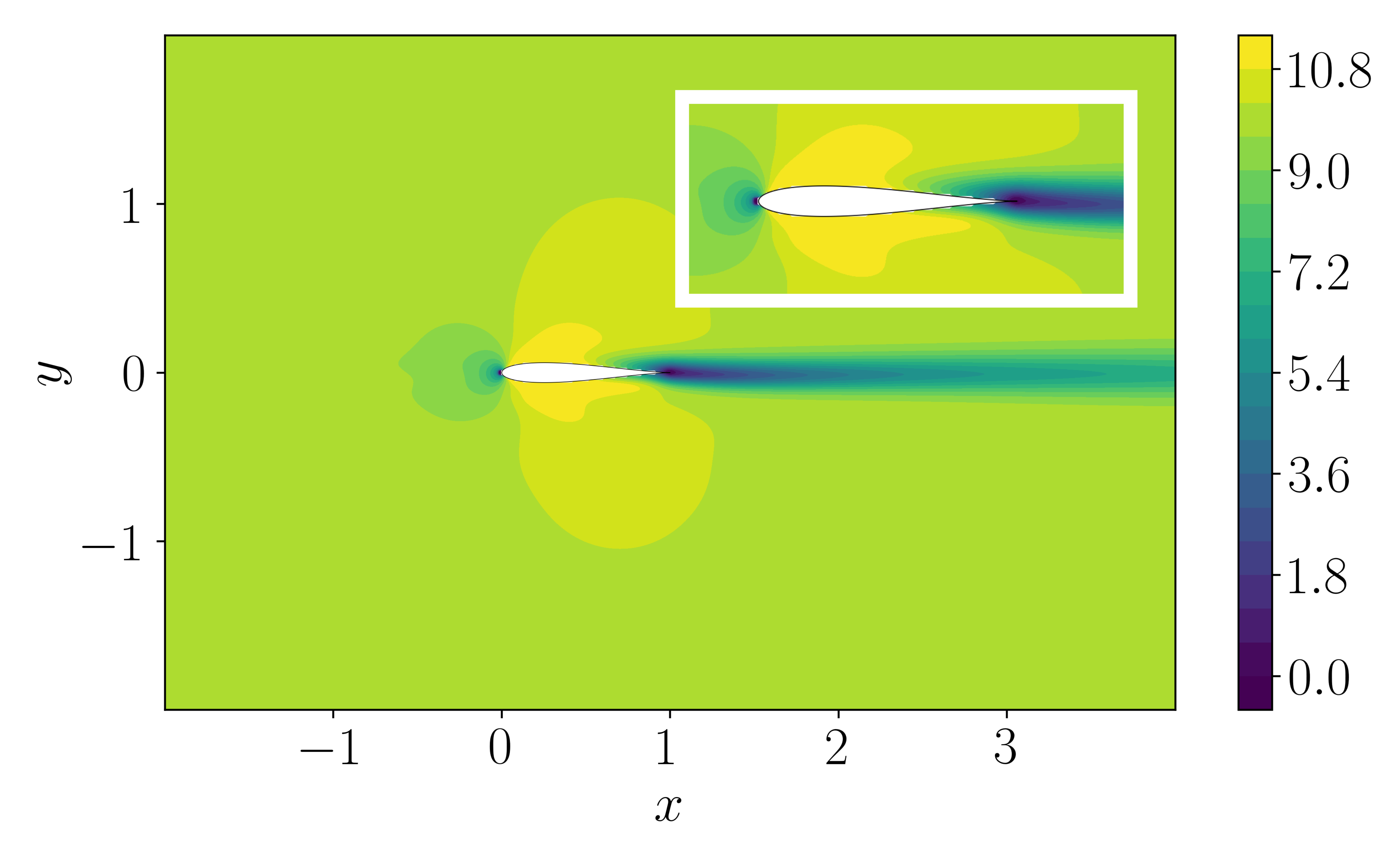}
        \caption{ML, uniform sampling}
    \end{subfigure}
\\
    \begin{subfigure}{0.35\textwidth}
        \includegraphics[width=\linewidth,trim={0 0 1.2in 0},clip]{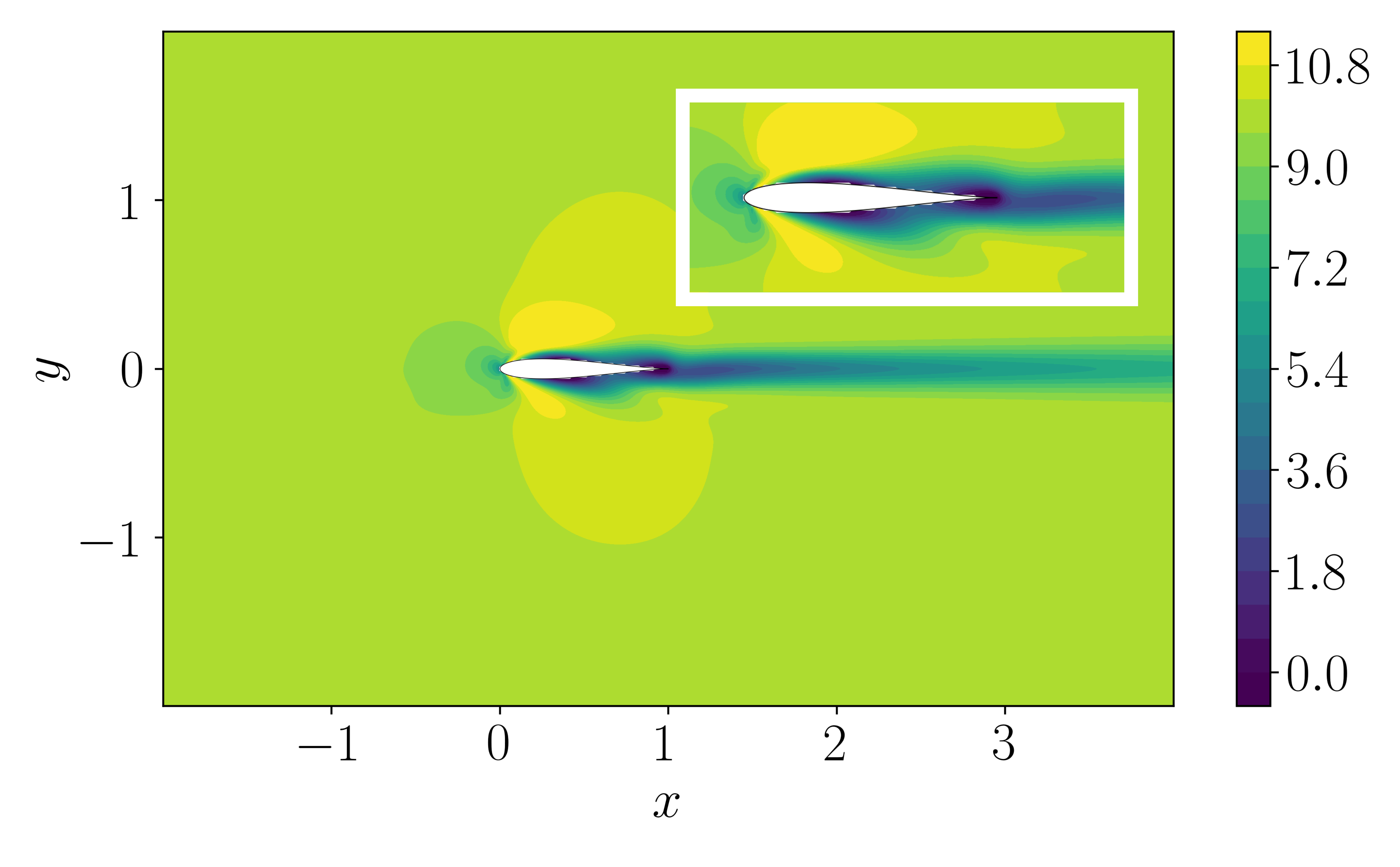}
        \caption{ML, random sampling}
    \end{subfigure}
    \begin{subfigure}{0.35\textwidth}
        \includegraphics[width=\linewidth,trim={0 0 1.2in 0},clip]{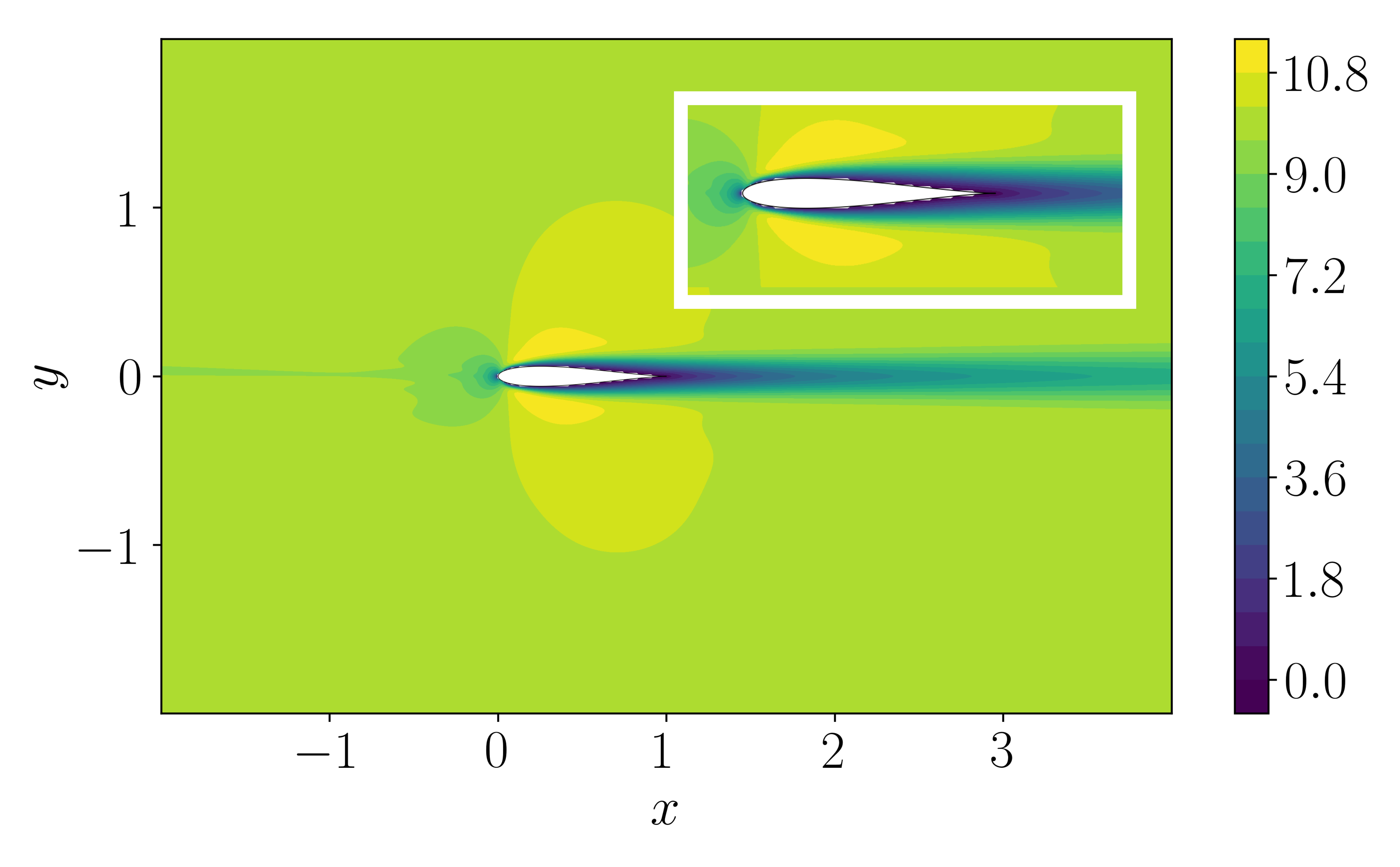}
        \caption{ML, quadtree sampling}
    \end{subfigure}
    \caption{ML-predicted $x-$component of the velocity at AoA=0\dg  (training point) is compared to the HF solution (a). The ML model training was carried out using (b) uniform, (c) random, and (d) quadtree sampling.}
  \label{fig:u-contour-airfoil}
\end{figure}

Comparisons of the field contour plots in Figs. \ref{fig:u-contour-airfoil} reveal the importance of a denser data distribution in the boundary layer. 
Both uniform and random distributions led to erroneous field distributions around the airfoil, though this was not indicated in the error plots that did not place a particular emphasis on the boundary layer prediction. The flow features in the boundary layers were replicated more accurately by the model trained based on the quadtree-sampled data. 

\begin{figure}[!ht]
    \centering
   \begin{subfigure}{0.32\textwidth}
        \includegraphics[width=\linewidth]{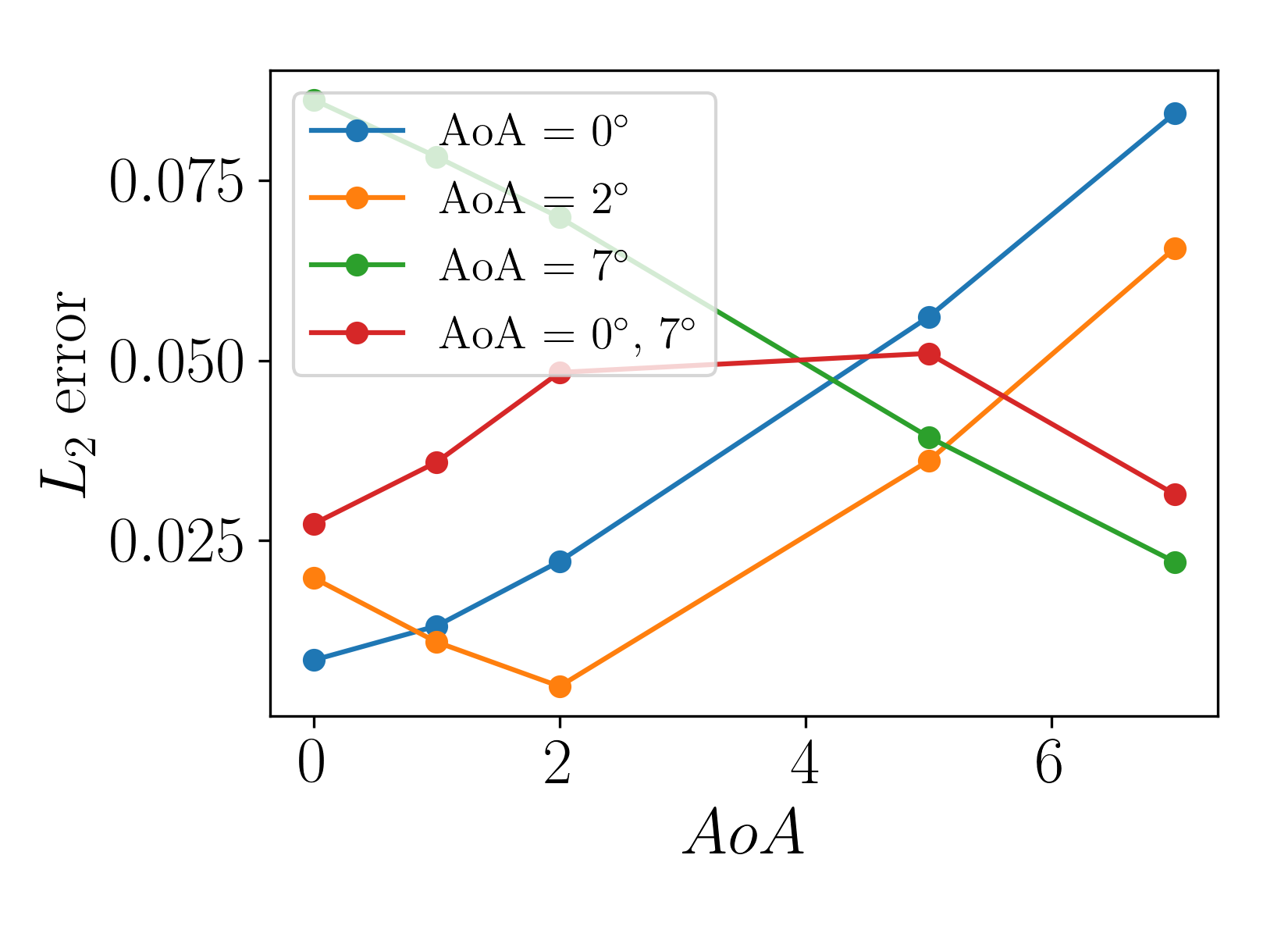}
        \caption{Velocity, x-component}
    \end{subfigure}
    \begin{subfigure}{0.32\textwidth}
        \includegraphics[width=\linewidth]{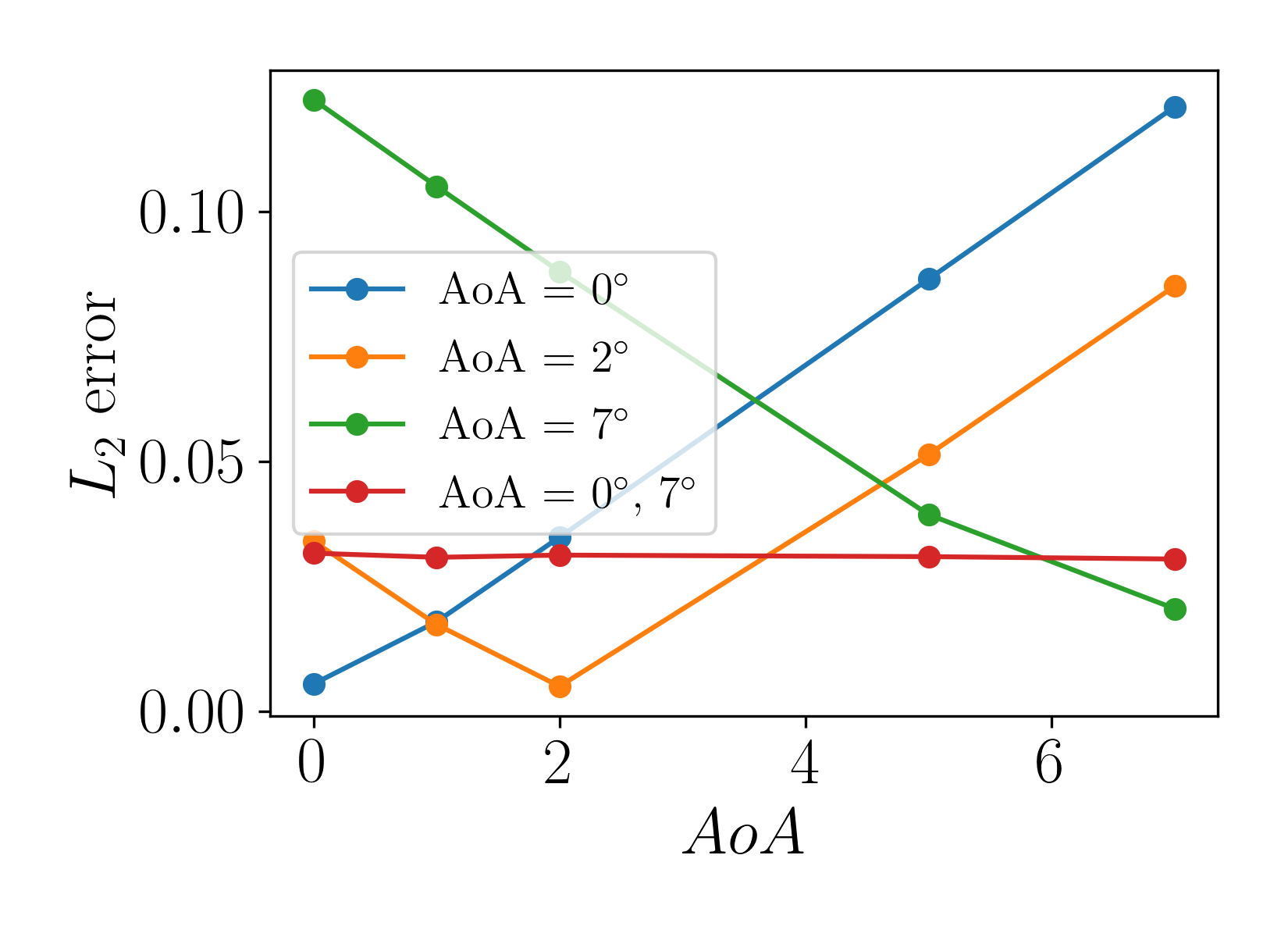}
        \caption{Velocity, y-component}
    \end{subfigure}
    \begin{subfigure}{0.32\textwidth}
        \includegraphics[width=\linewidth]{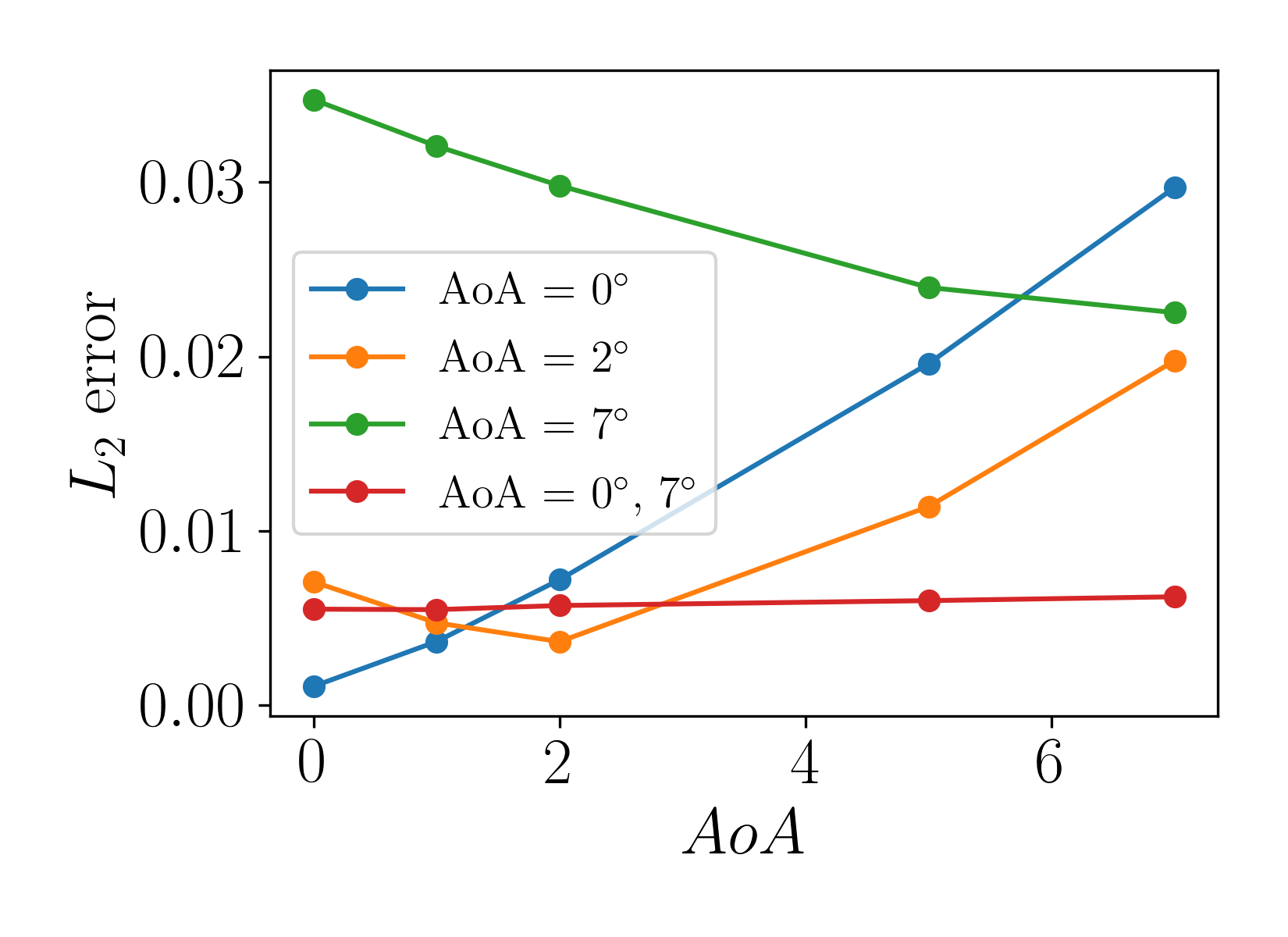}
        \caption{Pressure}
    \end{subfigure}
    \caption{Prediction errors for flow around a Joukowski airfoil based on ML models trained using various reference problems. Data from problems with AoA=0\dg  (blue), AoA=2\dg  (orange) AoA=7\dg  (green), and AoA=0\dg  and 7\dg  (red), were included in the ML model training. Data points distributed using quadtree sampling were used.}
    \label{fig:airfoil-ref-prob-error-comp}
\end{figure}

Another set of ML models were trained using different training data sets from problems corresponding to various reference AoA values. In the error plots in Fig. \ref{fig:airfoil-ref-prob-error-comp}, blue, orange and green curves represent prediction errors of models trained using 1002, 1000, and 1000 data points from the AoA=0\dg , 2\dg , and 7\dg  cases, respectively. Data points were distributed using the quadtree sampling method in each case. The lowest error is observed at the training AoA value in each case, and errors monotonically increase away from the training AoA value. The model represented by the red curve used data from two problems, at AoA=0\dg  and 7\dg . This model exhibits a better overall prediction performance throughout the AoA range of interest, although the cost of data generation and training are higher. To prepare data, two high-fidelity CFD runs (at AoA=0\dg  and 7\dg ) were required, and the total of 2002 data points following the quadtree sampling were used in the training. 

While the ML model was general enough to predict flow fields outside of the reference problems used for the model training, the model was not accurate enough to predict postprocessed features (e.g., surface forces) correctly. The accuracy of the ML model could be improved through various strategies, including the use of larger data size, redesigning of the neural network architecture, or other activities involving hyperparameter optimization. However, repeated resource allocations on costly hyperparameter optimization may be required for each flow problem of interest, as the most effective set of hyperparameters tends to be problem dependent.
Instead, we propose the use of trained ML models for CFD initialization to warm-start high-fidelity runs, which is less demanding of the accuracy of the trained ML models.

\subsection{CFD Initialization for Joukowski Airfoil}\label{sec:ex-airfoil-init}

\begin{figure}[!ht]
  \centering
  \begin{subfigure}{0.45\textwidth}
        \includegraphics[width=\linewidth,trim={0 0 0in 0},clip]{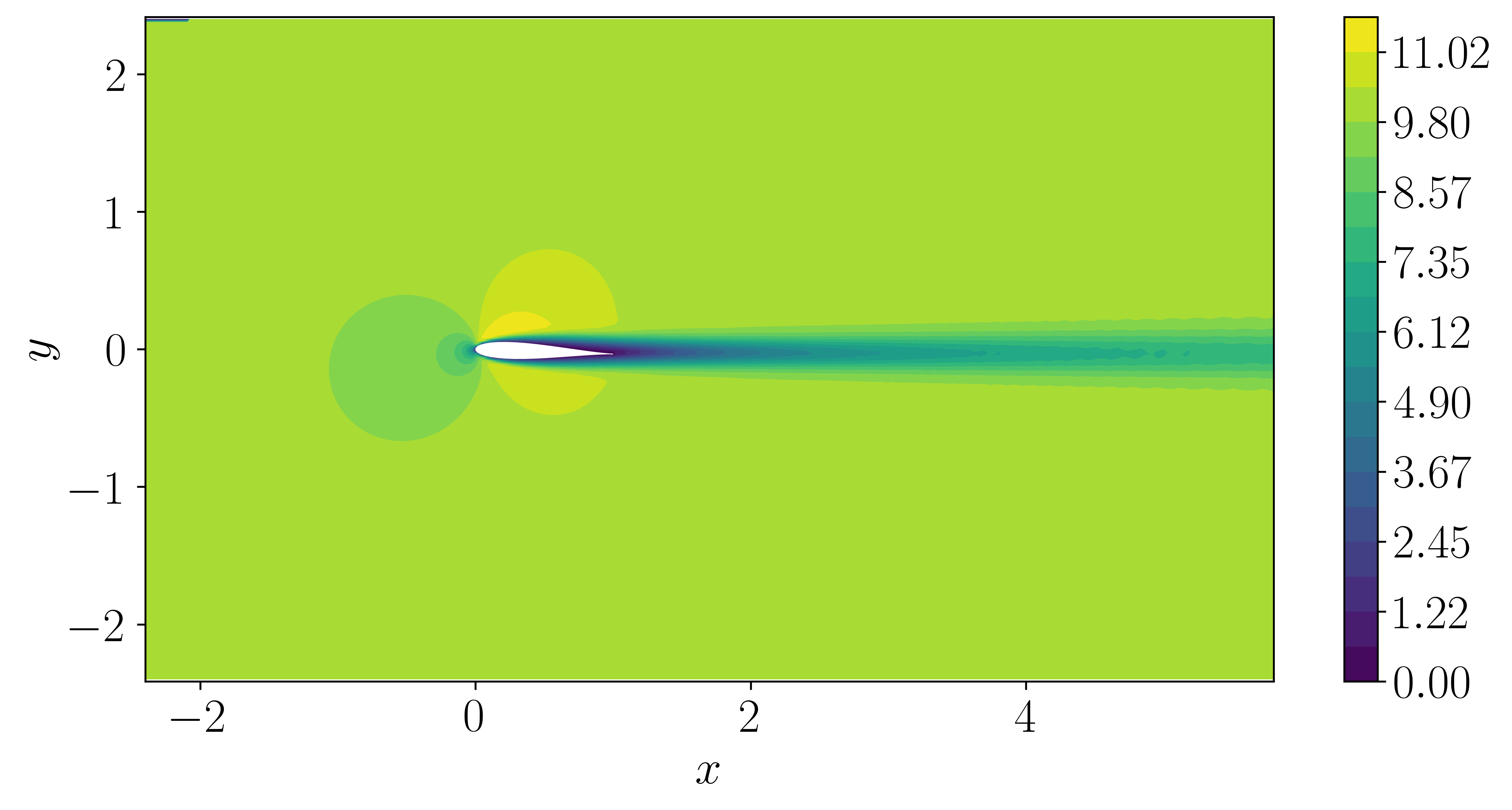}
        \caption{HF solution, $\left\|\bm{u}\right\|$}
   \end{subfigure}
  \begin{subfigure}{0.45\textwidth}
        \includegraphics[width=\linewidth,trim={0 0 0in 0},clip]{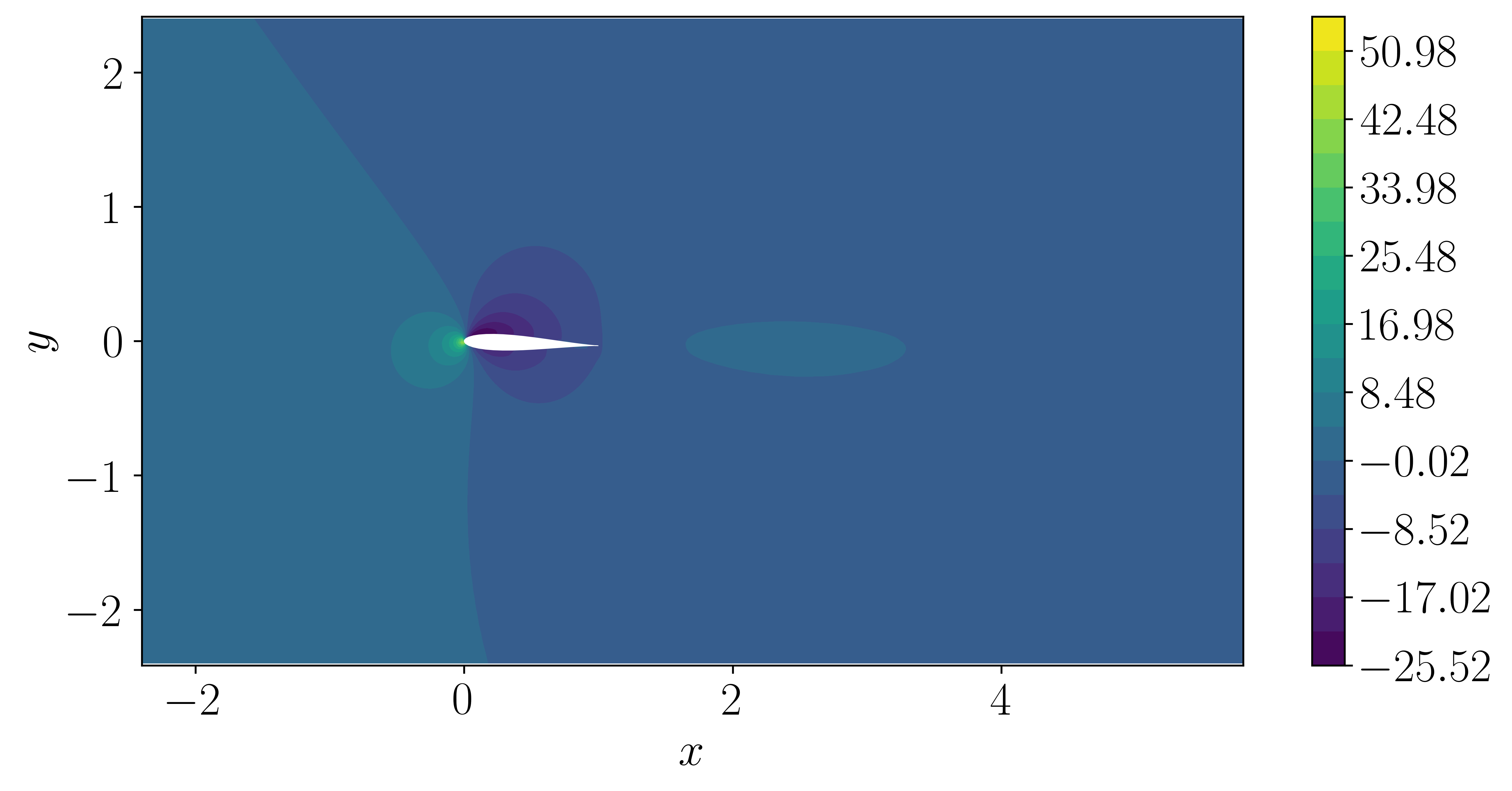}
        \caption{HF solution, $p$}
   \end{subfigure}
    \\
  \begin{subfigure}{0.45\textwidth}
        \includegraphics[width=\linewidth,trim={0 0 0in 0},clip]{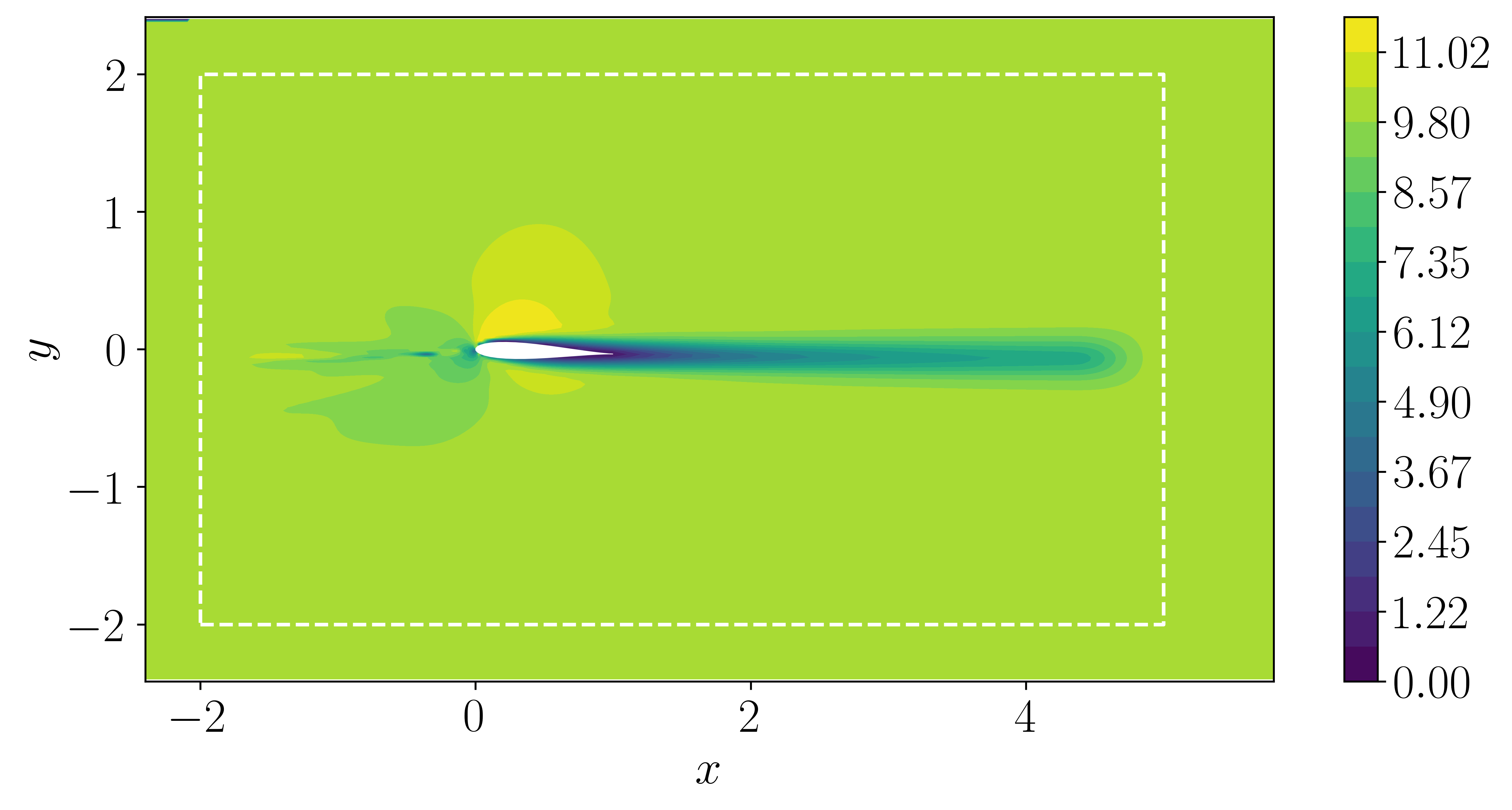}
        \caption{ML prediction $\left\|\tilde{\bm{u}}\right\|$, with POFU}
   \end{subfigure}
  \begin{subfigure}{0.45\textwidth}
        \includegraphics[width=\linewidth,trim={0 0 0in 0},clip]{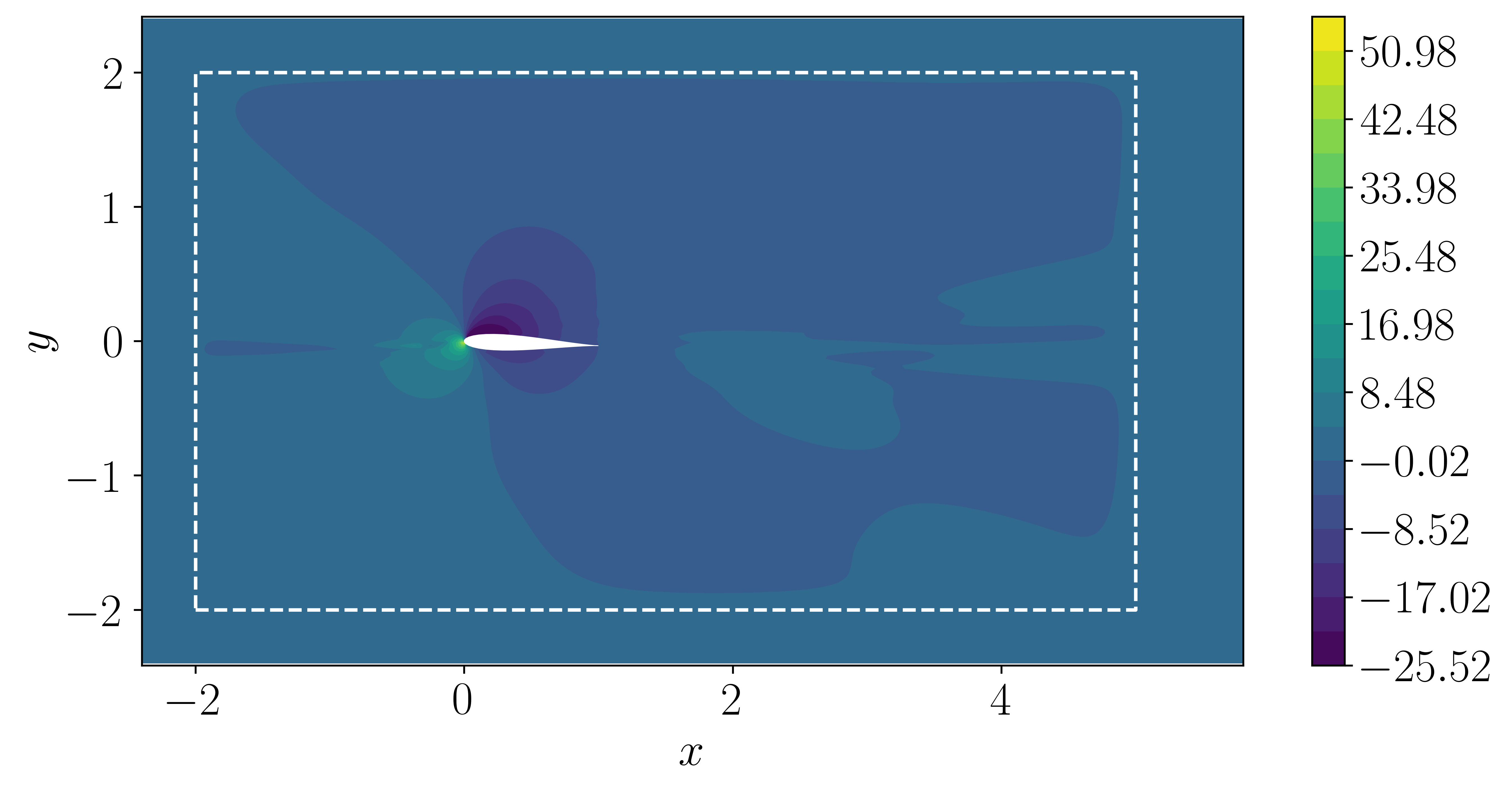}
        \caption{ML prediction $\tilde{p}$, with POFU}
   \end{subfigure}
    \caption{HF solution (a,b) and ML-predicted fields based on the model trained using AoA=0\dg, evaluated at AoA=2\dg  (c,d). POFU extension was used to smoothly transition from the ML prediction to freestream fields at the training window boundary (white dotted box) with transition widths $\left(s_{x}, s_{y}\right) = \left(0.1 l_{x},0.1 l_{y}\right)$, to prepare data for CFD initialization.}
  \label{fig:pofu-airfoil-AoA2}
\end{figure}

\begin{figure}[!ht]
  \centering
  \begin{subfigure}{0.45\textwidth}
        \includegraphics[width=\linewidth,trim={0 0 0in 0},clip]{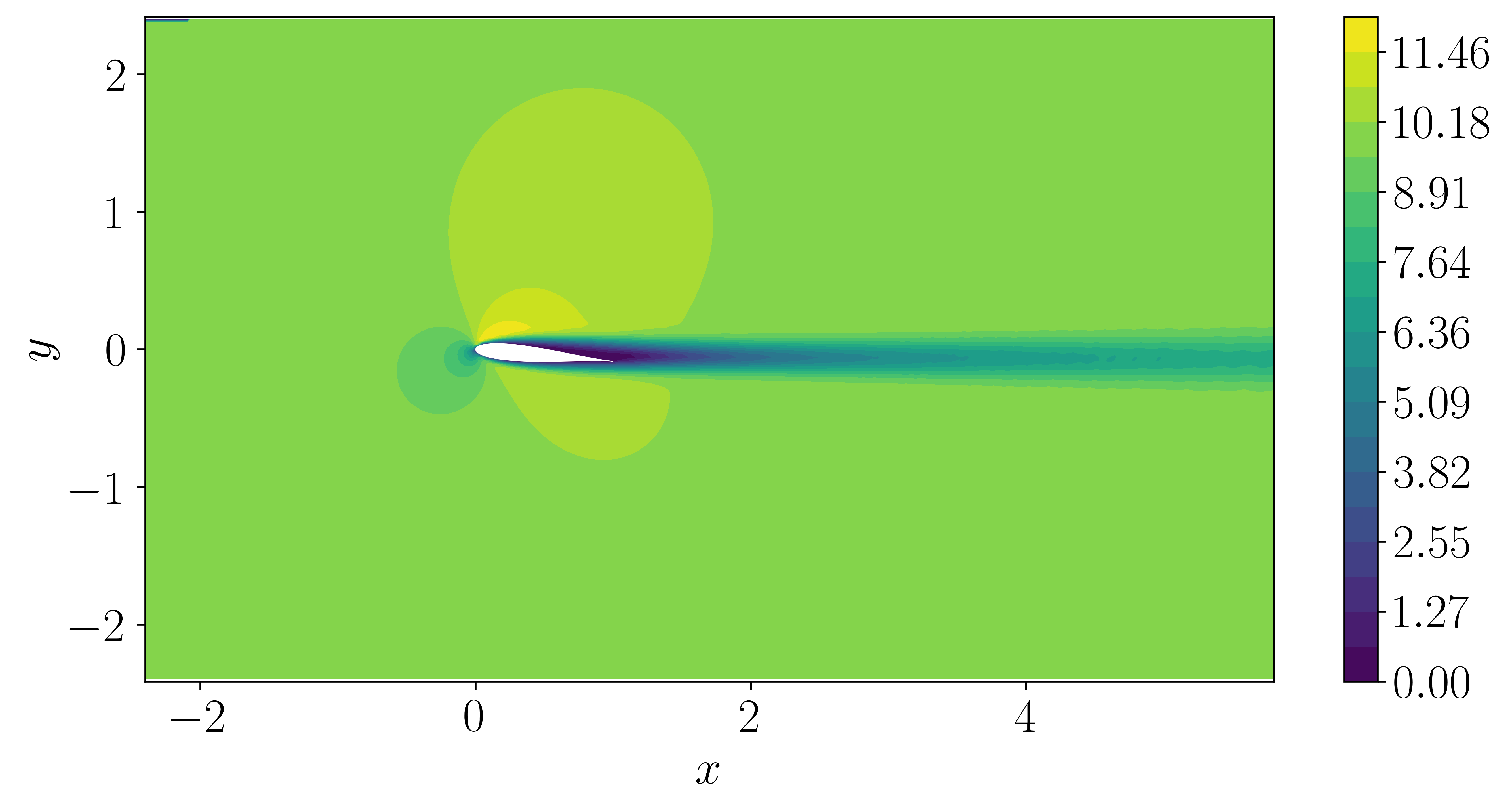}
        \caption{HF solution, $\left\|\bm{u}\right\|$}
   \end{subfigure}
  \begin{subfigure}{0.45\textwidth}
        \includegraphics[width=\linewidth,trim={0 0 0in 0},clip]{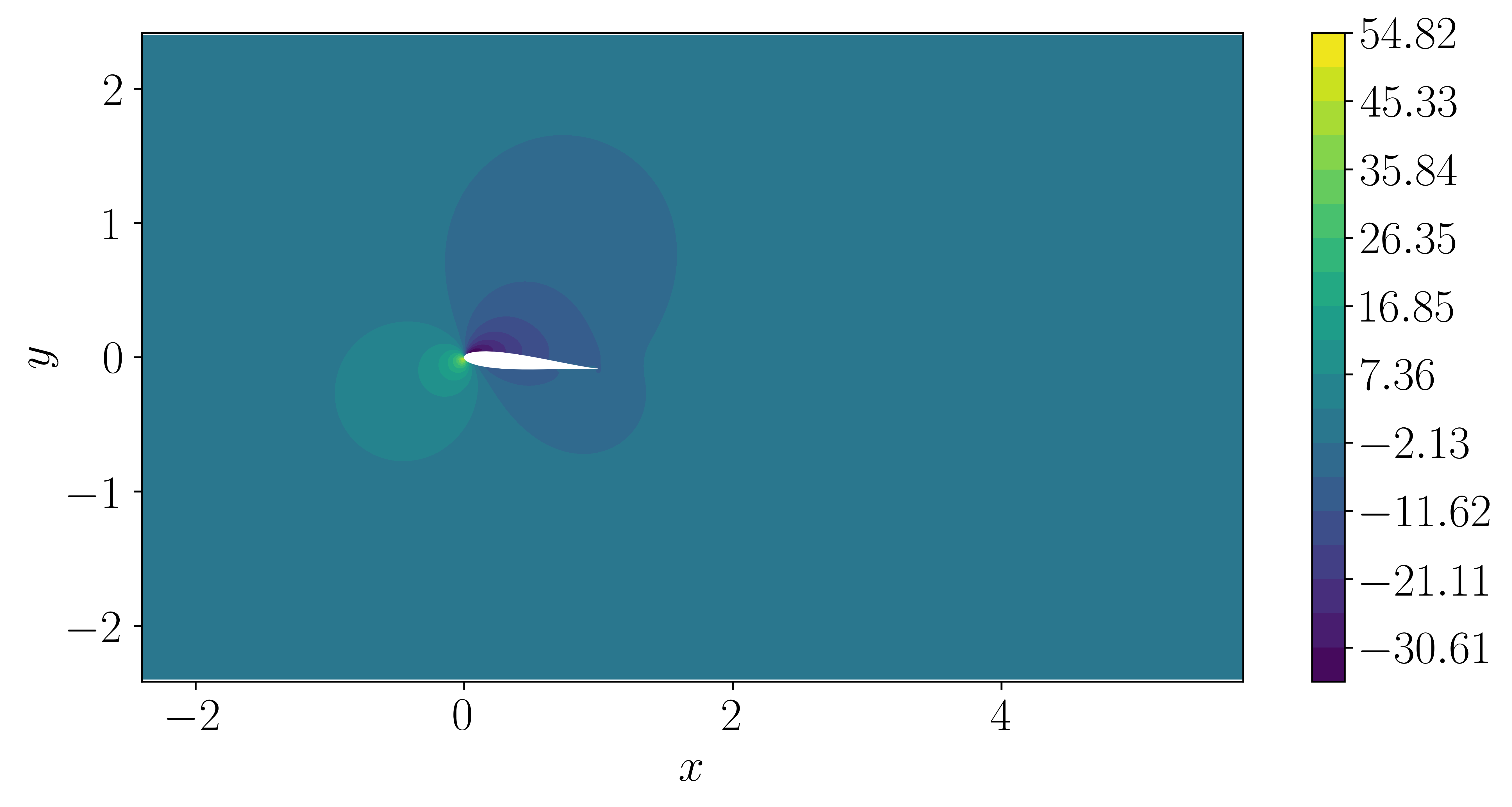}
        \caption{HF solution, $p$}
   \end{subfigure}
    \\
  \begin{subfigure}{0.45\textwidth}
        \includegraphics[width=\linewidth,trim={0 0 0in 0},clip]{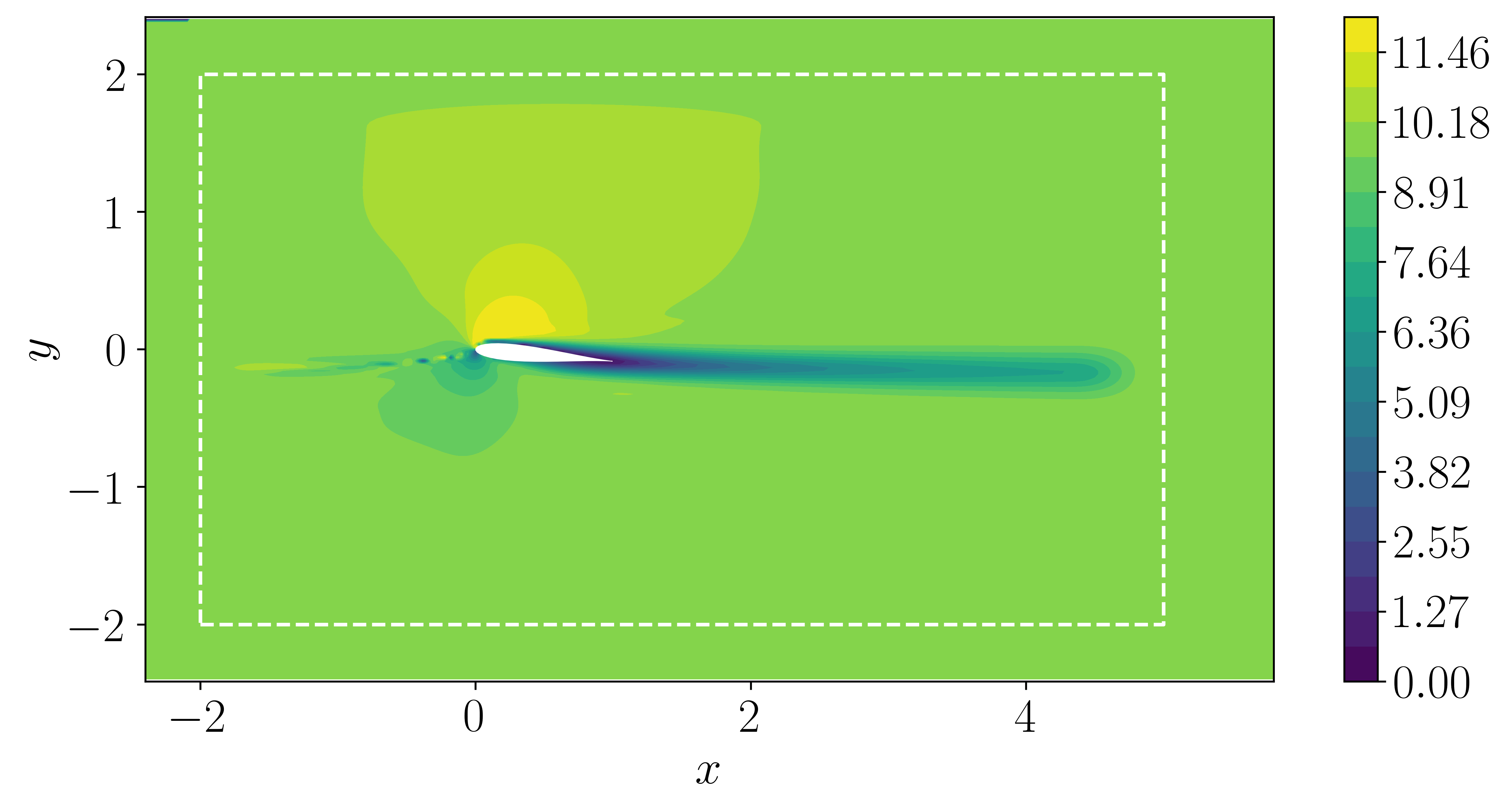}
        \caption{ML prediction, $\left\|\tilde{\bm{u}}\right\|$ with POFU}
   \end{subfigure}
  \begin{subfigure}{0.45\textwidth}
        \includegraphics[width=\linewidth,trim={0 0 0in 0},clip]{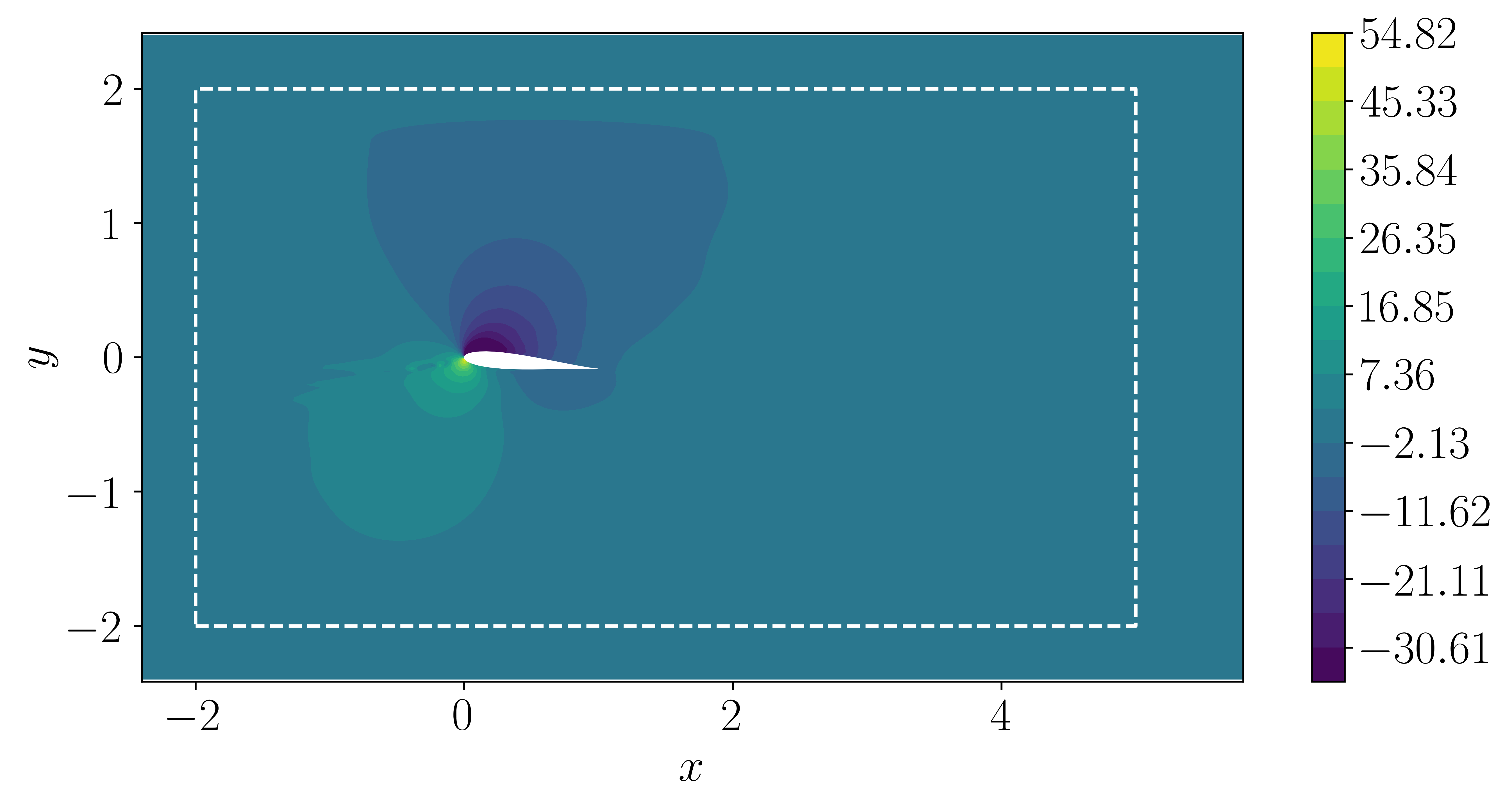}
        \caption{ML prediction, $\tilde{p}$ with POFU}
   \end{subfigure}
    \caption{HF solution (a,b) and ML-predicted fields based on the model trained using AoA=0\dg, evaluated at AoA=5\dg  (c,d). The same POFU extension parameters were used as in Fig. \ref{fig:pofu-airfoil-AoA2}.}
  \label{fig:pofu-airfoil-AoA5}
\end{figure}

CFD initialization using ML prediction is demonstrated in this section. The flow problem of the previous section, an incompressible Navier-Stokes flow around a Joukowski airfoil, was considered. An ML model was trained using 1602 data points from the AoA=0\dg case. The training data points were distributed using the quadtree sampling method, starting from 800 uniformly distributed points followed by three adaptive refinement passes. In order to provide the flow field prediction with an improved accuracy for the CFD initialization, Sobolev training was turned on in this case (i.e., $\losslgrad$ in Eq. (\ref{eqn:data_driven_loss_function_grad}) was included in the loss function). 
CFD meshes were generated as described earlier, and ML-predicted field quantities were evaluated at the mesh nodes and supplied to the solver as the initial guess. The POFU extension method was used to compute the field quantities over the full analysis domain that extended well beyond the ML training window. The algorithm for the ML initialization for CFD runs is provided in Algorithm \ref{alg:mlinit}, Appendix \ref{sec:appendix-alg}. The high-fidelity solution and the ML-predicted fields with POFU extension are shown in Figs. \ref{fig:pofu-airfoil-AoA2} and \ref{fig:pofu-airfoil-AoA5} for the AoA=2\dg  and 5\dg  cases, respectively. The white dotted boxes in Figs. \ref{fig:pofu-airfoil-AoA2} (c,d) and \ref{fig:pofu-airfoil-AoA5} (c,d) indicate the training windows, where the fields transition to the freestream conditions. For the baseline CFD runs, freestream conditions were used to initialize the solution process. The initialization test was repeated for three mesh refinement levels, and ML initialization and baseline results were compared in each case.

\begin{figure}[!ht]
    \centering
   \begin{subfigure}{0.4\textwidth}
        \includegraphics[width=\linewidth]{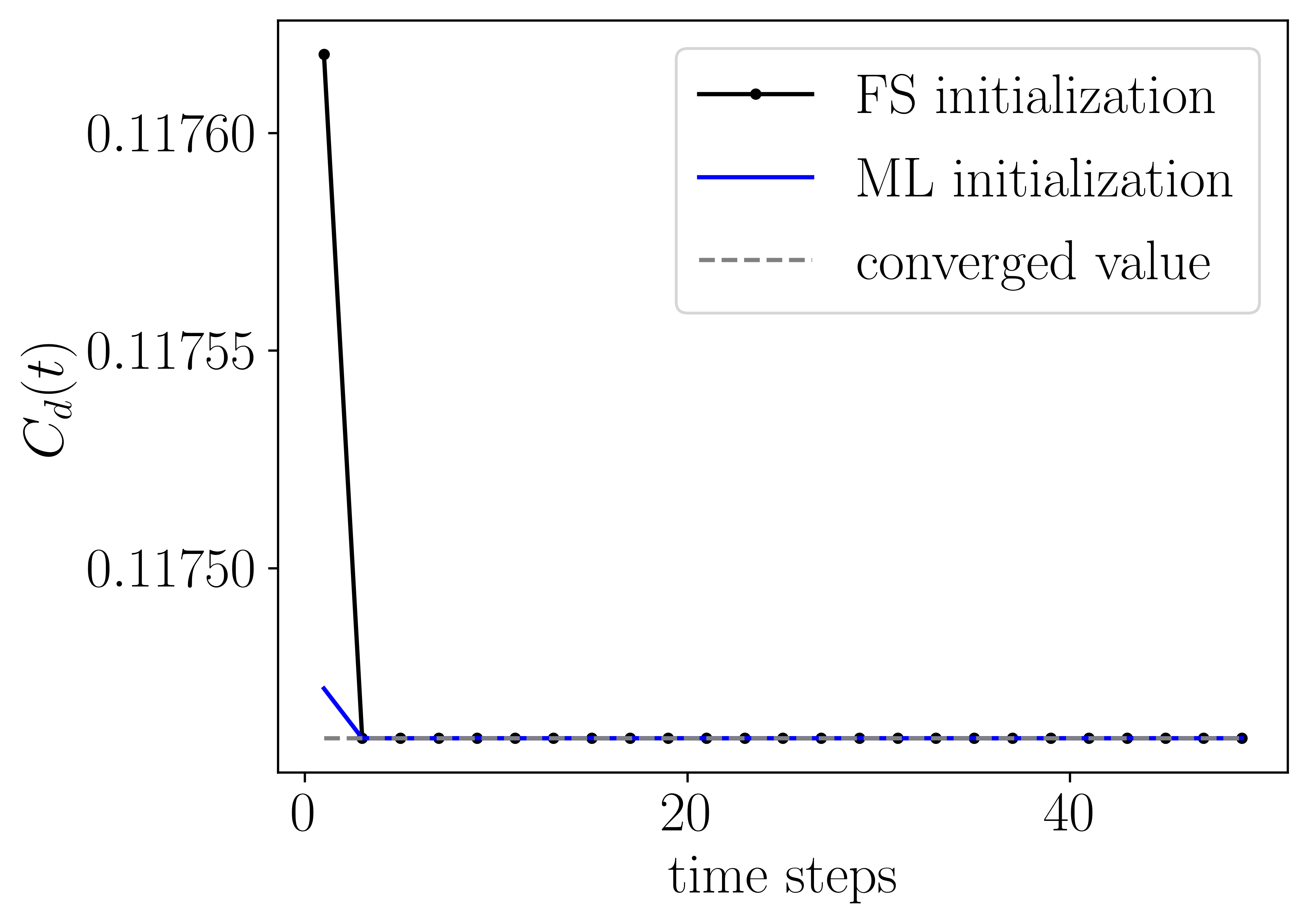}
        \caption{Drag coefficient}
    \end{subfigure}
    \begin{subfigure}{0.4\textwidth}
        \includegraphics[width=\linewidth]{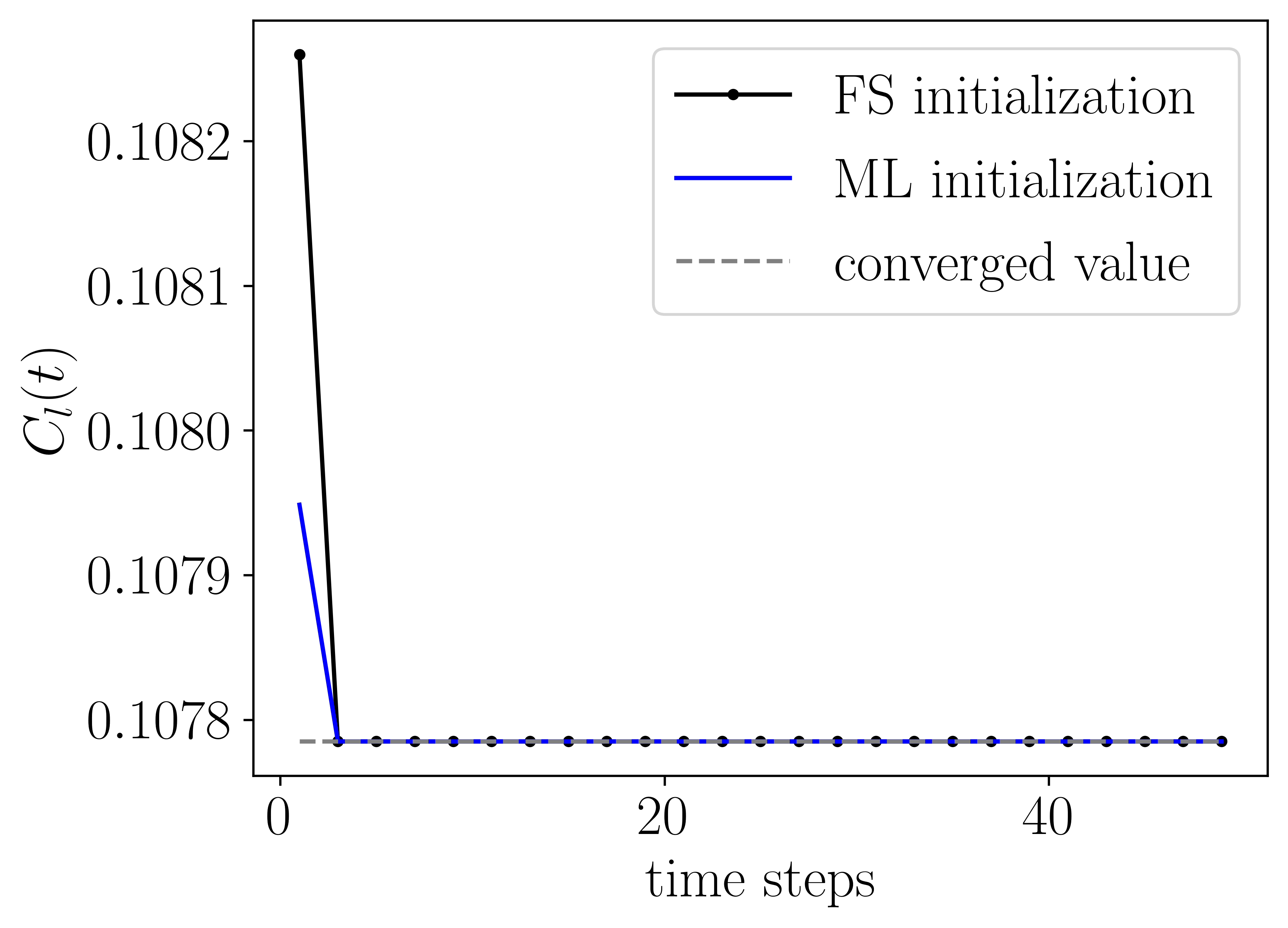}
        \caption{Lift coefficient}
    \end{subfigure}
    \caption{Comparison of forces computed using freestream and ML-predicted fields for CFD initialization for Joukowski airfoil at AoA=2\dg .}
    \label{fig:airfoil-init-AoA2}
\end{figure}

The CFD simulations were carried out using the largest time steps allowed in each case (determined empirically) and a sufficient number of time steps to reach convergence.  The computed drag and lift coefficients, $C_{d}$ and $C_{l}$, over the time steps using the finest mesh are plotted for the AoA=2\dg case in Fig. \ref{fig:airfoil-init-AoA2}. 
In this case, the same time step was used for both runs with ML and freestream initialization, and the convergence plot shows both runs reaching the converged values in just one step. The walltime required for each run was computed by finding the time step at which both drag and lift reached within $1.0\times10^{-6}$ of the converged values (one time step in this case) and reporting the run time. The results of the initialization study across different meshes are summarized in Tab. \ref{tab:acceleration-AoA2} \footnote{A desktop machine with Intel Core i7-4790K processor at 4GHz and 32GB RAM was used for all computations.}. 

In this case, the solver allowed for a large time step, and the simulation converged in one time step in all simulations. The ML initialization led to modest acceleration due to faster convergence within the single nonlinear solve. The drag and lift coefficients from the ML-initialized and baseline runs converge to the same values at each mesh refinement level. 

It is worth reporting that the maximum time step size allowed in CFD runs with ML initialization is dependent on the accuracy of the ML prediction. 
The visual comparison between the high-fidelity solution and the ML prediction in Fig. \ref{fig:pofu-airfoil-AoA2} suggests that a perfect representation of the ML predicted fields is not a requirement to achieve CFD acceleration. However, a number of models we have tested included inaccurate ML field predictions in critical flow regions (i.e., boundary layer) and resulted in slower CFD runs compared to the baseline. A priori determination of how effective a certain ML model may be for CFD acceleration is challenging. The development of an algorithmic method for such determination would require insight into the physics of the flow problem, solver implementations, and the ML model behavior.

\begin{figure}[!ht]
    \centering
   \begin{subfigure}{0.4\textwidth}
        \includegraphics[width=\linewidth]{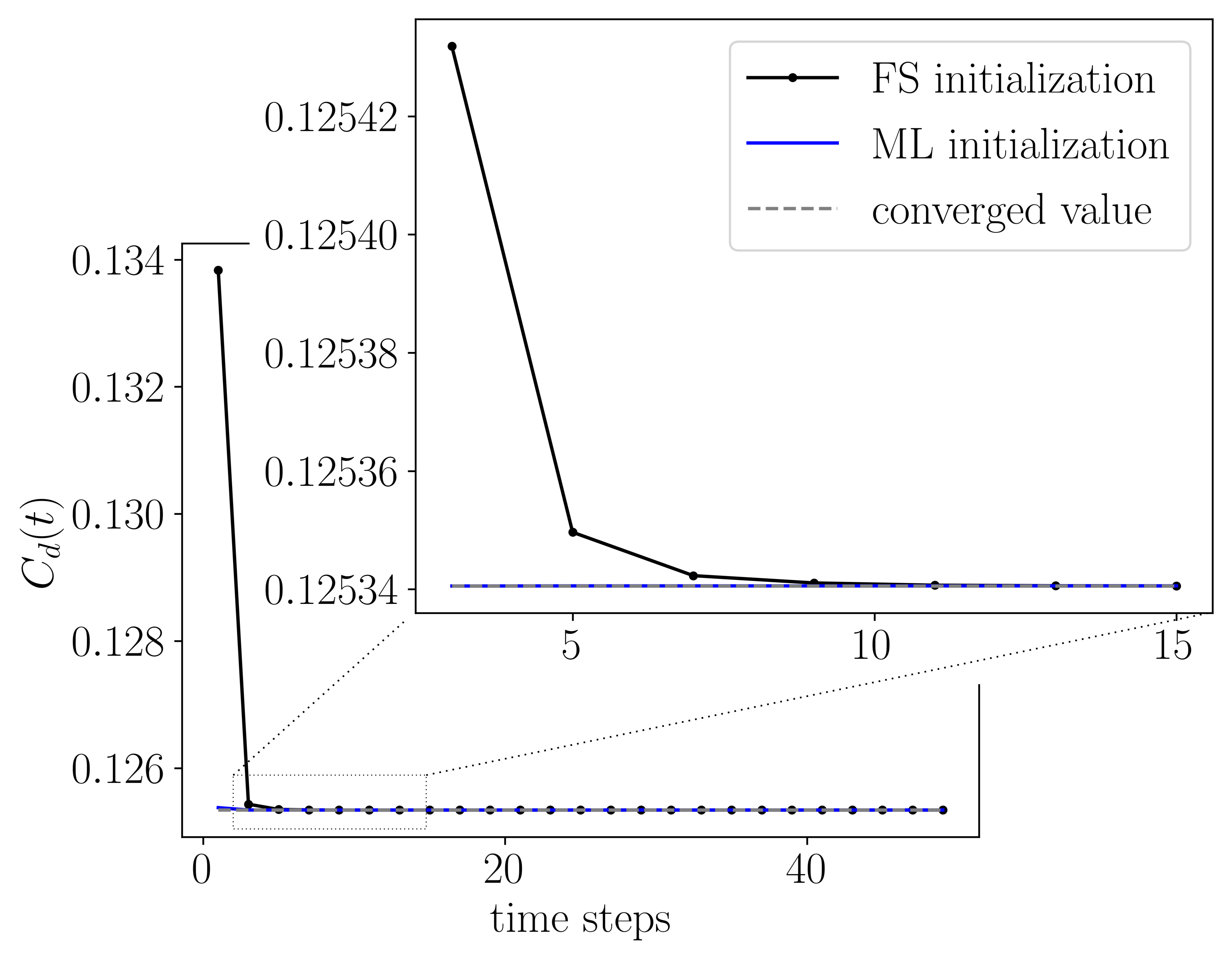}
        \caption{Drag coefficient}
    \end{subfigure}
    \begin{subfigure}{0.4\textwidth}
        \includegraphics[width=\linewidth]{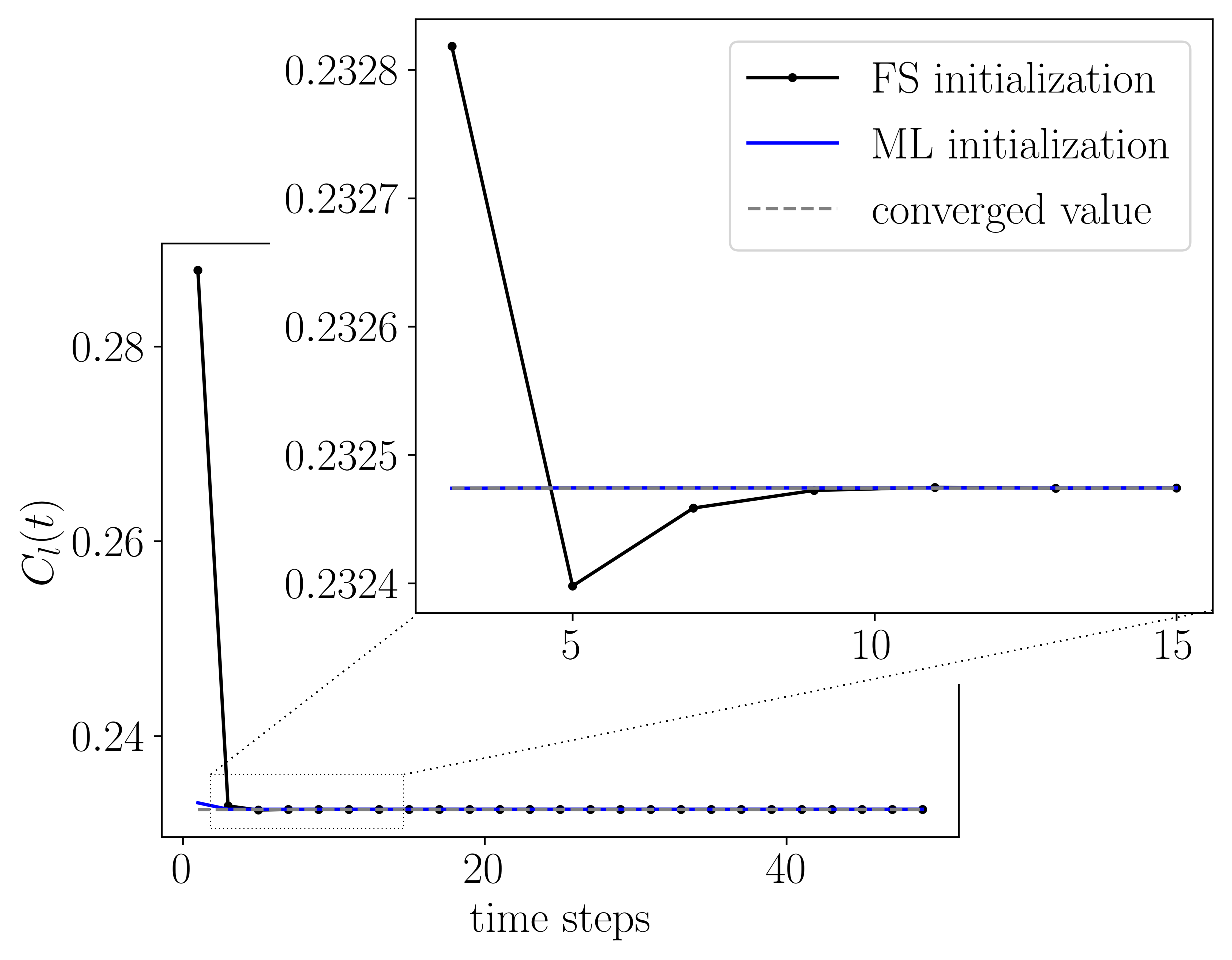}
        \caption{Lift coefficient}
    \end{subfigure}
    \caption{Comparison of forces computed using freestream and ML-predicted fields for CFD initialization for Joukowski airfoil at AoA=5\dg .}
    \label{fig:airfoil-init-AoA5}
\end{figure}

The comparison between the ML and freestream initialized CFD runs for the AoA=5\dg case is shown in Fig. \ref{fig:airfoil-init-AoA5}. These plots refer to the simulations using the finest mesh. 
In this case, the freestream initialization required a smaller time step and a larger number of time steps, resulting in slower convergence compared to the ML initialization. The zoom-in insets in Fig, \ref{fig:airfoil-init-AoA5} highlight the different convergence behaviors. These results along with the results for the two coarser meshes are summarized in Tab. \ref{tab:acceleration-AoA5}. 
Notably, the walltime of the CFD run on the finest mesh using the ML initialization is 42\% that of the freestream initialization run. It is also important to note that the ML initialization for the coarsest mesh led to a longer run time compared to the baseline, due to the small time step size required. We have observed that the ML initialization process is generally more robust at finer mesh.
This may be because ML initialization using a coarse mesh is prone to picking up and exaggerating local errors in the ML prediction. A possible remedy for this issue may be to postprocess the ML prediction to smooth out the flow fields before supplying them to the solver.
More broadly, development of approaches to achieve a highly robust ML initialization process demands a comprehensive consideration of not only the initialization data processing but also mesh, solver parameters, and other factors specific to each solver, and is a focus of our future investigations. 
Nevertheless, the current results demonstrate the utility of an ML model trained using data from a single CFD simulation for the AoA=0\dg case to provide acceleration of simulations at AoA=2\dg and 5\dg.

\begin{table}[!ht]
    \small
    \begin{center}
    \begin{tabular}{ |p{0.7in} |p{0.8in}||p{1in}|p{1in}|p{1in}|  }
    \hline
    initialization & & mesh refinement 1 & mesh refinement 2 & mesh refinement 3\\
    \hline
    {}    & walltime (s) & 1.8 & 6.7 & 31.6 \\
    {}    & time step size & 100 & 100 & 100 \\
    {}    & time steps & 1 & 1 & 1 \\
    {ML}    & $C_{d}$ & 0.1264 & 0.1187 & 0.1175 \\
    {}    & $C_{l}$ & 0.08792 & 0.09637 & 0.1078 \\
    {}    & $C_{d}$ error (\%) & 7.62 & 1.024 & 0.0\\
    {}    & $C_{l}$ error (\%) & 18.43 & 10.59 & 0.0\\
    \hline
    {}    & walltime (s) & 2.5 & 8.7 & 40.5 \\
    {}    & time step size & 100 & 100 & 100 \\
    {}    & time steps & 1 & 1 & 1 \\
    {freestream}    & $C_{d}$ & 0.1264 & 0.1187 & 0.1175 \\
    {}    & $C_{l}$ & 0.08792 & 0.09637 & 0.1078 \\
    {}    & $C_{d}$ error (\%) & 7.62 & 1.024 & 0.0\\
    {}    & $C_{l}$ error (\%) & 18.43 & 10.59 & 0.0\\
    \hline
    \end{tabular}
    \end{center}
        \caption{Comparison for CFD runs using ML and freestream initialization at AoA=2\dg. Relative erros in $C_{d}$ and $C_{l}$ are computed using the freestream-initialized solution at mesh refinement level 3 as reference.}
    \label{tab:acceleration-AoA2}
\end{table}

\begin{table}[!ht]
    \small
    \begin{center}
    \begin{tabular}{ |p{0.7in} |p{0.8in}||p{1in}|p{1in}|p{1in}|  }
    \hline
    initialization & & mesh refinement 1 & mesh refinement 2 & mesh refinement 3\\
    \hline
    {}    & walltime (s) & 12.0 & 7.5 & 36.3 \\
    {}    & time step size & 0.026 & 100 & 100 \\
    {}    & time steps & 24 & 1 & 1 \\
    {ML}    & $C_{d}$ & 0.136 & 0.1271 & 0.1253 \\
    {}    & $C_{l}$ & 0.2034 & 0.2092 & 0.2325 \\
    {}    & $C_{d}$ error (\%) & 8.528 & 1.393 & 0.0\\
    {}    & $C_{l}$ error (\%) & 12.52 & 9.997 & 1.721e-05\\
    \hline
    freestream    & walltime (s) & 2.5 & 8.9 & 85.8 \\
    {}    & time step size & 100 & 100 & 2.1 \\
    {}    & time steps & 1 & 1 & 6 \\
    {}    & $C_{d}$ & 0.1367 & 0.1271 & 0.1253 \\
    {}    & $C_{l}$ & 0.1975 & 0.2092 & 0.2325 \\
    {}    & $C_{d}$ error (\%) & 9.069 & 1.393 & 0.0 \\
    {}    & $C_{l}$ error (\%) & 15.05 & 9.997 & 0.0 \\
    \hline
    \end{tabular}
    \end{center}
    \caption{Comparison for CFD runs using ML and freestream initialization at AoA=5\dg .}
    \label{tab:acceleration-AoA5}
\end{table}

\section{Conclusions}\label{sec:conclusions}
In this work, we have presented an ML approach to learning steady flow fields. Our method constructs input features from the solutions of elliptic BVPs, which gives the ML model natural generalization to nearby geometries, and constructs a multi-fidelity model with an inexpensive low-fidelity solution and an ML additive discrepancy model. Our proposed method, MF-LEIF, was applied to make flow field predictions for the initialization of CFD simulations and demonstrated accelerated solver convergence.

We strived to develop an approach that allowed for ML model training with a small data requirement, both in terms of the number of high-fidelity simulations and the training data size. In this pursuit, pointwise evaluations of high-fidelity solution at one or two reference problems were used to generate the training data. Feature-based quadtree sampling method was used to efficiently distribute a small number of data points within a training window of the flow domain. Sobolev training was adopted to take advantage of the rich information available from the high-fidelity solution and to improve the model accuracy without increasing the data size (i.e., the number of data points). POFU extension was then applied to extend the flow field to a full flow domain for CFD initialization.

Challenges in the deployment of MF-LEIF to more realistic problems include the application of the method to compressible and turbulent flows, which may have further associated demands in CFD initialization. Important future considerations toward the extension of the method to 3D geometries and flows include training cost management with increased data size and accurate and efficient computation of suitable elliptic input features.

\section*{Acknowledgement}
This work was sponsored by the Air Force Office of Scientific Research Computational Mathematics Program (Dr. Fariba Fahroo, Program Officer).

\printbibliography 
\section*{Appendix}\label{sec:appendix}
\appendix
\section{Algorithms}\label{sec:appendix-alg}
The algorithm for each component of the proposed methodology is outlined in this section. The algorithms are ordered here in the logical order of operations and referenced in relevant sections of the main text.

\begin{subalgorithms}
\captionof{algorithm}{Uniform data sampling}\label{alg:uniform}
\begin{algorithmic}[1]

\Input
\Desc{$x_{0},x_{1},y_{0},y_{1}$}{training window corner coordinates}
\Desc{$N_{x},N_{y}$}{number of requested sampling points in $x$- and $y$-directions}
\EndInput
\Output
\Desc{$\sample{\x}$}{coordinates of $N_{p}$ sampled points}
\EndOutput
\State Generate an $N_{x}\times N_{y}$ uniform grid of points inside the training window
\State Remove invalid (e.g., inside solid) points 
\State Return coordinates of valid points $\sample{\x}$ and number of valid sample points $N_{p}$
\end{algorithmic}

\captionof{algorithm}{Random data sampling}\label{alg:random}
\begin{algorithmic}[1]

\Input
\Desc{$x_{0},x_{1},y_{0},y_{1}$}{training window corner coordinates}
\Desc{$N_{p}'$}{number of requested sampling points}
\EndInput
\Output
    \Desc{$\sample{\x}$}{coordinates of $N_{p}$ sampled points}
\EndOutput
\State Prepare randomly distributed points inside the training window
\State Remove invalid (e.g., inside solid) points 
\State Return coordinates of valid points $\sample{\x}$ and number of valid sample points $N_{p}$
\end{algorithmic}

\captionof{algorithm}{Quadtree data sampling}\label{alg:qdtree}
\begin{algorithmic}[1]

\Input
\Desc{$\vecqhf$}{high-fidelity flowfield interpolator}
\Desc{$x_{0},x_{1},y_{0},y_{1}$}{training window corner coordinates}
\Desc{$N_{p}'$}{number of requested sampling points}
\EndInput
\Output
    \Desc{$\sample{\x}$}{coordinates of $N_{p}$ sampled points}
\EndOutput
    \State Initialize quadtree using the training window as a root box or uniform distribution via Algorithm \ref{alg:uniform}
\While{$m < N_{p}'$}
    \For{each valid leaf box $p$}
        \State Extract the box center coordinate $\bm{x}_{p}$ 
        \State Divide the box into four child boxes (quadrants)
        \State Compute the child box center coordinates $\bm{x}_{c1},\cdots,\bm{x}_{c4}$
        \State Compute the refinement/derefinement metrics $d_{c1},\cdots,d_{c4}$ via Eq. (\ref{eq:qdtree})
        \State Reject invalid child boxes (e.g., inside solid) 
    \EndFor
    \State Rank all valid child boxes based on $d_{c}$ values
    \State Refine top $R_{ref}$ percent child boxes and convert to leaf boxes, remove their parent boxes as leaf boxes
    \State Derefine parents of bottom $R_{de}$ percent child boxes and revert its parent to leaf box 
    \If {number of leaf nodes $m \ge N_{p}'$}
        \State {\bf{break}}
    \EndIf
    \State Collect all leaf boxes and report the current number of valid leaf boxes $m$ 
\EndWhile
\State Return coordinates of valid points $\sample{\x}$ from valid leaf box centers and number of valid sample points $N_{p}$
\end{algorithmic}
\end{subalgorithms}

\begin{algorithm}
\captionof{algorithm}{Computation of EIFs}\label{alg:eifs}
\begin{algorithmic}[1]
\Input
    \Desc{$\sample{\x}$}{coordinates of $N_{p}$ sampling points}
\EndInput
\Output
    \Desc{$\sample{\eifs}$}{elliptic input features}
\EndOutput
    \State Compute $\Phi_{i}= \Phi_{i} \atxi$, $\Psi_{i} = \Psi \atxi$, $U_{i}=U\atxi$ for $i-1,\cdots,N_{p}$ by solving Eq. (\ref{eq:eif-phi}-\ref{eq:eif-psi}) or via analytical solutions in Appendix \ref{sec:appendix-PF}
\end{algorithmic}
\end{algorithm}

\begin{algorithm}
\captionof{algorithm}{ML model training}\label{alg:ml-training}
\begin{algorithmic}[1]

\Input
    \Desc{$\Kcases{\sample{\eifs}}$}{elliptic input features from K cases}
    \Desc{$\Kcases{\sample{\delta \vecqhf}}$}{flowfield discrepancy from K cases}

\EndInput
\Output
\Desc{$\delta \vecqml$}{trained ML model}
\EndOutput

\State Stack input training data $\eifs ^{k}$ from all cases, $k=1,\cdots,K$
\State Stack output training data $\delta \vecqhf^{k} = \vecqhf^{k} - \vecqlf^{k}$ from all cases, $k=1,\cdots,K$
\State Apply data scaling to the training data (e.g., Eq. (\ref{eq:minmax-data-scaling}))
\If{$\lossl = \lossldata$}
\State Train ML model $\delta \vecqml$ using loss term defined in Eq. (\ref{eqn:data_driven_loss_function})
\ElsIf{$\lossl = \lossldata + \losslgrad$}
    \State Compute gradients $\nabla \vecqml$ via Eq. (\ref{eqn:spatial_gradient_chain_rule})
\State Train ML model $\delta \vecqml$ using loss term defined in Eq. (\ref{eqn:data_driven_loss_function_grad})
\EndIf

\end{algorithmic}
\end{algorithm}

\begin{algorithm}
\captionof{algorithm}{ML prediction of flowfield}\label{alg:prediction}
\begin{algorithmic}[1]

\Input
    \Desc{$\eval{\x}$}{coordinates of $N_{e}$ evaluation points}
\Desc{$\vecqlf$}{low-fidelity flowfield evaluator}
\Desc{$\delta \vecqml$}{trained ML model}
\EndInput
\Output
    \Desc{$\eval{\vecqml}$}{ML predicted flowfield at $N_{e}$ evaluation points}
\EndOutput
\State Define a flow problem
\State Compute EIFs $\eifs$ via Algorithm \ref{alg:eifs} 
\State Evaluate the ML model $\delta \vecqml \eifs$ 

\State Compute the ML predicted flowfield $\vecqml = \delta \vecqml + \vecqlf$
\end{algorithmic}
\end{algorithm}

\begin{algorithm}
\captionof{algorithm}{POFU extension}\label{alg:pofu}
\begin{algorithmic}[1]

\Input
\Desc{$\vecqml$}{ML predicted flowfield}
\Desc{$\vecqfs$}{vector of freestream values}
\Desc{$x_{0},x_{1},y_{0},y_{1}$}{POFU window corner coordinates}
\Desc{$s_{x},s_{y}$}{POFU transition widths}
\EndInput
\Output
\Desc{$\vecqpofu$}{POFU extended flowfield}
\EndOutput
\State Compute the transition function $T$ via Eq. (\ref{eq:transition-func})
\State Compute the window function $W\left(x,y\right)$ via Eq. (\ref{eq:window-func})
\State Compute the POFU extended flowfield $\vecqpofu$ via Algorithm \ref{alg:prediction} and Eq. (\ref{eqn:pofu_extension})
\end{algorithmic}
\end{algorithm}

\begin{algorithm}
\captionof{algorithm}{ML initialization for CFD run}\label{alg:mlinit}
\begin{algorithmic}[1]

\Input
\Desc{$P$}{flow problem definition}
    \Desc{$\bm{X}_{g}$}{CFD mesh}
\Desc{$\delta \vecqml$}{trained ML model}
\Desc{$\vecqlf$}{low-fidelity flowfield evaluator}
\EndInput
\Output
\Desc{$\vecqhf$}{high-fidelity flowfield}
\EndOutput
\State Compute the POFU extended flowfield $\vecqpofu$ via Algorithm \ref{alg:pofu}
\State Provide $\vecqpofu$ as initial guess to the solver
\State Run CFD to obtain the high-fidelity solution $\vecqhf$
\end{algorithmic}
\end{algorithm}

\section{Derivation of Potential Flow Solutions}\label{sec:appendix-PF}
\subsection{Circular cylinder}
The complex velocity potential corresponding to a uniform fluid flow at constant speed $U_{0}$

\begin{equation}
\Omega\left(z\right)=U_{0}z,
\end{equation}

\noindent
is conformally mapped to obtain the complex potential for a flow around a circular cylinder of radius $a$ as 

\begin{equation}
    \Omega= U_{0}\left(z+\frac{a^{2}}{z}\right) = \Phi + i\Psi.
\end{equation}

\noindent
where $z$ is the complex variable, $z=re^{i\theta}$, and the real and imaginary parts of $\Omega$ represent velocity potential $\Phi$ and streamline $\Psi$ of the flow, respectively. 

\subsection{Joukowski airfoil}
The complex potential of the flow around a symmetric Joukowski airfoil with angle of attack at $\alpha$ is \cite{abbott_theory_2010}:

\begin{equation}\label{eq:potential-joukowski}
\Omega \left(\eta\right) = U_{0}\left[\left(\eta+\epsilon\right)e^{-i\alpha}+\frac{\left(a+\epsilon\right)^{2}e^{i\alpha}}{\eta+\epsilon}\right] + \frac{i\Gamma}{2\pi}\ln{\frac{\left(\eta+\epsilon\right)e^{-i\alpha}}{a+\epsilon}},
\end{equation}

\noindent where

\begin{equation}
    \Gamma = 4\pi U_{0} \left(a + \epsilon \right)\sin{\alpha},
\end{equation}

\noindent
and

\begin{equation}
\xi=\eta+\frac{1}{\eta} ~\Rightarrow~ \eta = \frac{1}{2}\left[\xi+e^{\frac{1}{2}\ln{\left(\xi-2\right)}}e^{\frac{1}{2}\ln{\left(\xi+2\right)}}\right].
\end{equation}

\noindent
A coordinate transformation $\xi=\xi\left(z\right)$ in Eq. (\ref{eq:xi-transform}) is used to shift and scale the airfoil such that the leading and trailing edges are located at 0 and 1.

\begin{equation}\label{eq:xi-transform}
\xi = C_{\xi}z+\xi_{L},
\end{equation}

\noindent with
\begin{equation}
\begin{array}{l}
\xi_{L}=-a-2\epsilon-\frac{a^{2}}{a+2\epsilon}, \\
\xi_{R}=2a, \\
C_{\xi}=\xi_{R}-\xi_{L}=3a+2\epsilon+\frac{a^{2}}{a+2\epsilon}.
\end{array}
\end{equation}

\end{document}